\documentclass[sigconf, nonacm]{acmart}

\usepackage[ruled,vlined,linesnumbered]{algorithm2e}
\usepackage{xspace}
\usepackage{amsthm,amssymb,amsmath,epsfig,graphics,enumerate,float,subfigure,bm}
\usepackage{epstopdf}
\usepackage{svg}
\usepackage{balance}
\usepackage{color}
\usepackage{setspace}

\newcommand\vldbdoi{XX.XX/XXX.XX}
\newcommand\vldbpages{XXX-XXX}
\newcommand\vldbvolume{14}
\newcommand\vldbissue{1}
\newcommand\vldbyear{2020}
\newcommand\vldbauthors{\authors}
\newcommand\vldbtitle{\shorttitle} 
\newcommand\vldbavailabilityurl{URL_TO_YOUR_ARTIFACTS}
\newcommand\vldbpagestyle{plain} 

\newcommand{\ignore}[1]{}
\newcommand{\nop}[1]{}
\newcommand{\eat}[1]{}
\newcommand{\kw}[1]{{\ensuremath{\mathsf{#1}}}\xspace}

\newcommand{\stitle}[1]{\vspace{1ex} \noindent{\bf #1}}
\long\def\comment#1{}

\newtheorem{corollary}{Corollary}

\newcommand{\approxkemeny}{\kw{ApproxKemeny}}
\newcommand{\rw}{\kw{RW}}
\newcommand{\lewalk}{\kw{LEWalk}}
\newcommand{\spanningtree}{\kw{SpanTree}}
\newcommand{\diagonal}{\kw{Diagonal}}

\newcommand{ \degree}{\kw{Degree}}

\newcommand{\pagerank}{\kw{PageRank}}
\newcommand{ \bc}{\kw{BC}}
\newcommand{ \cc}{\kw{CC}}
\newcommand{ \ecc}{\kw{ECC}}

\newcommand{ \ba}{\kw{BA}}
\newcommand{ \er}{\kw{ER}}
\newcommand{ \hy}{\kw{HY}}

\newcommand{ \topstack}{\kw{Top}}

\newcommand{ \kc}{\kw{KC}}
\newcommand{ \ps}{\kw{ps}}
\newcommand{ \vis}{\kw{vis}}
\newcommand{ \fin}{\kw{fin}}

\newcommand{ \dfs}{\kw{DFS}}
\newcommand{ \bfs}{\kw{BFS}}

\newcommand{ \lepath}{\kw{LE}}
\newcommand{ \nextarray}{\kw{next}}
\newcommand{ \intree}{\kw{InTree}}
\newcommand{ \false}{\kw{false}}
\newcommand{ \true}{\kw{true}}
\newcommand{ \randomneighbor}{\kw{RandomNeighbor}}

\newcommand{\wordnet}{\kw{WordNet}}
\newcommand{\orkut}{\kw{Orkut}}
\newcommand{\youtube}{\kw{Youtube}}
\newcommand{\livejournal}{\kw{LiveJournal}}
\newcommand{\pokec}{\kw{Pokec}}

\newcommand{\dblp}{\kw{DBLP}}
\newcommand{\amazon}{\kw{Amazon}}

\newcommand{\hepth}{\kw{Hep\textrm{-}\xspace th}}
\newcommand{\astroph}{\kw{Astro\textrm{-}\xspace ph}}

\newcommand{\emailenron}{\kw{Email\textrm{-}\xspace enron}}

\begin{document}
\title{Scalable Algorithms for Laplacian Pseudo-inverse Computation}

\author{
    {Meihao Liao\texorpdfstring{$^\dagger$},,
    Rong-Hua Li\texorpdfstring{$^\dagger$},,
    Qiangqiang Dai\texorpdfstring{$^\dagger$},,
    Hongyang Chen\texorpdfstring{$^{\ddagger}$},,  
    Guoren Wang\texorpdfstring{$^\dagger$},}
}
\affiliation{
	{\texorpdfstring{$^\dagger$} BBeijing Institute of Technology}
	\country{China}
    {\texorpdfstring{$^\ddagger$} ZZhejiang Lab}
	\country{China}
}
\email{
	{mhliao@bit.edu.cn,
    lironghuabit@126.com,
    qiangd66@gmail.com,
    dr.h.chen@ieee.org,
    wanggrbit@gmail.com}
}

\begin{abstract}
The pseudo-inverse of a graph Laplacian matrix, denoted as $L^\dagger$, finds extensive application in various graph analysis tasks. Notable examples include the calculation of electrical closeness centrality, determination of Kemeny's constant, and evaluation of resistance distance. However, existing algorithms for computing $L^\dagger$ are often computationally expensive when dealing with large graphs. To overcome this challenge, we propose novel solutions for approximating $L^\dagger$ by establishing a connection with the inverse of a Laplacian submatrix $L_v$. This submatrix is obtained by removing the $v$-th row and column from the original Laplacian matrix $L$. The key advantage of this connection is that $L_v^{-1}$ exhibits various interesting combinatorial interpretations. We present two innovative interpretations of $L_v^{-1}$ based on spanning trees and loop-erased random walks, which allow us to develop efficient sampling algorithms. Building upon these new theoretical insights, we propose two novel algorithms for efficiently approximating both electrical closeness centrality and Kemeny's constant. We extensively evaluate the performance of our algorithms on five real-life datasets. The results demonstrate that our novel approaches significantly outperform the state-of-the-art methods by several orders of magnitude in terms of both running time and estimation errors for these two graph analysis tasks. To further illustrate the effectiveness of electrical closeness centrality and Kemeny's constant, we present two case studies that showcase the practical applications of these metrics.
\end{abstract}

\maketitle

\pagestyle{\vldbpagestyle}
\begingroup\small\noindent\raggedright\textbf{PVLDB Reference Format:}\\
\vldbauthors. \vldbtitle. PVLDB, \vldbvolume(\vldbissue): \vldbpages, \vldbyear.
\href{https://doi.org/\vldbdoi}{doi:\vldbdoi}
\endgroup
\begingroup
\renewcommand\thefootnote{}\footnote{\noindent
This work is licensed under the Creative Commons BY-NC-ND 4.0 International License. Visit \url{https://creativecommons.org/licenses/by-nc-nd/4.0/} to view a copy of this license. For any use beyond those covered by this license, obtain permission by emailing \href{mailto:info@vldb.org}{info@vldb.org}. Copyright is held by the owner/author(s). Publication rights licensed to the VLDB Endowment. \\
\raggedright Proceedings of the VLDB Endowment, Vol. \vldbvolume, No. \vldbissue\ %
ISSN 2150-8097. \\
\href{https://doi.org/\vldbdoi}{doi:\vldbdoi} \\
}\addtocounter{footnote}{-1}\endgroup

\ifdefempty{\vldbavailabilityurl}{}{
\vspace{.3cm}
\begingroup\small\noindent\raggedright\textbf{PVLDB Artifact Availability:}\\
The source code, data, and/or other artifacts have been made available at \url{https://anonymous.4open.science/r/KC-vldb-2223}.
\endgroup
}

\section{Introduction}\label{sec:intro}
Given an $n$ node graph $G=(V,E)$ with an adjacency matrix $A$ and a diagonal degree matrix $D$ (each diagonal element is a degree of a node), two types of widely-used Laplacians are defined: the combinatorial Laplacian $L=D-A$ and the normalized Laplacian $\mathcal{L}=I-D^{-\frac{1}{2}}AD^{-\frac{1}{2}}$. Since the inverses of $L$ and $\mathcal{L}$ do not exist, previous studies often employ the Moore-Penrose pseudo-inverses $L^\dagger$ and $\mathcal{L}^\dagger$ \cite{pseudoinverse17, BozzoF12,angriman2020approximation} as alternatives. These Laplacian pseudo-inverses have proven to be crucial in various graph analysis applications, enabling the computation of important quantities such as electrical closeness centrality \cite{BozzoF12}, Kemeny's constant \cite{whyconstant}, Kirchhoff index \cite{kirchhoff08}, resistance distance \cite{22resistance, ResistanceYang}, hitting time \cite{chung2023forest, TreeFormula18, chebotarev2002forest}, and generalized graph spectral distance \cite{SpectralDistance}. In the database community, these random walk-based quantities are also widely used in designing graph embedding system \cite{embedding-sigmod21,embeddability-vldb13}, clustering in geo-social networks \cite{density-clustering-sigmod14}, long tail recommendation \cite{long-tail-recommendation-vldb12} and anomalous detection in time varying graphs \cite{anomalous-sigmod14}.

Among the various applications of Laplacian pseudo-inverses, we mainly focus on two typical applications: graph centrality and graph invariants. Graph centrality is used to quantify the importance of nodes within a network. Noteworthy centrality measures include PageRank \cite{15pagerankbeyond,11google}, eigenvector centrality \cite{01eigenvector}, closeness centrality \cite{closeness08} and electrical closeness centrality \cite{04information, 13resistance}. Among these centrality metrics, computing electrical closeness centrality is equivalent to computing the diagonal elements of the combinatorial Laplacian pseudo-inverse $L^\dagger$ \cite{13information,13resistance}. Unlike centrality measures based on shortest path distance (e.g., closeness centrality), electrical closeness centrality incorporates resistance distance between nodes \cite{13information,13resistance}. Since it considers all paths between nodes, the resulting centrality metric is often more robust. Its effectiveness has been demonstrated in diverse applications such as power grid vulnerability \cite{10powergrid} and recommendation systems \cite{07recommendation}. It has also been employed to measure the graph variance and graph curvature from a geometric perspective \cite{GraphVariance,SimplexGeometry}. Graph invariant, on the other hand, characterizes global properties of the entire graph. A notable graph invariant is Kemeny's constant \cite{whyconstant}, which corresponds to the trace of the pseudo-inverse of the normalized Laplacian matrix $\mathcal{L}^\dagger$. Kemeny's constant captures the extent of inter-connectivity among nodes in a graph. It finds broad applications in domains including robot surveillance \cite{20robot} and web search \cite{02surfer}.

However, the computation of Laplacian pseudo-inverse poses significant challenges when dealing with large graphs. The direct method involves eigen-decomposition, which has a time complexity of $O(n^3)$. Previous studies have formulated $L^\dagger$ as $L^\dagger=(L+\frac{\vec{1}\vec{1}^T}{n})^{-1}-\frac{\vec{1}\vec{1}^T}{n}$ \cite{BozzoF12,pseudoinverse17}. However, $L+\frac{\vec{1}\vec{1}^T}{n}$ becomes dense, making the computation of its inverse challenging. Additionally, for each column of $L^\dagger$, the state-of-the-art (SOTA) method \cite{pseudoinverse17} have proved that the $s$-th column of $L^\dagger$ is the solution of the Laplacian linear system $Lx=e_s-\frac{\vec{1}}{n}$, where $e_s$ is the unit vector with a value of $1$ in the $s$-th position and zeros elsewhere, and $\vec{1}$ denotes an all-one vector. However, solving this Laplacian system remains computationally expensive for large networks. Moreover, computing electrical closeness centrality and Kemeny's constant requires solving $n$ Laplacian systems, which is extremely expensive for large graphs.

In this paper, we address the challenge of computing Laplacian pseudo-inverse by establishing a connection with $L_v^{-1}$, the inverse of a submatrix obtained by removing the $v$-th row and column from the Laplacian matrix $L$. The formulation of $L_v$ offers three advantages: (i) $L_v$ preserves the sparsity of $L$, enabling efficient computation of its inverse; (ii) While there is a lack of intuitive combinatorial interpretations for the elements of $L^\dagger$ or $\mathcal{L}^\dagger$ in the existing literature, several intriguing combinatorial interpretations exist for the elements of $L_v^{-1}$ \cite{bapat2010graphs,22resistance,chaiken1982combinatorial}; (iii) As both $L^\dagger$ and $\mathcal{L}^\dagger$ remain invariants when varying the choice of $v$, we can heuristically select an optimal landmark node $v$ to further enhance the efficiency of the computation. Specifically, we present new formulas to express the elements of the Laplacian pseudo-inverse, $L^\dagger$ and $\mathcal{L}^\dagger$, in terms of $L_v^{-1}$. Additionally, we propose two novel combinatorial interpretations for the elements of $L_v^{-1}$ based on spanning trees and loop-erased walks. Since spanning trees can be sampled using loop-erased walks, these new interpretations of $L_v^{-1}$ enable the estimation of the elements of $L^\dagger$ and $\mathcal{L}^\dagger$ within the time required to sample a set of loop-erased walks.

We apply our newly-developed techniques to tackle two graph analysis tasks: electrical closeness centrality computation, specifically targeting the approximation of the diagonal elements of the pseudo-inverse of the combinatorial Laplacian matrix $(L^\dagger)_{uu}$ for each node $u\in V$, and Kemeny's constant computation, specifically aiming to approximate the trace of the pseudo-inverse of the normalized Laplacian matrix, denoted as $Tr(\mathcal{L}^\dagger)$. For both computational tasks, we propose two novel algorithms based on spanning tree sampling and loop-erased walk sampling. These algorithms demonstrate superior efficiency when compared to the SOTA approaches for approximating both electrical closeness centrality and Kemeny's constant.

We conduct extensive experiments on five real-world networks to evaluate the performance of the proposed algorithms for the two computational tasks. The experimental results demonstrate that our algorithms outperform the SOTA approaches in terms of both running time and estimation error. Regarding the approximation of electrical closeness centrality, our spanning tree sampling algorithm exhibits competitive and slightly better performance compared to the SOTA methods, while our loop-erased walk sampling algorithm achieves much lower execution times with similar estimation error. For the computation of Kemeny's constant, our loop-erased walk based algorithms demonstrate significant improvements. They are around two orders of magnitude faster than the SOTA algorithms, while maintaining a comparable level of estimation error. Additionally, we conduct case studies to demonstrate the effectiveness of employing Laplacian pseudo-inverse in graph centrality and graph invariant measures. Our contributions are summarized as follows:

\stitle{New theoretical results.} We present new formulas that express $L^\dagger$ and $\mathcal{L}^\dagger$ in terms of $L_v^{-1}$. Additionally, we introduce two novel combinatorial interpretations of $L_v^{-1}$ based on spanning trees and loop-erased walks. These theoretical results provide valuable insights and can be of independent interests to the research community.

\stitle{Novel algorithms.} We propose two innovative algorithms for electrical closeness centrality approximation, employing spanning tree sampling and loop-erased walk sampling techniques. Similarly, we develop two novel algorithms for approximating Kemeny's constant, utilizing spanning tree and loop-erased walk sampling approaches. Notably, the time complexity of drawing a sample in all the proposed algorithms is $Tr((I-P_v)^{-1})$, where $P_v$ is the matrix obtained by removing the $v$-th row and column of the probability transition matrix $P=D^{-1}A$. As verified in our experiments, $Tr((I-P_v)^{-1})$ is $O(n)$ for real-world graphs, thus our sampling approaches are very efficient. Furthermore, we provide rigorous analysis of these algorithms to demonstrate their efficiency and effectiveness.

\stitle{Extensive experiments.} We conduct extensive experiments using five real-life datasets to evaluate the performance of our algorithms. The results show that our algorithms significantly outperform the SOTA approaches in terms of both running time and estimation error. For example, for computing Kemeny's constant on a million-node graph \youtube, to achieve a relative error $10^{-4}$, the SOTA method requires $34,637$ seconds, while our algorithm \lewalk takes only $1,615$ seconds, resulting in a $22\times$ speed improvement. Additionally, we present two case studies to showcase the effectiveness of electrical closeness centrality and Kemeny's constant. The results demonstrate that electrical closeness centrality serves as a reliable and robust centrality metric. Furthermore, Kemeny's constant proves to be capable of distinguishing different types of graphs and offering insights for network structure design. For reproducibility purpose, we make the source code of our work available at \url{https://anonymous.4open.science/r/KC-vldb-2223}.

\section{Preliminaries}\label{sec:preliminaries}
Consider an unweighted, undirected connected graph denoted as $G = (V, E)$, where $|V| = n$ denotes the number of nodes and $|E| = m$ represents the number of edges. Let $A$ be the adjacency matrix of $G$, where $(A)_{ij} = 1$ if there exists an edge between nodes $i$ and $j$. Additionally, we define the degree matrix $D$ as a diagonal matrix with $(D)_{ii} = d_i$, representing the degree of node $i$. To define a simple random walk on $G$, we use the probability transition matrix $P = D^{-1}A$. In each step, the random surfer jumps to a neighbor of the current node $u$ with a probability of $\frac{1}{d_u}$, where $d_u$ is the degree of node $u$. It is well-known that the stationary distribution of such a random walk can be represented by the vector $\bm{\pi}$, where $\bm{\pi}(i) = \frac{d_i}{\sum_{u \in V} d_u} = \frac{d_i}{2m}$. This stationary distribution satisfies the properties of a probability distribution, namely, $\sum_{u \in V} \bm{\pi}(u) = 1$. Furthermore, it satisfies the equation $\bm{\pi}^T P = \bm{\pi}^T$.

In our analysis, we consider two types of Laplacian matrices: the combinatorial Laplacian $L = D - A$ and the normalized Laplacian $\mathcal{L} = I - D^{-\frac{1}{2}}AD^{-\frac{1}{2}}$. Both Laplacian matrices are singular matrices of rank $n-1$. By employing eigen-decomposition, we can express $L$ as $L = \sum_{i=1}^n{\lambda_iu_iu_i^T}$, where $\lambda_1 \leq \cdots \leq \lambda_n$, and $\mathcal{L}$ as $\mathcal{L} = \sum_{i=1}^n{\sigma_i\tilde{u}i\tilde{u}i^T}$, where $\sigma_1 \leq \cdots \leq \sigma_n$. It is important to note that both matrices have $\lambda_1 = \sigma_1 = 0$. Consequently, the inverses of the Laplacian matrices do not exist, and instead, we can alternatively use Laplacian pseudo-inverses. Among various Laplacian pseudo-inverses, the Moore-Penrose pseudo-inverse is of significant importance, as it is unique for a given Laplacian matrix. Specifically, it is defined as $L^\dagger = \sum_{i=2}^n{\frac{1}{\lambda_i}u_iu_i^T}$ for $L$, and $\mathcal{L}^\dagger = \sum_{i=2}^n{\frac{1}{\sigma_i}\tilde{u}_i\tilde{u}_i^T}$ for $\mathcal{L}$. For convenience, in the rest of this paper, we refer to the Moore-Penrose pseudo-inverse of Laplacian as Laplacian pseudo-inverse. The Laplacian pseudo-inverses find wide applications. In this paper, we mainly focus on two applications: electrical closeness centrality computation and Kemeny's constant computation, while other Laplacian pseudo-inverse related quantities (e.g., Kirchhoff index \cite{kirchhoff08}, normalized random walk betweenness centrality \cite{RandomWalkBetweeness}) can also be directly approximated by the proposed techniques (See discussions at the end of Section~\ref{sec:kcc}).

\stitle{Electrical closeness centrality (\ecc).} Resistance distance measures the distance between nodes in a network by treating it as an electrical system. The resistance distance between node $s$ and $t$ is defined as $r(s,t)=(L^\dagger)_{ss}+(L^\dagger)_{tt}-2(L^\dagger)_{st}$. It is a distance metric on graphs. Compared to the shortest path distance, resistance distance takes into account all paths between nodes, making it more robust. The electrical closeness centrality $\mathbf{c}(u)$, then, is defined as $\frac{n-1}{\sum_{v\in V}r(u,v)}=\frac{n-1}{Tr(L^\dagger)+n(L^\dagger)_{uu}}$ \cite{BozzoF12}. The term $Tr(L^\dagger)$ is the same for all nodes $u\in V$. Consequently, the larger the value of $(L^\dagger){uu}$, the smaller the value of $c(u)$. Therefore, computing the diagonal elements $(L^\dagger)_{uu}$ is a key problem in determining the electrical closeness centrality.

\stitle{Kemeny's constant (\kc).} The hitting time $h(s,u)$ between node $s$ and $u$ is the expected length for a random walk starts from $s$ to hit $u$. Kemeny's constant, denoted as $\kappa(G)$, is then defined as $\kappa(G)=\sum_{u\in V}{h(s,u)\bm{\pi}(u)}$. In other words, Kemeny's constant represents the expected length of a random walk from a source node to a randomly-chosen target node in a graph \cite{whyconstant}. It is well-known that the Kemeny's constant is a constant for every source node $s$ and can be computed as $Tr(\mathcal{L}^\dagger)=\sum_{i=2}^n{\frac{1}{\sigma_i}}=\sum_{i=2}^n{\frac{1}{1-\lambda_i}}$ \cite{KDD21Kemeny}. Kemeny's constant provides important insights into the connectivity of a graph. A smaller Kemeny's constant implies closer connectivity between nodes in the graph.
\section{New theoretical results}\label{sec:new-results}
\comment{
\begin{figure}[t!] \vspace*{-0.2cm}
	\color{blue}
	\begin{center}
		\includegraphics[width=0.74\columnwidth, height=2.6cm]{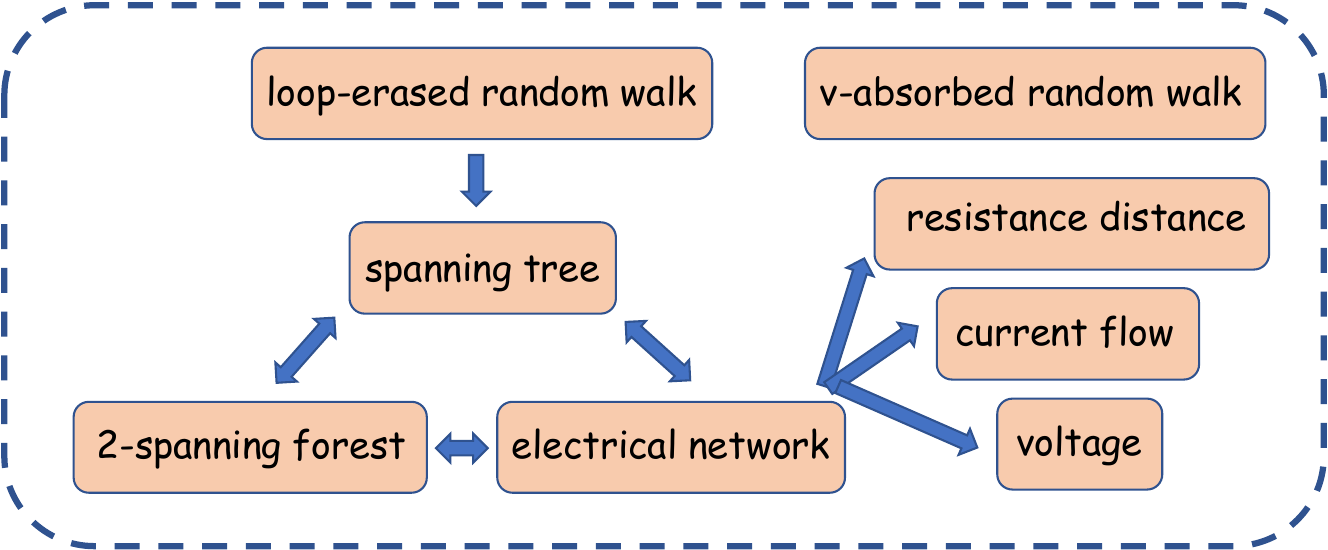}
	\end{center}
	\vspace*{-0.3cm}
	\caption{A colorful map of Laplacian pseudo-inverse.}
	\vspace*{-0.3cm}
	\label{fig:proof_outline}
\end{figure}}
In this section, we propose a series of new theoretical results. First, we give new formulas of $L^\dagger$ and $\mathcal{L}^\dagger$ by relating them to $L_v^{-1}$. The results show that in order to compute $L^\dagger$ and $\mathcal{L}^\dagger$, it suffices to compute $L_v^{-1}$. Thus, we focus on studying $L_v^{-1}$. We review two existing theoretical interpretations of $L_v^{-1}$ in terms of $v$-absorbed walk and electrical network. However, it is very costly to sample $v$-absorbed walks and compute an electrical system. To overcome this challenge, we propose two novel combinatorial interpretations of $L_v^{-1}$ in terms of spanning trees and loop-erased walks, which are easier to sample. The new point view of $L_v^{-1}$ is utilized to design efficient algorithms for approximating \ecc and \kc in Section~\ref{sec:ecc} and Section~\ref{sec:kcc} respectively.
\subsection{Relating Laplacian pseudo-inverses to $L_v^{-1}$}
Since the inverse of $L$ does not exist, the analysis of a Moore-Penrose pseudo-inverse is much harder than a normal matrix inverse. Previous studies are often based on the formula $L^\dagger=(L+\frac{\vec{1}\vec{1}^T}{n})^{-1}-\frac{\vec{1}\vec{1}^T}{n}$ \cite{BozzoF12}. However, $L+\frac{\vec{1}\vec{1}^T}{n}$ becomes a dense matrix that it is hard to obtain its inverse, and there is no obvious physical meaning of $(L+\frac{\vec{1}\vec{1}^T}{n})^{-1}$. To address this issue, we consider another important matrix $L_v$ to study the properties of the Laplacian pseudo-inverse. Specifically,  $L_v$ is a submatrix of $L$ obtained by removing the $v$-th row and column from $L$, which preserves the sparsity of $L$. It is important to note that unlike $L$, the inverse of $L_v$ exists. Moreover, we find that $L_v^{-1}$ has several interesting combinatorial interpretations, and $L^\dagger$ can also be formulated by $L_v^{-1}$. Since $L^\dagger$ is independent of the choice of $v$, we can heuristically choose $v$ as an easy-to-hit landmark node (e.g., the highest-degree node) to accelerate the computation of $L^\dagger$. Due to the space limit, all the missing proofs can be found in the full version of the paper \cite{fullversion}.

\begin{theorem}\label{theorem:L_dagger_L_v}
	Let $\vec{1}$ be an $(n-1)\times1$ all-one vector, we have: 
	\begin{align*}
		L^\dagger=\left(\begin{matrix}
			I-\frac{1}{n}\vec{1}\vec{1}^T \\
			-\frac{1}{n}\vec{1}^T
		\end{matrix}\right)L_v^{-1}\left(\begin{matrix}
			I-\frac{1}{n}\vec{1}\vec{1}^T & -\frac{1}{n}\vec{1}
		\end{matrix}\right).
	\end{align*}
\comment{$L^\dagger=\left(\begin{matrix}
		(L^\dagger)_v & -\Vec{\mathbf{l}_v} \\
		\Vec{\mathbf{l}_v} & (L^\dagger)_{vv}
	\end{matrix}\right)=\left(\begin{matrix}
		I-\frac{1}{n}\vec{1}\vec{1}^T \\
		-\frac{1}{n}\vec{1}^T
	\end{matrix}\right)L_v^{-1}\left(\begin{matrix}
		I-\frac{1}{n}\vec{1}\vec{1}^T & -\frac{1}{n}\vec{1}
	\end{matrix}\right)$.}
\comment{
	\begin{proof}
		Since $L\vec{1}=0$, every column and row of $L$ sums up to $1$. Thus, the $v$-th row of $L$ can be represented as $[-L_v\vec{1}; \vec{1}^TL_v\vec{1}]^T$. As a result, $L$ can also be written as $L=\left(\begin{matrix}
			I \\
			-\vec{1}^T
		\end{matrix}\right)L_v\left(\begin{matrix}
			I & -\vec{1}
		\end{matrix}\right)$. It is easy to verify that the matrix satisfies the four properties of the unique Moore-Penrose pseudo-inverse: (in the following formulas, the dimension of all-one matrix and identity matrix are adapted to their representations)
		
		(i) $(L^\dagger L)^T=L^\dagger L$; (ii) $(LL^\dagger)^T=LL^\dagger$:
		\begin{align*}
			L^\dagger L&=\left(\begin{matrix}
				I-\frac{1}{n}\vec{1}\vec{1}^T \\
				-\frac{1}{n}\vec{1}^T
			\end{matrix}\right)L_v^{-1}\left(\begin{matrix}
				I-\frac{1}{n}\vec{1}\vec{1}^T & -\frac{1}{n}\vec{1}
			\end{matrix}\right)\left(\begin{matrix}
				I \\
				-\vec{1}^T
			\end{matrix}\right)L_v\left(\begin{matrix}
				I & -\vec{1}
			\end{matrix}\right)\\
			&=\left(\begin{matrix}
				I-\frac{1}{n}\vec{1}\vec{1}^T \\
				-\frac{1}{n}\vec{1}^T
			\end{matrix}\right)\left(\begin{matrix}
			I & -\vec{1}
		\end{matrix}\right)=I-\frac{1}{n}\vec{1}\vec{1}^T,
		\end{align*}
		\begin{align*}
			LL^\dagger&=\left(\begin{matrix}
				I \\
				-\vec{1}^T
			\end{matrix}\right)L_v\left(\begin{matrix}
				I & -\vec{1}
			\end{matrix}\right)\left(\begin{matrix}
				I-\frac{1}{n}\vec{1}\vec{1}^T \\
				-\frac{1}{n}\vec{1}^T
			\end{matrix}\right)L_v^{-1}\left(\begin{matrix}
				I-\frac{1}{n}\vec{1}\vec{1}^T & -\frac{1}{n}\vec{1}
			\end{matrix}\right)\\
			&=\left(\begin{matrix}
				I \\
				-\vec{1}^T
			\end{matrix}\right)\left(\begin{matrix}
			I-\frac{1}{n}\vec{1}\vec{1}^T & -\frac{1}{n}\vec{1}
		\end{matrix}\right)=I-\frac{1}{n}\vec{1}\vec{1}^T.
		\end{align*}
		
		Since $LL^\dagger=L^\dagger L=I-\frac{1}{n}\vec{1}\vec{1}^T$, the conditions (i) and (ii) are both satisfied;
		
		(iii) $LL^\dagger L=L$:
		\begin{align*}
			LL^\dagger L&=\left(\begin{matrix}
				I \\
				-\vec{1}^T
			\end{matrix}\right)L_v\left(\begin{matrix}
				I & -\vec{1}
			\end{matrix}\right)\left(\begin{matrix}
				I-\frac{1}{n}\vec{1}\vec{1}^T \\
				-\frac{1}{n}\vec{1}^T
			\end{matrix}\right)L_v^{-1}\left(\begin{matrix}
				I-\frac{1}{n}\vec{1}\vec{1}^T & -\frac{1}{n}\vec{1}
			\end{matrix}\right)\\ &\quad\left(\begin{matrix}
			I \\
			-\vec{1}^T
		\end{matrix}\right)L_v\left(\begin{matrix}
		I & -\vec{1}
	\end{matrix}\right)\\
	&=\left(\begin{matrix}
		I \\
		-\vec{1}^T
	\end{matrix}\right)L_v\left(\begin{matrix}
		I & -\vec{1}
	\end{matrix}\right)=L;
		\end{align*}
	
		(iv) $L^\dagger LL^\dagger=L^\dagger$:
		\begin{align*}
			L^\dagger LL^\dagger&=\left(\begin{matrix}
				I-\frac{1}{n}\vec{1}\vec{1}^T \\
				-\frac{1}{n}\vec{1}^T
			\end{matrix}\right)L_v^{-1}\left(\begin{matrix}
				I-\frac{1}{n}\vec{1}\vec{1}^T & -\frac{1}{n}\vec{1}
			\end{matrix}\right)\left(\begin{matrix}
				I \\
				-\vec{1}^T
			\end{matrix}\right)L_v\left(\begin{matrix}
				I & -\vec{1}
			\end{matrix}\right)\\
			&\quad\left(\begin{matrix}
				I-\frac{1}{n}\vec{1}\vec{1}^T \\
				-\frac{1}{n}\vec{1}^T
			\end{matrix}\right)L_v^{-1}\left(\begin{matrix}
				I-\frac{1}{n}\vec{1}\vec{1}^T & -\frac{1}{n}\vec{1}
			\end{matrix}\right)\\
			&=\left(\begin{matrix}
				I-\frac{1}{n}\vec{1}\vec{1}^T \\
				-\frac{1}{n}\vec{1}^T
			\end{matrix}\right)L_v^{-1}\left(\begin{matrix}
				I-\frac{1}{n}\vec{1}\vec{1}^T & -\frac{1}{n}\vec{1}
			\end{matrix}\right)=L^\dagger.
		\end{align*}
		
		As a result, the matrix defined satisfies four properties of Moore-Penrose of $L$. Since the Moore-Penrose inverse is unique, the theorem is established.
	\end{proof}}
\end{theorem}
\comment{
According to Theorem~\ref{theorem:L_dagger_L_v}, $L^\dagger$ can be represented by $L_v^{-1}$. By multiplying the elements, we represent $L^\dagger$ as:
\begin{equation}\label{equation:L_dagger}
		\left(\begin{matrix}
		L_v^{-1}-\frac{1}{n}\vec{1}\vec{1}^TL_v^{-1}-\frac{1}{n}L_v^{-1}\vec{1}\vec{1}^T+\frac{1}{n^2}\vec{1}\vec{1}^TL_v^{-1}\vec{1}\vec{1}^T & \frac{1}{n}L_v^{-1}\vec{1}+\frac{1}{n^2}\vec{1}\vec{1}^TL_v^{-1}\vec{1} \\
		-\frac{1}{n}\vec{1}^TL_v^{-1}+\frac{1}{n^2}\vec{1}^TL_v^{-1}\vec{1}\vec{1}^T & \frac{1}{n^2}\vec{1}^TL_v^{-1}\vec{1}
	\end{matrix}\right)
\end{equation}
Although the formula seems complex at the first glance, we have represented $L^\dagger$ by elements of $L_v^{-1}$. If we can compute $L_v^{-1}$, we can compute $L^\dagger$ through this formula with little additional cost. We can further give the formula for computing the $s,t$-th element from $L_v^{-1}$. The elements can lie at the $v$-th row and column or not, and when the elements are the diagonal elements, the formula can further be simplified. As a result, there are four situations:}
Based on Theorem~\ref{theorem:L_dagger_L_v}, we can further obtain the element-wise representation of $L^\dagger$. There are four situations:
\begin{equation}\label{equation:l-ss}
	(L^\dagger)_{st}=\mathbf{e}_s^T L_v^{-1}\mathbf{e}_t-\frac{1}{n}\mathbf{e}_s^TL_v^{-1}\Vec{1}-\frac{1}{n}\mathbf{e}_t^TL_v^{-1}\Vec{1}+\frac{1}{n^2}\Vec{1}^TL_v^{-1}\Vec{1},\quad s,t\neq v
\end{equation}
\begin{equation}\label{equation:l-st}
	(L^\dagger)_{ss}=\mathbf{e}_s^T L_v^{-1}\mathbf{e}_s-2\frac{1}{n}\mathbf{e}_s^TL_v^{-1}\Vec{1}+\frac{1}{n^2}\Vec{1}^TL_v^{-1}\Vec{1},\quad s,t\neq v
\end{equation}
\begin{equation}\label{equation:l-sv}
	(L^\dagger)_{sv}=-\frac{1}{n}\mathbf{e}_s^TL_v^{-1}\Vec{1}+\frac{1}{n^2}\Vec{1}^TL_v^{-1}\Vec{1},\quad s\neq v
\end{equation}
\begin{equation}\label{equation:l-vv}
	(L^\dagger)_{vv}=\frac{1}{n^2}\Vec{1}^TL_v^{-1}\Vec{1},
\end{equation}
\comment{
To intuitively understand these results, firstly, for the $v$-th row and column of $L^\dagger$, we divide them into two cases: the diagonal elements $(L^\dagger)_{vv}$ and non-diagonal elements $(L^\dagger)_{sv}$ with $s\neq v$. Generally, $(L^\dagger)_{vv}=\frac{1}{n^2}\vec{1}^TL_v^{-1}\vec{1}$, the average of all elements in $L_v^{-1}$. Also, $(L^\dagger)_{sv}=\frac{1}{n^2}\mathbf{1}^TL_v^{-1}\mathbf{1}-\frac{1}{n}\mathbf{e}_s^TL_v^{-1}\mathbf{1}$, the average of all elements in $L_v^{-1}$ minus the average of the $j$-th column of $L_v^{-1}$. The two formulas can interpret $(L^\dagger)_{ij}$ by $L_i^{-1}$ or $L_j^{-1}$, which means that we represent $(L^\dagger)_{ij}$ with different $i,j$ by different Laplacian minors $L_i^{-1}$ or $L_j^{-1}$. Secondly, for other rows and columns (except $v$-th). For $s,t\neq v$, we have $(L^\dagger)_{st}=(L_v^{-1})_{st}-\frac{1}{n}\mathbf{e}_s^TL_v^{-1}\vec{1}-\frac{1}{n}\mathbf{e}_t^TL_v^{-1}\vec{1}+\frac{1}{n^2}\vec{1}^TL_v^{-1}\vec{1}$, which is $(L_v^{-1})_{st}$ minus the average of the $s$-th column of $L_v^{-1}$ and the average of the $t$-th column of $L_v^{-1}$, plus the average of all elements of $L_v^{-1}$. For the diagonal elements, we have $(L^\dagger)_{ss}=(L_v^{-1})_{ss}-\frac{2}{n}\mathbf{e}_s^TL_v^{-1}\vec{1}+\frac{1}{n^2}\vec{1}^TL_v^{-1}\vec{1}$.}
These formulas show that we can represent all elements of $L^\dagger$ by $L_v^{-1}$. It is worth stressing that the node $v$ can be chosen arbitrarily, while the elements of $L^\dagger$ maintain as invariants. Thus, in practice, we can select a "good" landmark node to speed up the computation.

In addition, we can also represent $\mathcal{L}^\dagger$ in terms of $L_v^{-1}$. The normalized Laplacian $\mathcal{L}$ is only different from $L$ in terms of degree normalized terms. Thus, similar theorem can also be obtained for $\mathcal{L}^\dagger$.
\begin{theorem}\label{theorem:l_n_dagger}
	Let $\bm{\pi}_v$ be an $(n-1)\times1$ vector obtained by deleting the $v$-th element of the stationary distribution vector $\bm{\pi}$, $\sqrt{\bm{\pi}_v}$ is the vector obtained by applying square root on each elements of $\bm{\pi}_v$, $\sqrt{\bm{\pi}_v}(u)=\sqrt{\bm{\pi}_v(u)}$. $\mathcal{L}_v$ is the submatrix of $\mathcal{L}$ obtained by deleting the $v$-th row and column. We have:
	\begin{align*}
		\mathcal{L}^\dagger=\left(\begin{matrix}
			I-\sqrt{\bm{\pi}_v}\sqrt{\bm{\pi}_v}^T \\
			-\sqrt{\bm{\pi}(v)}\sqrt{\bm{\pi}_v}^T
		\end{matrix}\right)\mathcal{L}_v^{-1}\left(\begin{matrix}
			I-\sqrt{\bm{\pi}_v}\sqrt{\bm{\pi}_v}^T & -\sqrt{\bm{\pi}(v)}\sqrt{\bm{\pi}_v}
		\end{matrix}\right).
	\end{align*}
	\comment{
	\begin{align*}
		\mathcal{L}^\dagger &= \left(\begin{matrix}
			(\mathcal{L}^\dagger)_v & -\Vec{\mathbf{nl}_v} \\
			-\Vec{\mathbf{nl}_v}^T & (\mathcal{L}^\dagger)_{vv}
		\end{matrix}\right)\\
		&=\left(\begin{matrix}
			I-\sqrt{\bm{\pi}_v}\sqrt{\bm{\pi}_v}^T \\
			-\sqrt{\bm{\pi}(v)}\sqrt{\bm{\pi}_v}^T
		\end{matrix}\right)\mathcal{L}_v^{-1}\left(\begin{matrix}
			I-\sqrt{\bm{\pi}_v}\sqrt{\bm{\pi}_v}^T & -\sqrt{\bm{\pi}(v)}\sqrt{\bm{\pi}_v}
		\end{matrix}\right).
	\end{align*}}
\comment{
	\begin{proof}
		Since $\mathcal{L}=D^{-\frac{1}{2}}LD^{-\frac{1}{2}}$, each element of $\mathcal{L}$ can be represented as $(\mathcal{L})_{st}=\frac{(L)_{st}}{\sqrt{d_sd_t}}$. The $s$-th element of $\mathcal{L}\sqrt{\bm{\pi}}$ is $\mathcal{L}\sqrt{\bm{\pi}}(s)=\sum_{u\in V}{\frac{(L)_{su}}{\sqrt{d_sd_u}}}\sqrt{\frac{d_u}{2m}}=0$. Thus, we have $\mathcal{L}\sqrt{\bm{\pi}}=0$. The $v$-th row of $\mathcal{L}$ can be represented as $[-\frac{\mathcal{L}_v\sqrt{\bm{\pi}_v}}{\bm{\pi}(v)};\quad \frac{1}{\bm{\pi}(v)}\sqrt{\bm{\pi}_v}^T\mathcal{L}_v\sqrt{\bm{\pi}_v}]^T$. As a result, $\mathcal{L}=\left(\begin{matrix}
			I \\
			-\frac{1}{\sqrt{\bm{\pi}(v)}}\sqrt{\bm{\pi}_v}^T
		\end{matrix}\right)\mathcal{L}_v\left(\begin{matrix}
			I & -\frac{1}{\sqrt{\bm{\pi}(v)}}\sqrt{\bm{\pi}_v}
		\end{matrix}\right)$. Next, we verify that the matrix satisfies four properties of the unique Moore-Penrose pseudo-inverse: (in the following formulas, the dimension of identity matrix are adapted to their representations)
		
		(i) $(\mathcal{L}^\dagger \mathcal{L})^T=\mathcal{L}^\dagger \mathcal{L}$; (ii) $(\mathcal{L}\mathcal{L}^\dagger)^T=\mathcal{L}\mathcal{L}^\dagger$:
		\begin{align*}
			\mathcal{L}^\dagger \mathcal{L}&=\left(\begin{matrix}
				I-\sqrt{\bm{\pi}_v}\sqrt{\bm{\pi}_v}^T \\
				-\sqrt{\bm{\pi}(v)}\sqrt{\bm{\pi}_v}^T
			\end{matrix}\right)\mathcal{L}_v^{-1}\left(\begin{matrix}
				I-\sqrt{\bm{\pi}_v}\sqrt{\bm{\pi}_v}^T & -\sqrt{\bm{\pi}(v)}\sqrt{\bm{\pi}_v}
			\end{matrix}\right)\\
			&\quad \left(\begin{matrix}
				I \\
				-\frac{1}{\sqrt{\bm{\pi}(v)}}\sqrt{\bm{\pi}_v}^T
			\end{matrix}\right)\mathcal{L}_v\left(\begin{matrix}
				I & -\frac{1}{\sqrt{\bm{\pi}(v)}}\sqrt{\bm{\pi}_v}
			\end{matrix}\right)\\
			&=\left(\begin{matrix}
				I-\sqrt{\bm{\pi}_v}\sqrt{\bm{\pi}_v}^T \\
				-\sqrt{\bm{\pi}(v)}\sqrt{\bm{\pi}_v}^T
			\end{matrix}\right)\left(\begin{matrix}
			I & -\frac{1}{\sqrt{\bm{\pi}(v)}}\sqrt{\bm{\pi}_v}
		\end{matrix}\right)=I-\sqrt{\bm{\pi}}\sqrt{\bm{\pi}}^T,
		\end{align*}
		\begin{align*}
			\mathcal{L}\mathcal{L}^\dagger &=\left(\begin{matrix}
				I \\
				-\frac{1}{\sqrt{\bm{\pi}(v)}}\sqrt{\bm{\pi}_v}^T
			\end{matrix}\right)\mathcal{L}_v\left(\begin{matrix}
				I & -\frac{1}{\sqrt{\bm{\pi}(v)}}\sqrt{\bm{\pi}_v}
			\end{matrix}\right)\left(\begin{matrix}
				I-\sqrt{\bm{\pi}_v}\sqrt{\bm{\pi}_v}^T \\
				-\sqrt{\bm{\pi}(v)}\sqrt{\bm{\pi}_v}^T
			\end{matrix}\right)\\
			&\quad \mathcal{L}_v^{-1}\left(\begin{matrix}
				I-\sqrt{\bm{\pi}_v}\sqrt{\bm{\pi}_v}^T & -\sqrt{\bm{\pi}(v)}\sqrt{\bm{\pi}_v}
			\end{matrix}\right)\\
			&=\left(\begin{matrix}
				I \\
				-\frac{1}{\sqrt{\bm{\pi}(v)}}\sqrt{\bm{\pi}_v}^T
			\end{matrix}\right)\left(\begin{matrix}
				I-\sqrt{\bm{\pi}_v}\sqrt{\bm{\pi}_v}^T & -\sqrt{\bm{\pi}(v)}\sqrt{\bm{\pi}_v}
			\end{matrix}\right)\\
			&=I-\sqrt{\bm{\pi}}\sqrt{\bm{\pi}}^T
		\end{align*}
		
		Since $\mathcal{L}\mathcal{L}^\dagger=\mathcal{L}^\dagger\mathcal{L}=I-\sqrt{\bm{\pi}}\sqrt{\bm{\pi}}^T$, the conditions (i) and (ii) are both satisfied;
		
		(iii) $\mathcal{L}\mathcal{L}^\dagger \mathcal{L}=\mathcal{L}$:
		\begin{align*}
			\mathcal{L}\mathcal{L}^\dagger \mathcal{L}&=\left(\begin{matrix}
				I \\
				-\frac{1}{\sqrt{\bm{\pi}(v)}}\sqrt{\bm{\pi}_v}^T
			\end{matrix}\right)\mathcal{L}_v\left(\begin{matrix}
				I & -\frac{1}{\sqrt{\bm{\pi}(v)}}\sqrt{\bm{\pi}_v}
			\end{matrix}\right)\left(\begin{matrix}
				I-\sqrt{\bm{\pi}_v}\sqrt{\bm{\pi}_v}^T \\
				-\sqrt{\bm{\pi}(v)}\sqrt{\bm{\pi}_v}^T
			\end{matrix}\right)\\
			&\quad \mathcal{L}_v^{-1}\left(\begin{matrix}
				I-\sqrt{\bm{\pi}_v}\sqrt{\bm{\pi}_v}^T & -\sqrt{\bm{\pi}(v)}\sqrt{\bm{\pi}_v}
			\end{matrix}\right)\left(\begin{matrix}
			I \\
			-\frac{1}{\sqrt{\bm{\pi}(v)}}\sqrt{\bm{\pi}_v}^T
		\end{matrix}\right)\\
			&\quad\mathcal{L}_v\left(\begin{matrix}
		I & -\frac{1}{\sqrt{\bm{\pi}(v)}}\sqrt{\bm{\pi}_v}
	\end{matrix}\right)\left(\begin{matrix}
	I-\sqrt{\bm{\pi}_v}\sqrt{\bm{\pi}_v}^T \\
	-\sqrt{\bm{\pi}(v)}\sqrt{\bm{\pi}_v}^T
\end{matrix}\right)\\
			&=\left(\begin{matrix}
				I \\
				-\frac{1}{\sqrt{\bm{\pi}(v)}}\sqrt{\bm{\pi}_v}^T
			\end{matrix}\right)\mathcal{L}_v\left(\begin{matrix}
				I & -\frac{1}{\sqrt{\bm{\pi}(v)}}\sqrt{\bm{\pi}_v}
			\end{matrix}\right)\left(\begin{matrix}
				I-\sqrt{\bm{\pi}_v}\sqrt{\bm{\pi}_v}^T \\
				-\sqrt{\bm{\pi}(v)}\sqrt{\bm{\pi}_v}^T
			\end{matrix}\right)\\
			&=\mathcal{L};
		\end{align*}
		
		(iv) $\mathcal{L}^\dagger \mathcal{L}\mathcal{L}^\dagger=\mathcal{L}^\dagger$:
		\begin{align*}
			\mathcal{L}^\dagger \mathcal{L}\mathcal{L}^\dagger&=\left(\begin{matrix}
				I-\sqrt{\bm{\pi}_v}\sqrt{\bm{\pi}_v}^T \\
				-\sqrt{\bm{\pi}(v)}\sqrt{\bm{\pi}_v}^T
			\end{matrix}\right)\mathcal{L}_v^{-1}\left(\begin{matrix}
				I-\sqrt{\bm{\pi}_v}\sqrt{\bm{\pi}_v}^T & -\sqrt{\bm{\pi}(v)}\sqrt{\bm{\pi}_v}
			\end{matrix}\right)\\
			&\quad \left(\begin{matrix}
				I \\
				-\frac{1}{\sqrt{\bm{\pi}(v)}}\sqrt{\bm{\pi}_v}^T
			\end{matrix}\right)\mathcal{L}_v\left(\begin{matrix}
				I & -\frac{1}{\sqrt{\bm{\pi}(v)}}\sqrt{\bm{\pi}_v}
			\end{matrix}\right)\left(\begin{matrix}
			I-\sqrt{\bm{\pi}_v}\sqrt{\bm{\pi}_v}^T \\
			-\sqrt{\bm{\pi}(v)}\sqrt{\bm{\pi}_v}^T
		\end{matrix}\right)\\
		&\quad\mathcal{L}_v^{-1}\left(\begin{matrix}
		I-\sqrt{\bm{\pi}_v}\sqrt{\bm{\pi}_v}^T & -\sqrt{\bm{\pi}(v)}\sqrt{\bm{\pi}_v}
	\end{matrix}\right)\\
			&=\left(\begin{matrix}
				I-\sqrt{\bm{\pi}_v}\sqrt{\bm{\pi}_v}^T \\
				-\sqrt{\bm{\pi}(v)}\sqrt{\bm{\pi}_v}^T
			\end{matrix}\right)\mathcal{L}_v^{-1}\left(\begin{matrix}
				I-\sqrt{\bm{\pi}_v}\sqrt{\bm{\pi}_v}^T & -\sqrt{\bm{\pi}(v)}\sqrt{\bm{\pi}_v}
			\end{matrix}\right)\\
			&=\mathcal{L}^\dagger.
		\end{align*}
		
		As a result, the matrix defined satisfies four properties of Moore-Penrose of $\mathcal{L}$. Since the Moore-Penrose inverse is unique, the theorem is established.
	\end{proof}
}
\end{theorem}
According to Theorem~\ref{theorem:l_n_dagger}, $\mathcal{L}^\dagger$ can be represented by $\mathcal{L}_v^{-1}$. Suppose $\Vec{\mathbf{d}}_v$ is the degree vector that deleting the $v$-th element and $\Vec{\mathbf{d}}_v(u)=d_u$. Since $\mathcal{L}_v^{-1}=D_v^{\frac{1}{2}}L_v^{-1}D_v^{\frac{1}{2}}$, we can further obtain similar results for element-wise representation of $\mathcal{L}^\dagger$ in terms of $L_v^{-1}$:
\comment{
\begin{equation}\label{equation:ln-st-lvn}
	\begin{split}
		(\mathcal{L}^\dagger)_{st}&=(\mathcal{L}_v^{-1})_{st}-\sqrt{\bm{\pi}(t)}\mathbf{e}_s^T\mathcal{L}_v^{-1}\sqrt{\bm{\pi}_v}-\sqrt{\bm{\pi}(s)}\mathbf{e}_t^T\mathcal{L}_v^{-1}\sqrt{\bm{\pi}_v}\\
		&\quad+\sqrt{\bm{\pi}(s)\bm{\pi}(t)}\sqrt{\bm{\pi}_v}^T\mathcal{L}_v^{-1}\sqrt{\bm{\pi}_v},\quad s,t\neq v
	\end{split}
\end{equation}
\begin{equation}\label{equation:ln-sv-lvn}
	(\mathcal{L}^\dagger)_{sv}=-\sqrt{\bm{\pi}(v)}\mathbf{e}_s^T\mathcal{L}_v^{-1}\sqrt{\bm{\pi}_v}+\sqrt{\bm{\pi}(s)\bm{\pi}(v)}\sqrt{\bm{\pi}_v}^T\mathcal{L}_v^{-1}\sqrt{\bm{\pi}_v},\quad s\neq v
\end{equation}
\begin{equation}\label{equation:ln-vv-lvn}
	(\mathcal{L}^\dagger)_{vv}=\bm{\pi}(v)(\sqrt{\bm{\pi}_v}^T\mathcal{L}_v^{-1}\sqrt{\bm{\pi}_v}),
\end{equation}
Moreover, since $L_v^{-1}=(I-P_v)^{-1}D_v^{-1}$ and $\mathcal{L}_v^{-1}=D_v^{\frac{1}{2}}(I-P_v)^{-1}D_v^{-\frac{1}{2}}$, $\mathcal{L}_v^{-1}=D_v^{\frac{1}{2}}L_v^{-1}D_v^{\frac{1}{2}}$. For element-wise representation, $(\mathcal{L}_v^{-1})_{ij}=(L_v^{-1})_{ij}\sqrt{d_id_j}$. Suppose $\Vec{\mathbf{d}}_v$ is the degree vector that deleting the $v$-th element and $\Vec{\mathbf{d}}_v(u)=d_u$, we can further give the element-wise formula for $\mathcal{L}^\dagger$ in terms of $L_v^{-1}$:}
\begin{equation}\label{equation:ln-st-lv}
	\begin{split}
		(\mathcal{L}^\dagger)_{st}&=\sqrt{\bm{\pi}(s)\bm{\pi}(t)}(2m\mathbf{e}_s^T L_v^{-1}\mathbf{e}_t-\mathbf{e}_s^TL_v^{-1}\Vec{\mathbf{d}}_v-\mathbf{e}_t^TL_v^{-1}\Vec{\mathbf{d}}_v\\
		&\quad+\frac{1}{2m}\Vec{\mathbf{d}}_v^TL_v^{-1}\Vec{\mathbf{d}}_v),\quad s,t\neq v
	\end{split}
\end{equation}
\begin{equation}\label{equation:ln-ss-lv}
	(\mathcal{L}^\dagger)_{ss}=\bm{\pi}(s)(2m\mathbf{e}_s^T L_v^{-1}\mathbf{e}_s-2\mathbf{e}_s^TL_v^{-1}\Vec{\mathbf{d}}_v+\frac{1}{2m}\Vec{\mathbf{d}}_v^TL_v^{-1}\Vec{\mathbf{d}}_v),\quad s,t\neq v
\end{equation}
\begin{equation}\label{equation:ln-sv-lv}
	(\mathcal{L}^\dagger)_{sv}=\sqrt{\bm{\pi}(s)\bm{\pi}(v)}(-\mathbf{e}_s^TL_v^{-1}\Vec{\mathbf{d}}_v+\frac{1}{2m}\Vec{\mathbf{d}}_v^TL_v^{-1}\Vec{\mathbf{d}}_v),\quad s\neq v
\end{equation}
\begin{equation}\label{equation:lv-vv-lv}
	(\mathcal{L}^\dagger)_{vv}=\bm{\pi}(v)(\frac{1}{2m}\Vec{\mathbf{d}}_v^TL_v^{-1}\Vec{\mathbf{d}}_v),
\end{equation}

\stitle{Discussion.} From these equations, the elements of both combinatorial Laplacian pseudo-inverse $L^\dagger$ and normalized Laplacian pseudo-inverse $\mathcal{L}^\dagger$ can be represented by the elements of $L_v^{-1}$. There are three quantities of special interest: (i) the elements of $L_v^{-1}$, $(L_v^{-1})_{ij}$; (ii) the sum (weighted sum) of a column of $L_v^{-1}$, $L_v^{-1}\vec{1}$ ($L_v^{-1}\Vec{\mathbf{d}}_v$); (iii) the sum (weighted sum) of all elements of $L_v^{-1}$, $\vec{1}^TL_v^{-1}\vec{1}$ ($\Vec{\mathbf{d}}_v^TL_v^{-1}\Vec{\mathbf{d}}_v$). Once we have computed these three quantities, the computation of Laplacian pseudo-inverse can be completed simply by adding or subtracting these quantities. Next, we discuss the theoretical interpretations of $L_v^{-1}$, which implies that both $L$ and $\mathcal{L}$ can also have similar interesting explanations.

\subsection{Existing interpretations of $L_v^{-1}$}\label{subsec:existing-interpretation}
In this subsection, we review two existing interpretations of $L_v^{-1}$ in terms of $v$-absorbed walk and electrical network. Since $L^\dagger$ and $\mathcal{L}^\dagger$ can be formulated by $L_v^{-1}$, they can also be interpreted by these objects. First, $v$-absorbed walk is a type of random walk that stops when hitting the node $v$. $L_v^{-1}$ can be intuitively interpreted by $v$-absorbed walks.
\begin{lemma}\label{lemma:v-absorbed-walk}("$v$-absorbed walk" interpretation of $L_v^{-1}$ \cite{22resistance})
	Let $\tau_v[s,u]$ denote the expected number of walk steps that passes $u$ in a $v$-absorbed walk starts from $s$, $\tilde{\tau}_v[s,u]=\frac{\tau_v[s,u]}{d_u}$. We have: $(L_v^{-1})_{su}=\tilde{\tau}_v[s,u]$.
\end{lemma}
The $v$-absorbed random walk interpretation of $L_v^{-1}$ is very useful for computing resistance distance \cite{22resistance}. For resistance distance $r(s,t)$, sampling two $v$-absorbed walks separately from $s$ and $t$ consumes $(h(s,v)+h(t,v))$ time, where $h(s,v)$ is the hitting time from $s$ to $v$. It is fast for real-life large graphs since there often exists a node $v$ (e.g. the highest-degree node) that is easy to hit so that $h(s,v)$ and $h(t,v)$ are small \cite{22resistance}. However, for $L^\dagger$ and $\mathcal{L}^\dagger$, we need to compute $\vec{1}^TL_v^{-1}\vec{1}$ (or $\Vec{\mathbf{d}}_v^TL_v^{-1}\Vec{\mathbf{d}}_v$), which requires the value of all elements of $L_v^{-1}$. In this case, we need to sample $v$-absorbed walks from all nodes, which takes $\sum_{u\in V}h(u,v)$ for each sample. Clearly, the time cost is expensive for large graphs even when we choose $v$ as an easy-to-hit node.
\comment{
A spanning tree is a connected subgraph of $G$ with no cycle. Similarly, a $2$-spanning forest is a subgraph of $G$ with no cycle that has $2$ tree components. Let $|\cdot|$ denote the cardinality of a set. According to the all-minors matrix-forest theorem \cite{chaiken1982combinatorial}, we can express the elements of $L_v^{-1}$ by the number of $2$-spanning forests.
\begin{lemma}\label{lemma:2-spanning-forest}("$2$-spanning forest" interpretation of $L_v^{-1}$ \cite{chaiken1982combinatorial})
	Let $\mathcal{T}$ denote the set of spanning trees. Let $\mathcal{F}_{v|s,u}$ denote the set of $2$-spanning forests that $v$ is in one tree component, $s$ and $u$ are both in the other tree component. We have: $(L_v^{-1})_{su}=\frac{|\mathcal{F}_{v|s,u}|}{|\mathcal{T}|}$.
\end{lemma}
According to the $2$-spanning forest interpretation, a direct method for approximating Laplacian pseudo-inverses is to count the number of $2$-spanning forests by sampling. However, sampling a $2$-spanning forest is a hard problem and no efficient algorithm for sampling $2$-spanning forests is known \cite{RandomForestandAnalysis,WilsonThesis}.}

Additionally, $L_v^{-1}$ is also related to electrical networks.
\comment{We consider a graph $G$ as an electrical network, where each node is a junction and edge is a unit resistor, the voltages and currents within the network must adhere to various physical laws. Kirchhoff's current law states that the currents flowing into and out of a junction are equal. Similarly, Kirchhoff's voltage law asserts that the voltage drop across an edge is equal to the current flowing through that edge multiplied by its resistance. Furthermore, Ohm's law describes the relationship between voltage, current, and resistance, stating that the effective resistance between two nodes is the voltage difference divided by the current flowing through the edge connecting those nodes. The energy of an electrical system can be calculated by multiplying the voltage between the input node and the output node with the currents flowing into and out of the system.} When considering the graph $G$ as an electrical network, where each node is a junction and each edge is a unit resistor, based on the classic physical laws (details can be found in \cite{fullversion}), when sending flows into the electrical network, the voltages, currents, effective resistances as well as the energy of the electrical system can all be represented by $L_v^{-1}$ \cite{bollobas1998modern, pseudoinverse17}.
\begin{lemma}\label{lemma:electrical-network}("electrical network" interpretation of $L_v^{-1}$ \cite{bollobas1998modern})
	Consider an electrical network that unit flow comes in through node $s$, and unit flow comes out through node $v$. Then $(L_v^{-1})_{su}$ is the voltage at node $u\in V$. Specifically, $(L_v^{-1})_{ss}$ is the effective resistance between node $s$ and $v$. According to the Ohm's law,  $(L_v^{-1})_{se_1}-(L_v^{-1})_{se_2}$ is the current flow through edge $(e_1,e_2)$. As a by-product, the energy of the electrical network is also $(L_v^{-1})_{ss}$.
\end{lemma}
Computing the corresponding quantities in an electrical system requires solving a linear system \cite{pseudoinverse17}, which is costly when applied for large graphs. In practice, all diagonal elements of $L_v^{-1}$ are often required. This usually involves solving $n$ linear systems, which is much more expensive.

In the following, we propose two novel interpretations of $L_v^{-1}$ in terms of spanning trees and loop-erased walks. The two interesting combinatorial interpretations allow us to develop efficient sampling algorithms.
\subsection{Spanning tree interpretation of $L_v^{-1}$}\label{subsec:new-interpretation-st}
There exist many studies that relating $L_v^{-1}$ to spanning trees \cite{chaiken1982combinatorial,SpanningEdgeCentrality,angriman2020approximation,22resistance}. However, the SOTA results \cite{22resistance} only work for the diagonal elements $(L_v^{-1})_{uu}$, \comment{The classical matrix tree theorem represents $det(L_v)=|\mathcal{T}|$, the number of spanning trees. Based on this, Hayashi et al. \cite{SpanningEdgeCentrality} states that $(L_v^{-1})_{uu}=\frac{|\mathcal{T}_{uv}|}{|\mathcal{T}|}$ if there is an edge between $u$ and $v$, where $\mathcal{T}_{uv}$ denotes the set of spanning trees that containing the edge $(u,v)$. Recently, the results are generalized to $(L_v^{-1})_{uu}$ in \cite{22resistance, angriman2020approximation} where there is not necessary an edge between $u$ and $v$. However, } it is still unknown whether we can represent arbitrary elements of $L_v^{-1}$ in terms of spanning trees. In this subsection, we overcome this problem by considering paths in spanning trees.
\begin{lemma}\label{lemma:spanning-tree}("spanning tree" interpretation of $L_v^{-1}$)
	Let $\mathcal{T}$ denote the set of spanning trees. Let $\mathcal{T}^{s,v}_{e_1,e_2}$ denote the set of spanning trees where the unique path from $s$ to $v$ passes edge $e=(e_1,e_2)$. Let $\mathcal{P}^{u,v}$ be an arbitrary path from $u$ to $v$ in graph $G$. Then we have: $(L_v^{-1})_{su}=\sum_{(e_1,e_2)\in\mathcal{P}^{u,v}}\frac{|\mathcal{T}^{s,v}_{e_1,e_2}|-|\mathcal{T}^{s,v}_{e_2,e_1}|}{|\mathcal{T}|}$.
	\comment{
	\begin{proof}
		According to Lemma~\ref{lemma:electrical-network}, $(L_v^{-1})_{su}$ is the voltage at node $u$ in an electrical network that unit current flow comes in through node $s$ and unit current flow comes out through node $v$. In such an electrical network, let $I(s,v,e_1,e_2)$ denote the current flow along the edge $(e_1,e_2)$. According to Kirchhoff's current law and Ohm's law, the voltage $(L_v^{-1})_{su}=\sum_{(e_1,e_2)\in\mathcal{P}^{u,v}}{I(s,v,e_1,e_2)}$. It remains to prove that the current flows through $(e_1,e_2)$ is $\frac{|\mathcal{T}^{s,v}_{e_1,e_2}|-|\mathcal{T}^{s,v}_{e_2,e_1}|}{|\mathcal{T}|}$. According to Lemma~\ref{lemma:electrical-network}, the current is $(L_v^{-1})_{se_1}-(L_v^{-1})_{se_2}$. According to Lemma~\ref{lemma:2-spanning-forest}, the current can be written as number of $2$-spanning forests $\frac{|\mathcal{F}_{v|s,e_1}|-|\mathcal{F}_{v|s,e_2}|}{|\mathcal{T}|}$. We can partition  $\mathcal{F}_{v|s,e_1}$ into two sets: (i) $e_2$ in the component that $v$ belongs to ($\mathcal{F}_{v,e_2|s,e_1}$); and (ii) $e_2$ in the component that $s,e_1$ belong to ($\mathcal{F}_{v|s,e_1,e_2}$). We have $|\mathcal{F}_{v|s,e_1}|-|\mathcal{F}_{v|s,e_2}|=|\mathcal{F}_{v,e_2|s,e_1}|+|\mathcal{F}_{v|s,e_1,e_2}|-(|\mathcal{F}_{v,e_1|s,e_2}|+|\mathcal{F}_{v|s,e_1,e_2}|)=|\mathcal{F}_{s,e_1|e_2,v}|-|\mathcal{F}_{s,e_2|e_1,v}|$. Consider the number of $2$-spanning trees $|\mathcal{F}_{s,e_1|e_2,v}|$ that $s, e_1$ are in one component, $e_2, v$ are in the other component. There is a bijection between the spanning tree set $\mathcal{T}^{s,v}_{e_2,e_1}$ and the $2$-spanning forest set $\mathcal{F}_{s,e_1|e_2,v}$. This is because we can add $(e_1,e_2)$ in a $2$-spanning forest in $\mathcal{F}_{s,e_1|e_2,v}$, then we will get a spanning tree in $\mathcal{T}^{s,v}_{e_2,e_1}$, and the reverse is also true. As a result, we prove that the current is $\frac{|\mathcal{F}_{v|s,e_1}|-|\mathcal{F}_{v|s,e_2}|}{|\mathcal{T}|}$, which completes the proof.
	\end{proof}
}
\end{lemma}

Lemma~\ref{lemma:spanning-tree} represents $L_v^{-1}$ in terms of the number of spanning trees. To our knowledge, this is the first time that matrix tree theorems for an arbitrary element of $L_v^{-1}$ is given. Based on Lemma~\ref{lemma:spanning-tree}, we can sample a number of spanning trees, examine whether the path from $s$ to $v$ passes through the edges $(e_1,e_2)\in\mathcal{P}^{u,v}$, then we can build an unbiased estimator of $(L_v^{-1})_{sv}$. For an estimator of $L^\dagger$, we can further combine the number of spanning trees to estimate the elements of $L_v^{-1}$ to build an unbiased estimator of $L^\dagger$. In such cases, we are not only interesting in a single element of $L_v^{-1}$. By combining the results of Lemma~\ref{lemma:electrical-network} and Lemma~\ref{lemma:spanning-tree}, we can obtain the following corollary:
\begin{corollary}
	Suppose that $\Vec{f}\in\mathbf{R}^m$ is the flow in an electrical network, $\Vec{b}$ is the demand of the flow ($\Vec{b}(u)$ represents the flow comes out through node $u$) that satisfying $\Vec{1}^T\Vec{b}=0$. Let $\Vec{f}_T$ denote a flow on a spanning tree $T$ with the same demand $\Vec{b}$, we have: $\Vec{f}=\sum_{T\in\mathcal{T}}\frac{1}{|\mathcal{T}|}\Vec{f}_T$.
	\comment{
	\begin{proof}
		According to Lemma~\ref{lemma:electrical-network} and Lemma~\ref{lemma:spanning-tree}, $(L_v^{-1})_{su}$ is the voltage at node $u$ in an electrical network that unit current flow comes in through node $s$ and unit current flow comes out through node $v$. In such an electrical network, let $I(s,v,e_1,e_2)$ denote the current flow along the edge $(e_1,e_2)$. According to Kirchhoff's current law and Ohm's law, the voltage $(L_v^{-1})_{su}=\sum_{(e_1,e_2)\in\mathcal{P}^{u,v}}{I(s,v,e_1,e_2)}$, and $I(s,v,e_1,e_2)=\frac{|\mathcal{T}^{s,v}_{e_1,e_2}|-|\mathcal{T}^{s,v}_{e_2,e_1}|}{|\mathcal{T}|}$. Since if we send flows from $s$ to $v$ on a spanning tree $T$, then there is a current flow passes through $(e_1,e_2)$ if and only if $T\in\mathcal{T}^{s,v}_{e_1,e_2}$. The theorem is established because any electrical system is a linear combination of a simple electrical system in Lemma~\ref{lemma:electrical-network}.
	\end{proof}}
\end{corollary}
The corollary indicates that if we can represent the desired quantities (e,g. $L^\dagger$) as voltages or flows in an electrical system, then we can sample a number of spanning trees and sending flows on them. By performing some operations on the spanning trees, we can obtain an estimation of the electrical system, as well as the desired quantities. These ideas will be utilized in our algorithms in Section~\ref{sec:ecc} and Section~\ref{sec:kcc}. 

For real-world graphs, the Wilson algorithm \cite{wilson1996generating} is the best algorithm for sampling a spanning tree. The time complexity of the Wilson algorithm is $Tr((I-P_v)^{-1})$ \cite{HamiltonCycle}. To quantify how large $Tr((I-P_v)^{-1})$ is, notice that $Tr((I-P_v)^{-1})=\sum_{u\in V,\ n\neq v}{(I-P_v)^{-1}_{uu}}$. According to Lemma~\ref{lemma:v-absorbed-walk}, the element $(I-P_v)^{-1}_{uu}=(L_v^{-1})_{uu}d_u$ can be interpreted as the expected number of steps in a $v$-absorbed walk starts from $u$. Specifically, $(I-P_v)^{-1}_{uu}$ is the expected number of passes to the node $u$ itself. If we select $v$ as an easy-to-hit landmark node (e.g. the highest-degree node), the probability that a node $u$ passes itself twice is very small. As a result, $Tr((I-P_v)^{-1})$ is close to $O(n)$. This is also verified in our experiments on real-life large graphs (see Section~\ref{subsec:exp-setting}).

\comment{
The above results connect $L_v^{-1}$ to various objects related to $v$ (the $v$-absorbed random walk, the $2$-spanning forest that has separate $v$ and two selected nodes, the electrical network where the current flows out through $v$, the spanning tree that the unique path from $s$ to $v$ has some properties), the choice of $v$ is arbitrary. Sometimes, the desired quantity is not related to $v$ (e.g. the resistance distance $r(s,t)$ with $s,t\neq v$), while there is an invariant that $r(s,t)=(L_s^{-1})_{tt}=(L_v^{-1})_{ss}+(L_v^{-1})_{tt}-2(L_v^{-1})_{st}$. We can derive that $(L_{v^\prime}^{-1})_{su}=(L_v^{-1})_{su}+(L_v^{-1})_{v^\prime v^\prime}-(L_v^{-1})_{sv^\prime}-(L_v^{-1})_{uv^\prime}$. That is, Laplacian minors of a different $v^\prime\neq v$ can also be expressed by $L_v^{-1}$. The advantage of doing this is that we can compute some quantities easily. For example, we can compute all pairs of resistance distance if we only know $L_v^{-1}$ (instead of computing $L_s^{-1}$ or $L_t^{-1}$ for $r(s,t)$), this is the idea of \cite{22resistance}. For another example, by considering $n$ electrical networks where unit flow comes in a node $u\in V$ and comes out through node $v$, we can obtain all pair resistance distance $r(s,t)$ (instead of considering $n^2$ electrical networks with unit flow in $s$ and unit flow out $t$). In this paper, we additionally show that $L_v^{-1}$ can be obtained from one single loop-erased random walk trajectory with root $v$.}
\subsection{Loop-erased walk interpretation of $L_v^{-1}$}\label{subsec:new-interpretation-lewalk}
Loop-erased walk is a type of random walk where we obtain a loop-erased trajectory by erasing all the loops in the random walk trajectory. The Wilson algorithm \cite{wilson1996generating} is a famous algorithm for sampling a spanning tree that utilizing the loop-erased walk. Given a graph $G$ and a root node $v$, the Wilson algorithm maintains its loop-erased trajectory to construct a spanning tree $\mathcal{T}$. It starts with only the root node $v$ in $\mathcal{T}$, $\mathcal{T}=\{v\}$. By fixing an arbitrary ordering of $V\setminus \{v\}$, random walks are simulated from a node $u\notin\mathcal{T}$ (following the ordering) until it hits $\mathcal{T}$. Each time a random walk stops, the loop-erased trajectory of the random walk is added into $\mathcal{T}$. This process terminates when all nodes of $G$ are added into $\mathcal{T}$. In this paper, we denote such a process by $\lepath_v$, which is a complete execution of a loop-erased walk with root node $v$. It is well-known that $\mathcal{T}$ is a uniformly sampled spanning tree independent of the node ordering we fixed \cite{wilson1996generating}.

Although loop-erased walks are mainly used to generate spanning trees and spanning forests \cite{SpanningEdgeCentrality, GraphSignal21, inversetrace19}, we find that the distribution of loop-erased walks are less studied with little applications. \comment{Marchal et. al. \cite{HamiltonCycle} proved that the expected running time of $\lepath_v$ is $Tr(L_v^{-1}D_v)=Tr((I-P_v)^{-1})$.} Recently, Liao et al. \cite{22resistance} proves that $(L_v^{-1}D_v)_{uu}=((I-P_v)^{-1})_{uu}$ is the expected number of passes of node $u$ in $\lepath_v$. Their analysis only work for the diagonal elements $(L_v^{-1})_{uu}$ for $u\in V$, $u\neq v$, thus cannot be utilized for approximating Laplacian pseudo-inverse. To tackle this limitation, we derive several novel results based on the stack representation of $\lepath_v$.

The basic proof technique in \cite{wilson1996generating} to show why the Wilson algorithm produces a uniform spanning tree is a stack representation of $\lepath_v$. Wilson et al. \cite{wilson1996generating} proved that the following process is identical to $\lepath_v$: Given a graph $G$ and a node $v$, in the stack representation of $\lepath_v$, each node $u\in V$ maintains a stack $S[u]$. All the stacks are initialized empty and marked white. Then, we operate on the stacks following an ordering of $V\setminus\{v\}$ which is fixed before. Specifically, $S[v]$ is marked gray first. When a stack $S[u]$ is met, we (i) randomly select a node $w$ from the neighbors of $u$, and (ii) push $w$ on the top of $S[u]$. Finally, we (iii) switch to the stack $S[w]$. The operation continues until we switch to a stack that is marked gray. Each time there is a loop on top of the stacks, the loop will be popped. When a stack marked gray is met, we mark all stacks that are not empty as gray and continue the operations from a stack $S[u]$ that has not been marked gray. This process terminates until all nodes are marked gray. Wilson proved that the resulting stacks are independent of the node ordering we fixed. Denote $\topstack(S[u])$ the top element of $S[u]$, then the edges $(u, \topstack(S[u]))$ for $u\in V$ and $u\neq v$ form a uniformly sampled spanning tree with root $v$. 

\begin{figure}[t] \vspace*{-0.1cm}
	\begin{center}
		\begin{tabular}[t]{c}\hspace*{-0.3cm}
			\subfigure[]{
				\includegraphics[width=0.18\columnwidth, height=1.4cm]{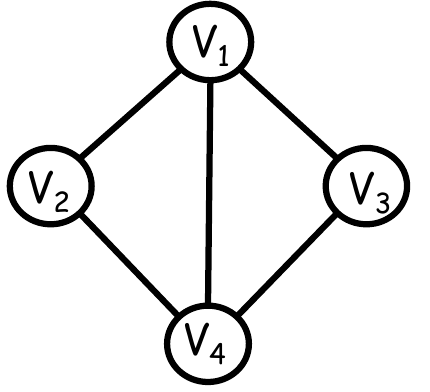}
			}\hspace{0.02\columnwidth}
			\subfigure[]{
				\includegraphics[width=0.18\columnwidth, height=1.4cm]{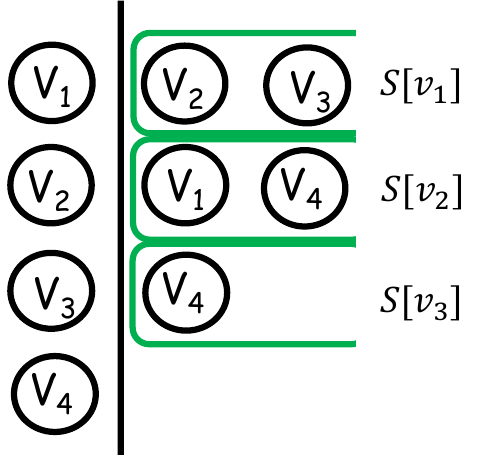}
			}\hspace{0.02\columnwidth}
			\subfigure[]{
				\includegraphics[width=0.18\columnwidth, height=1.4cm]{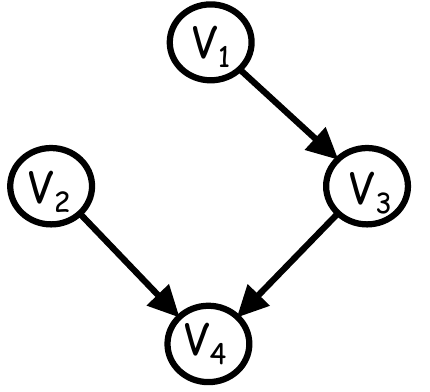}
			}\hspace{0.02\columnwidth}
			\subfigure[]{
				\includegraphics[width=0.3\columnwidth, height=1.2cm]{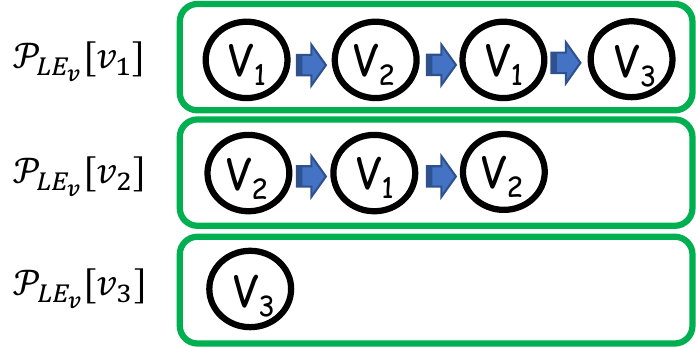}
			}
		\end{tabular}
	\end{center}
	\vspace*{-0.4cm}
	\caption{\small A loop-erased walk $\lepath_v$ with $v=v_4$ and its stack representation. (a) an example graph; (b) the stack representation of $\lepath_v$. When operating the stacks by the ordering $v_1$, $v_2$, $v_3$, the loop-erased walk trajectory is $v_1,v_2,v_1,v_3,v_4,v_2,v_4$. When the ordering is $v_2$, $v_3$, $v_1$, the trajectory is $v_2,v_1,v_2,v_4,v_3,v_4,v_1,v_3$; (c) The resulting spanning tree is independent of the node ordering; (d) the loop-included path $\mathcal{P}_{\lepath_v}[u]$ for $u=v_1, v_2, v_3$.}
	\label{fig:example-loop-erased-walk}
	\vspace*{-0.3cm}
\end{figure}

In this paper, we go one step further. When there is a loop on top of the stacks, instead of popping the nodes of the loop out of the stacks, we keep them in the stack. We discover that such a stack representation of $\lepath_v$ capture all information of $L_v^{-1}$. Fig.~\ref{fig:example-loop-erased-walk} illustrates a running example. Fig.~\ref{fig:example-loop-erased-walk}(a) is an example graph and Fig.~\ref{fig:example-loop-erased-walk}(b) is the stack representation of a sample of $\lepath_v$. It can be seen that the stack representation is independent of the fixed node ordering. When the fixed node ordering of $V\setminus\{v_4\}$ is $v_1$, $v_2$, $v_3$, the loop-erased walk trajectory (the stack process ordering) is $v_1,v_2,v_1,v_3,v_4,v_2,v_4$. When the ordering is $v_2$, $v_3$, $v_1$, the trajectory is $v_2,v_1,v_2,v_4,v_3,v_4,v_1,v_3$. Although the ordering is different, we can obtain a same stack representation of these loop-erased walks. The resulting spanning tree is also the same, as illustrated in Fig.~\ref{fig:example-loop-erased-walk}(c). This implies that when we have sampled a loop-erased walk $\lepath_v$, we can obtain $n!$ samples from the stack representation of $\lepath_v$ by using different node orderings to traverse the stack representation of $\lepath_v$.

We have seen above that the spanning tree maintained at the top of the stacks can interpret $L_v^{-1}$. Here, we can exploit more information from the stacks. Specifically, we define a loop-included path $\mathcal{P}_{\lepath_v}[s]$ from the stack representation of $\lepath_v$, which is constructed as follows: The start node of $\mathcal{P}_{\lepath_v}[s]$ is $u=s$, following the stack representation of $\lepath_v$, $u$ jumps to the node at the bottom of $S[u]$, until the root node $v$ is reached. Each time the path leaves a node $u$, the node at the bottom of $S[u]$ is removed from $S[u]$. As illustrated in Fig.~\ref{fig:example-loop-erased-walk}(d), there is a sample of $\mathcal{P}_{\lepath_v}[u]$ from all nodes $u=v_1,v_2,v_3$. We state that the defined loop-included path $\mathcal{P}_{\lepath_v}$ can interpret arbitrary element of $L_v^{-1}$.
\begin{lemma}\label{lemma:loop-erased-walk}("Loop-erased walk" interpretation of $L_v^{-1}$)
	Given a loop-erased walk $\lepath_v$, let $\tau_{\lepath_v}[s,u]$ be the expected number of existences of node $u$ in $\mathcal{P}_{\lepath_v}[u]$, $\tilde{\tau}_{\lepath_v}[s,u]=\frac{\tilde{\tau}_{\lepath_v}[s,u]}{d_u}$. We have: $(L_v^{-1})_{su}=\tilde{\tau}_{\lepath_v}[s,u]$. Specifically, let $\tilde{\tau}_{\lepath_v}[s,s]$ be the normalized expected number of existences of node $s$ in $\lepath_v$, we have: $(L_v^{-1})_{ss}=\tilde{\tau}_{\lepath_v}[s,s]$.
	\comment{
	\begin{proof}
		Given the stack representation of $\lepath_v$. Since the stacks are independent of the node ordering, each node can be the first node, while the same stack representation will be obtained. When we select $u$ as the first node to construct $\mathcal{P}_{\lepath_v}[u]$, the distribution of $\mathcal{P}_{\lepath_v}[u]$ is the same as the distribution of $\mathcal{P}_v[u]$, where $\mathcal{P}_v[u]$ denotes the path of a $v$-absorbed walk trajectory from $u$. From Lemma~\ref{lemma:v-absorbed-walk}, we have $\tilde{\tau}_{\lepath_v}[s,u]=\tilde{\tau}_v[s,u]=\frac{\tau_v[s,u]}{d_u}=(L_v^{-1})_{su}$. When a random walk from $u$ reaches $v$ in $\lepath_v$, $\lepath_v$ will never pass $u$ again. Thus $\tilde{\tau}_{\lepath_v}[s,s]=(L_v^{-1})_{ss}$ is also the expected number of passes of $u$ in the whole trajectory of $\lepath_v$.
	\end{proof}}
\end{lemma}
Lemma~\ref{lemma:loop-erased-walk} suggests that the distribution of $\mathcal{P}_{\lepath_v}[u]$ is the same as an $v$-absorbed walk trajectory from $u$. Thus, the expected length of $\mathcal{P}_{\lepath_v}[u]$ is the hitting time $h(u,v)$, the expected number of nodes in the stacks is $Tr((I-P_v)^{-1})$. The advantage of the "loop-erased walk" interpretation for sampling is apparently. In order to approximate all elements of $L_v^{-1}$, we need to sample $v$-absorbed walks from all nodes $u\in V$, which consumes $\sum_{u\in V}h(u,v)=\vec{1}^T(I-P_v)^{-1}\vec{1}$ time. For loop-erased walk sampling, we only need to sample one loop-erased walk with node $v$ in expected running time $Tr((I-P_v)^{-1})$, which is much lower than $\vec{1}^T(I-P_v)^{-1}\vec{1}$.

Similar to spanning tree sampling, the loop-erased walk sampling also requires a landmark node $v$ as input. When $v$ is an easy-to-hit node, $Tr((I-P_v)^{-1})$ is small, which makes the sampling techniques more efficient. For convenience, we set $v$ as the highest-degree node in graph. We observe that such a choice is near optimal for real-life graphs, as evidenced in our experiments (see Section~\ref{subsec:exp-performance}).
\comment{
\begin{algorithm}[t]
	\scriptsize
	\caption{cycle popping algorithm}
	\label{algo:cycle-pop}
	\LinesNumbered
	\KwIn{Graph $G$, a landmark node $v$}
	\KwOut{$\mathcal{C}_u$ for each node $u\in V\setminus\{v\}$}
	$\mathcal{T}\leftarrow\emptyset$;\\
	\For{each node $u\in V\setminus\{v\}$}{$\mathcal{C}_u\leftarrow\emptyset$;\\}
	\For{each node $s\in V$}{
		$u\leftarrow s$, $\mathcal{P}_s\leftarrow\emptyset$;\\
		\While{$u\notin\mathcal{T}$}{
			$next\leftarrow RandomNeighbor(u)$;\\
			$\mathcal{P}_s\leftarrow\mathcal{P}_s\cup\{(u,next)\}$;\\
			$u\leftarrow next$;\\
			\If{$u$ forms a cycle $c_u$}{
				$\mathcal{P}_s\leftarrow\mathcal{P}_s\setminus c_u$;\\
				$\mathcal{C}_u\leftarrow\mathcal{C}_u\cup \{c_u\}$;\\
				\For{each node $w\in c_u$}{
					$\mathcal{C}_w\leftarrow\cup_{z\in c_u}\mathcal{C}_z$;
				}
			}
		}
		$\mathcal{T}\leftarrow\mathcal{T}\cup\mathcal{P}_s$;
	}
	\For{each node $u\in V$}{
			\For{each node $w$ that is $u$'s child in $\mathcal{T}$}{
				$\mathcal{C}_u\leftarrow\mathcal{C}_u\cup \{C_w\}$;
			}
	}
	\Return $\mathcal{C}_u$ for each node $u\in V\setminus\{v\}$;
\end{algorithm}}
\comment{
\subsection{New results on $L^\dagger$}
We can represent $L^\dagger$ in terms of $L_v^{-1}$. In this section, we give theoretical interpretation of elements of the combinatorial Laplacian pseudo-inverse $L^\dagger$. The results are based on the fact that $L^\dagger$ is a matrix whose row and column sum up to $1$. Thus, $L^\dagger$ can be represented by $L_v^{-1}$.

According to this formula, we can easily derive the $v$-absorbed random walk interpretation of $L^\dagger$.
\begin{lemma}("$v$-absorbed random walk" interpretation of $L^\dagger$)
	Let $\tau_v[s,u]$ denote the expected number of walk steps that passes $u$ in an $v$-absorbed random walk starts from $s$, $\tilde{\tau}_v[s,u]=\frac{\tau_v[s,u]}{d_u}$. We have: $(L^\dagger)_{vv}=\frac{1}{n^2}\sum_{i\in V}\sum_{j\in V}\tilde{\tau}_v[i,j]$, which is $\frac{1}{n}$ multiplied by the weighted length of a $v$-absorbed random walk starts from a uniformly randomly chosen node $u$. $(L^\dagger)_{vs}=\frac{1}{n^2}\sum_{i\in V}\sum_{j\in V}\tilde{\tau}_v[i,j]-\frac{1}{n}\sum_{u\in V}\tilde{\tau}[s,u]$, which is $\frac{1}{n}$ multiplied by the weighted length of a $v$-absorbed random walk starts from a uniformly randomly chosen node $u$ minus the length from $s$.
\end{lemma}
According to matrix-forest theorem, $(L_v^{-1})_{st}=\frac{|\mathcal{F}_{v|s,t}|}{|\mathcal{T}|}$, which is the number of $2$-spanning forests where $v$ is in one tree component, $s,t$ are in the other tree component, divided by the number of spanning trees. Through this interpretation, we can easily derive the tree formulas for $L^\dagger$ (the Green function). Note that our proofs are much easier than the original proofs which utilizes the expansion of the determinant $det(L+\frac{\mathbf{1}\mathbf{1}^T}{n})$ \cite{chung2023forest}.
\begin{lemma}("$2$-spanning forest" interpretation of $L^\dagger$)
	Let $\mathcal{T}$ denote the set of spanning trees. Let $\mathcal{F}_{2}$ denote the set of $2$-spanning forests. We have: $(L^\dagger)_{vv}=\frac{1}{n^2|\mathcal{T}|}\sum_{T_1\cup T_2\in \mathcal{F}_{2}}{|T_2|^2}$. Let $\mathcal{F}_{u,v|\cdot}$ denote the set of $2$-spanning forests that $u,v$ are in the same tree component. Let $\mathcal{F}_{u|v}$ denote the set of $2$-spanning forests that $u,v$ are in different same tree components. We have: $(L^\dagger)_{vs}=\frac{1}{n^2|\mathcal{T}|}(\sum_{T_1\cup T_2\in \mathcal{F}_{u,v|\cdot},\ s,v\in T_1}{|T_2|^2}-\sum_{T_1\cup T_2\in \mathcal{F}_{u|v},\ u\in T_1,v\in T_2}{|T_1||T_2|})$.
\end{lemma}
The electrical network can be regarded as a Laplacian linear system $Lx=b$, the solution is $x=L^\dagger b$. For example, $Lx=e_i-e_j$ represents the electrical network that unit flow comes through node $i$ and comes out through node $j$, the solution $(e_i-e_j)^Tx=(e_i-e_j)^TL^\dagger(e_i-e_j)$ is the resistance distance. Additionally, a column of $L^\dagger$ can be obtained by solving the linear system $Lx=e_i-\frac{1}{n}$ \cite{pseudoinverse17}. Generally, $b$ can be any vector that is orthogonal to all $1$ vector $\mathbf{1}$. Thus, we have the following result:
\begin{lemma}("electrical network" interpretation of $L^\dagger$)
	Consider a electrical network that unit flow comes in through node $s$, and $\frac{1}{n}$ flow comes out through each node $v\in V$. Then $(L^\dagger)_{su}$ is the voltage at node $u\in V$. According to Ohm's law,  $(L^\dagger)_{se_1}-(L^\dagger)_{se_2}$ is the current flow through edge $(e_1,e_2)$. As a by-product, the energy of the electrical network is also $(L^\dagger)_{ss}$.
\end{lemma}
Again, sampling a $2$-spanning forest or solving an electrical system is hard, we can convert the two results into a "spanning tree" interpretation. Notice that in the "electrical network" interpretation, the linear system $Lx=e_s-\frac{1}{n}$ can be regarded as a linear combination of $n$ linear systems $Lx=e_s-e_u$ with $u\in V$. In a spanning tree $T$, each node $u\in T$ has a unique path $\mathcal{P}_T[u]$ to the root $v$. We denote the number of times that all these paths passes node $u$ as $level(u)$, we have $level(u)=\sum_{s\in V}{\mathbf{1}_{\mathcal{P}_T[s] \text{passes $u$}}}$. Then we can give a "spanning tree" interpretation of $L^\dagger$.
\begin{lemma}("spanning tree" interpretation of $L^\dagger$)
	Let $\mathcal{T}$ denote the set of spanning trees. Let $\mathcal{T}^{s,v}_{e_1,e_2}$ denote the set of spanning trees where the unique path from $s$ to $v$ passes edge $e=(e_1,e_2)$. Let $\mathcal{P}$ be the set of arbitrary paths from $s$ to each node $v\in V$. Then we have: $L^\dagger_{ss}=\sum_{(e_1,e_2)\in\mathcal{P}}\sum_{v\in V}\frac{|\mathcal{T}^{s,v}_{e_1,e_2}|-|\mathcal{T}^{s,v}_{e_2,e_1}|}{n|\mathcal{T}|}$.
\end{lemma}
\begin{figure}[t!]
	\begin{center}
		\vspace*{-0.2cm}
		\begin{tabular}[t]{c}
			\hspace*{0cm}
			\subfigure[$G$, an $s$-flow]{
				\raisebox{0.15\height}{
					\includegraphics[width=0.205\columnwidth, height=2.6cm]{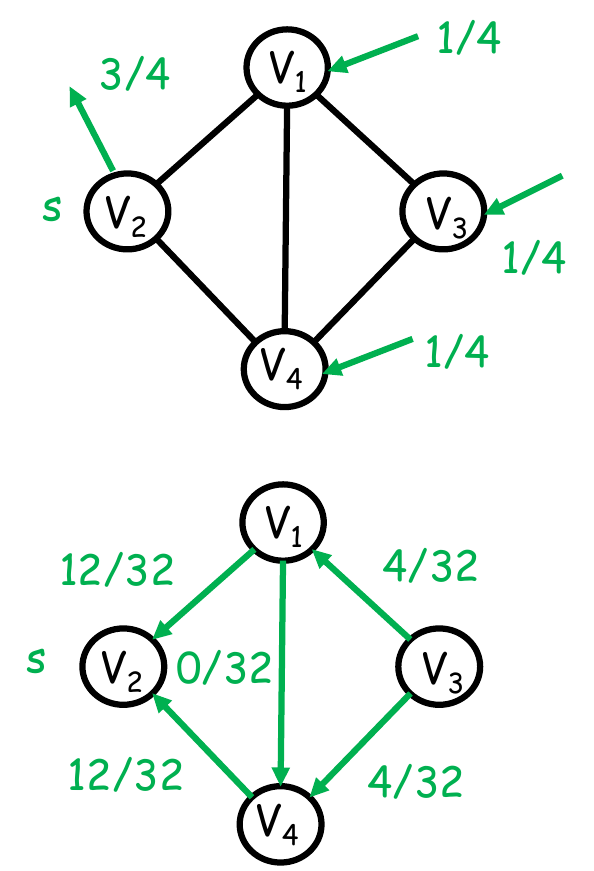}}
			}\hspace{0.08cm}
			\subfigure[$s$-flow on spanning trees]{
				\raisebox{0.1\height}{
					\includegraphics[width=0.7\columnwidth, height=2.6cm]{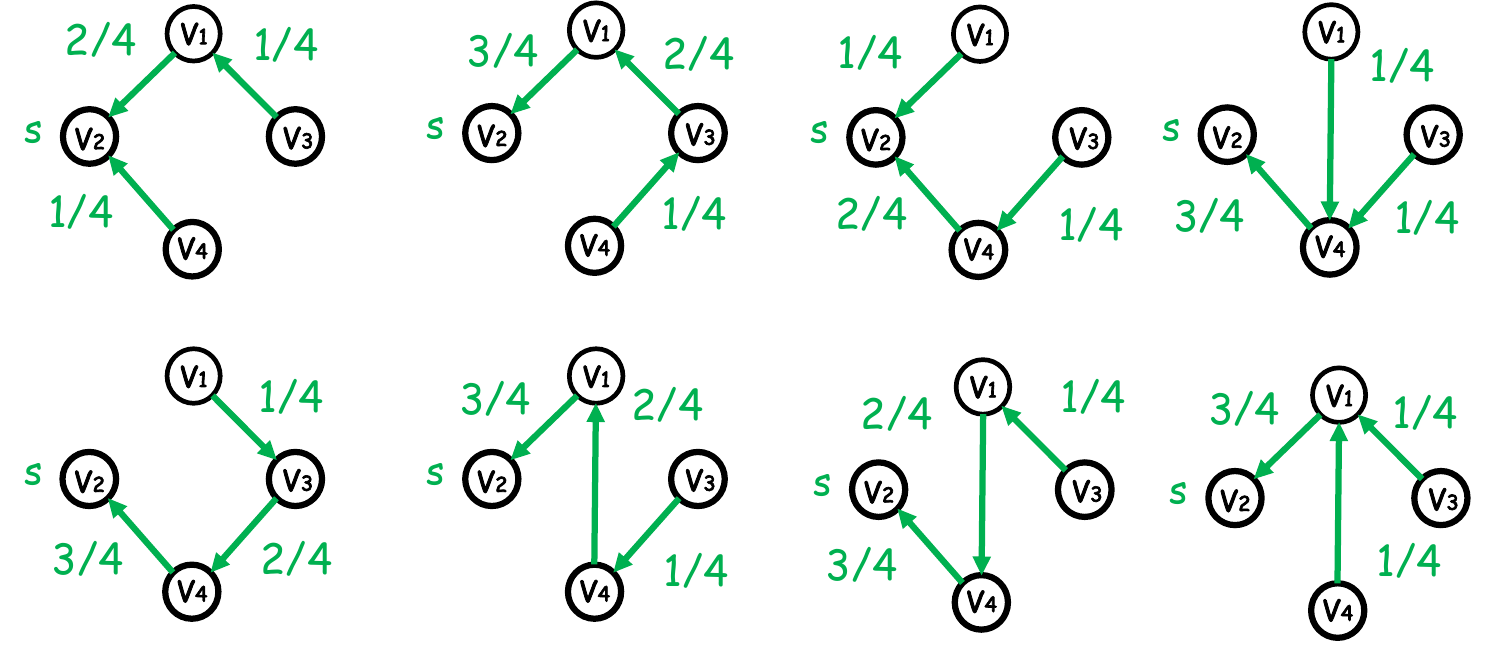}}
			}
		\end{tabular}
	\end{center}
	\vspace*{-0.4cm}
	\caption{Illustration of sending flows on spanning trees. (a) A graph $G$ and an $s$-flow when a unit current flows in $v_2$ and $1/n$ current flows out every node $u\in V$; (b) There are $8$ spanning trees. The current flows along an edge $(u, v)$ in $G$ equals the average current flows on $(u, v)$ on each spanning tree.}
	\vspace*{-0.6cm}
	\label{fig:flow-example}
\end{figure}
Fig.~\ref{fig:flow-example} illustrates a running example of sending flows on spanning trees. The resulting $s$-flow is an average of flows on spanning trees. We can also prove the theorem by constructing bijection between spanning trees and $2$-spanning forests. Similar to resistance distance, we can also use a landmark technique. So all the elements of $L^\dagger$ can be expressed by a fixed $L_v^{-1}$. The Laplacian pseudo-inverse can be read from a single loop-erased walk trajectory.
\begin{lemma}("Loop-erased walk" interpretation of $L^\dagger$)
	Let $\tau_l[s,u]$ be the expected number of existences of node $u$ in $\mathcal{P}_{LE}[u]$ in a loop-erased walk with root $v$, $\tilde{\tau}_l[s,u]=\frac{\tilde{\tau}_l[s,u]}{d_u}$. We can represent the diagonal item as $(L^\dagger)_{vv}=\frac{1}{n^2}\sum_{i,j\in V}\tilde{\tau}_l[i,j]$. For non-diagonal items, we have: $(L^\dagger)_{vs}=\frac{1}{n^2}\sum_{i,j\in V}\tilde{\tau}_l[i,j]-\frac{1}{n}\sum_{u\in V}\tilde{\tau}_l[s,u]$.
\end{lemma}
Now we have given interpretation of each element of $L^\dagger$ and $\mathcal{L}^\dagger$. Many quantities of interest are function of $L^\dagger$ ($\mathcal{L}^\dagger$) such as $Tr(\mathcal{L}^\dagger)$, the diagonal elements $(L^\dagger)_{uu}$. Next, we focus on two examples, the electrical closeness centrality ($(L^\dagger)_{uu}$) and the Kemeny's constant ($Tr(\mathcal{L}^\dagger)$), we design efficient algorithms based on our new interpretation and the landmark based approach.}

\section{Algorithms for ECC computation}\label{sec:ecc}
In this section, we study the problem of approximating electrical closeness centrality (\ecc). Recall that in order to compute \ecc, it suffices to compute $(L^\dagger)_{uu}$ for $u\in V$. Then, based on the formula $\mathbf{c}(u)=\frac{n-1}{n(L^\dagger)_{uu}+Tr(L^\dagger)}$, the \ecc $\mathbf{c}(u)$ for all $u\in V$ can be computed in linear time if we have known $(L^\dagger)_{uu}$ for all $u\in V$. According to Equation~(\ref{equation:l-ss}) and Equation~(\ref{equation:l-vv}), $(L^\dagger)_{uu}=(L_v^{-1})_{uu}-2\mathbf{e}_u^TL_v^{-1}\vec{1}+\frac{1}{n^2}\vec{1}^TL_v^{-1}\vec{1}$ for $u\in V$ and $u\neq v$; and $(L^\dagger)_{vv}=\frac{1}{n^2}\vec{1}^TL_v^{-1}\vec{1}$. The computation of $(L^\dagger)_{uu}$ for $u\in V$ can be decomposed into two stages: (i) computation of $(L_v^{-1})_{uu}$ for all $u\in V$ and $u\neq v$; (ii) computation of $\mathbf{e}_u^TL_v^{-1}\vec{1}$ for $u\in V$ and $u\neq v$. Then, $\vec{1}^TL_v^{-1}\vec{1}=\sum_{u\in V,\ u\neq v}{\mathbf{e}_u^TL_v^{-1}\vec{1}}$, the whole diagonal of $L^\dagger$ can be obtained. In the following, we first review 
 existing algorithms for computing \ecc. Then, we propose two novel algorithms for approximating $(L^\dagger)_{uu}$ for $u\in V$ based on sampling spanning trees and sampling loop-erased walks. We give comprehensive analysis on the proposed algorithms.
 
\subsection{Existing solutions}
As we discussed before, the problem of computing \ecc is identical to computing $(L^\dagger)_{uu}$. A straightforward approach is to apply eigen-decomposition on $L$ which costs $O(n^3)$ time. Bozzo and Franceschet \cite{BozzoF12} computes $L^\dagger$ by the formula $L^\dagger=(L+\frac{\vec{1}\vec{1}^T}{n})^{-1}-\frac{\vec{1}\vec{1}^T}{n}$. However, the all-one matrix makes the first part a dense matrix, and it is hard to compute the inverse of a dense matrix, thus the method is not scalable to large real-life graphs. Recently, Angriman et al. approximate $(L^\dagger)_{uu}$ for $u\in V$ by spanning tree sampling \cite{angriman2020approximation}. They first solve a linear system $L\mathbf{x}=\mathbf{e}_s-\frac{\vec{1}}{n}$ to obtain the $s$-th column of $L^\dagger$. Then they use spanning tree sampling to approximate single-source resistance distance $r(s,u)$ for all $u\in V$. Finally, $(L^\dagger)_{uu}$ for $u\in V$ is obtained by $(L^\dagger)_{uu}=r(s,u)+2(L^\dagger)_{su}-(L^\dagger)_{ss}$. This algorithm is the state-of-the-art algorithm to compute \ecc. However, the cost of accurately solving a Laplacian system is still hard for large graphs. Compared to their work, we show that there is no need to solve the linear system as it can also be approximated via spanning tree sampling. Our algorithm directly approximates the elements of $L^\dagger$, thus gives a "pure" spanning tree sampling algorithm. Moreover, our loop-erased walk sampling algorithm further improves the efficiency of \ecc approximation.

\subsection{A spanning tree sampling algorithm}\label{subsec:ecc-spanningtree}
In this subsection, we propose a novel Monte Carlo algorithm for approximating \ecc based on spanning tree sampling. Firstly, according to the "electrical system" interpretation of $L_v^{-1}$, $(L_v^{-1})_{uu}$ for $u\in V$ and $u\neq v$ is the single-source resistance distance $r(v,u)$ from node $v$ to all nodes $u$ in graph. We can use the spanning tree sampling algorithms proposed in \cite{22resistance, angriman2020approximation} to approximate $(L_v^{-1})_{uu}$ for $u\in V$. However, it is not enough to approximate the diagonal elements of $L_v^{-1}$ ($\mathbf{e}_u^TL_v^{-1}\vec{1}$ for $u\in V$ and $u\neq v$ should also be approximated). To tackle this problem, we first interpret the $v$-th column of $L^\dagger$, the $(L^\dagger)_{vu}$ for $u\in V$ in terms of electrical voltages of a certain electrical system. From the “spanning tree" interpretation of $L_v^{-1}$, sending current flows on the graph is identical to sending current flows on all spanning trees of the graph and take the average. Then, we approximate the electrical voltages by sampling a number of spanning trees. After that, we can approximate $\mathbf{e}_u^TL_v^{-1}\vec{1}$ for $u\in V$ and $u\neq v$ from $(L^\dagger)_{vu}$ for $u\in V$ based on Equation~(\ref{equation:l-sv}) and Equation~(\ref{equation:l-vv}). Thus, we give a "pure" spanning tree sampling algorithm for approximating $(L^\dagger)_{uu}$ for $u\in V$ by combining the two stages. Finally, we give analysis on the proposed algorithm.

\stitle{Interpreting $L^\dagger$ by electrical voltages.} In Section~\ref{subsec:existing-interpretation}, $(L_v^{-1})_{su}$ for $u\in V$ and $u\neq v$ is the voltage at node $u$ in an electrical system where a unit current flow comes in through node $s$ and a unit current flow comes out through node $v$. Based on Equation~(\ref{equation:l-sv}) and Equation~(\ref{equation:l-vv}), due to the fact that electrical systems are linear objects, we can combine $n$ such electrical systems to interpret $(L^\dagger)_{vu}$ for $u\in V$:
\begin{lemma}\label{lemma:l-dagger-electrical-system}
	Consider an electrical system that $\frac{1}{n}$ current flow comes in through every node $u\in V$, and $\frac{1}{n}$ current flow comes out through node $v$ in graph $G$. Then, in such an electrical system, $-(L^\dagger)_{vu}$ is the voltage at node $u\in V$. According to the Ohm's law, $(L^\dagger)_{ve_2}-(L^\dagger)_{ve_1}$ is the current flow through edge $(e_1,e_2)$. As a by-product, the energy of the electrical network is also $(L^\dagger)_{vv}$.
	\comment{
	\begin{proof}
		Consider the electrical system as a linear combination of $n$ electrical systems that $\frac{1}{n}$ flow comes in through node $u\in V$, comes out through node $v$. According to Lemma~\ref{lemma:electrical-network}, the voltage at node a node $s\neq v$ is $\frac{1}{n}(L_v^{-1})_{us}$. When combining $n$ such electrical systems, the voltage at node $u$ is $\mathbf{x}(u)=\frac{1}{n}\sum_{s\in V}{(L_v^{-1})_{us}}=\frac{1}{n}\mathbf{e}_u^TL_v^{-1}\vec{1}$, this is based on the setting $\mathbf{x}(v)=0$. Since the column of $L^\dagger$ sums up to $0$, when setting the voltages to satisfy all voltages sum up to $0$, $-\mathbf{x}(v)$ then is the average of such voltages. We have $\mathbf{x}(v)=-\frac{1}{n^2}\vec{1}^TL_v^{-1}\vec{1}=-(L^\dagger)_{vv}$ (based on Equation~(\ref{equation:l-vv})). The voltage at node $u$ is $\mathbf{x}(u)=-\frac{1}{n^2}\vec{1}^TL_v^{-1}\vec{1}+\frac{1}{n}\mathbf{e}_s^TL_v^{-1}\vec{1}=-(L^\dagger)_{vu}$ (based on Equation~(\ref{equation:l-sv})). According to Ohm's law, the current flow along edge $(e_1,e_2)$ is the difference of voltages, $(L^\dagger)_{ve_2}-(L^\dagger)_{ve_1}$. Finally, the energy of the system is the product of current and voltage, which is $\sum_{u\in V}\frac{1}{n}\frac{1}{n}\mathbf{e}_u^TL_v^{-1}\vec{1}=\frac{1}{n^2}\vec{1}^TL_v^{-1}\vec{1}=(L^\dagger)_{vv}$. Notice that the energy also equals the square sum of the currents on all edges.
	\end{proof}}
\end{lemma}
\begin{figure}[t!]
	\begin{center}
		\vspace*{-0.3cm}
		\begin{tabular}[t]{c}
			\hspace*{0cm}
			\subfigure[$G$, network flow]{
				\raisebox{0.15\height}{
					\includegraphics[width=0.205\columnwidth, height=2.6cm]{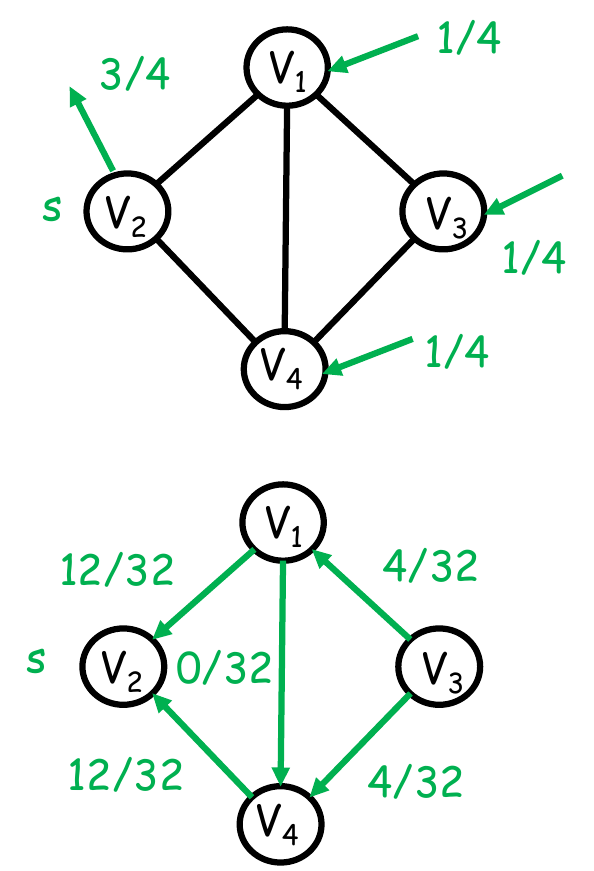}}
			}\hspace{0.08cm}
			\subfigure[flows on spanning trees]{
				\raisebox{0.1\height}{
					\includegraphics[width=0.7\columnwidth, height=2.6cm]{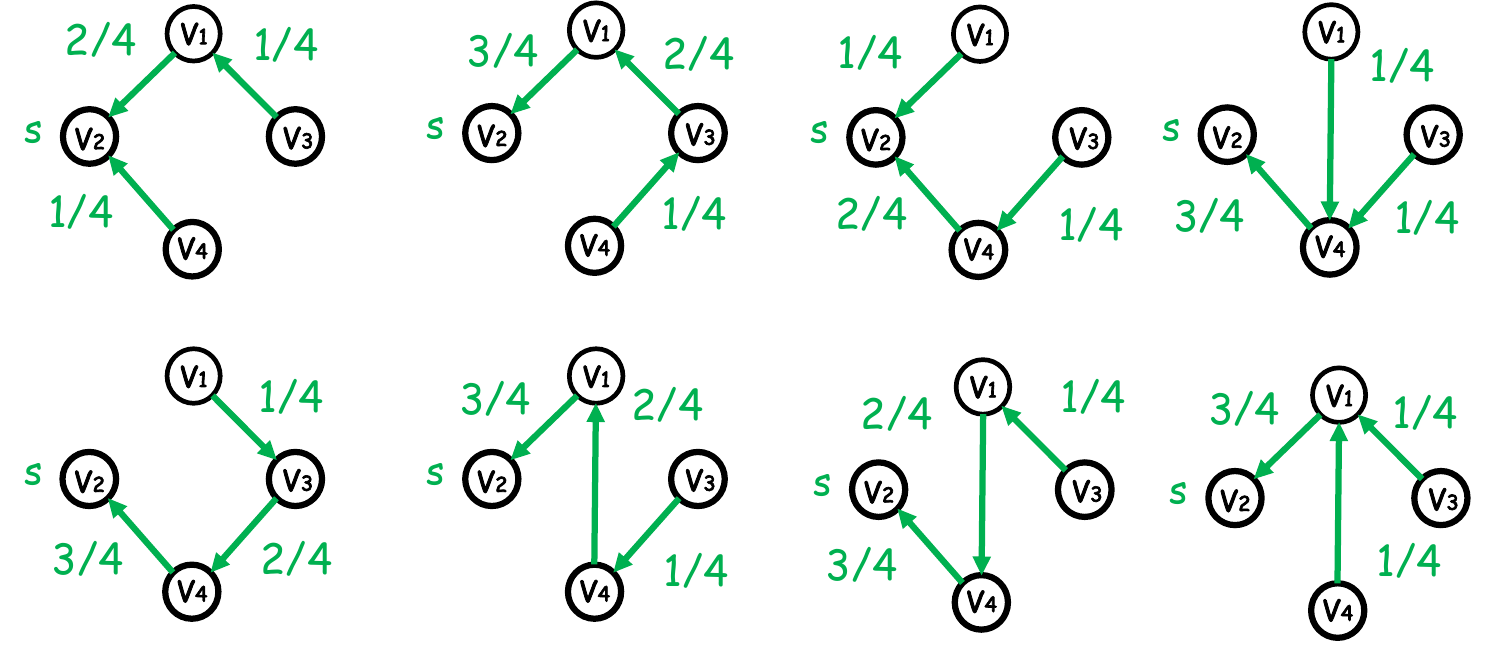}}
			}
		\end{tabular}
	\end{center}
	\vspace*{-0.5cm}
	\caption{\small Illustration of approximating $L^\dagger$ by spanning trees. (a) A graph $G$ and an electrical network flow where $\frac{1}{n}$ current flow comes in through every node $u\in V$ and a unit current flow comes out through $v_2$; (b) There are $8$ spanning trees. The current flows along an edge $(u, v)$ in $G$ equals the average current flows through $(u, v)$ on all spanning trees.}
	\vspace*{-0.4cm}
	\label{fig:flow-example}
\end{figure}
To better understand the result, we give an illustrative example. In Fig.~\ref{fig:flow-example}(a), there is an electrical system where $\frac{1}{n}$ ($\frac{1}{4}$) current flow comes in through every node in $G$ and $1$ current flow comes out through $v_2$. The current flow on each edge is also depicted in Fig.~\ref{fig:flow-example}(a). As we can see, the energy of the system is $(\frac{12}{32})^2+(\frac{12}{32})^2+(\frac{4}{32})^2+(\frac{4}{32})^2=\frac{5}{16}=(L^\dagger)_{22}$; $-(L^\dagger)_{23}$ is the voltage at node $v_3$ which is $\frac{3}{16}$. Thus, $(L^\dagger)_{23}=-\frac{3}{16}$. We can obtain the voltages by solving a Laplacian linear system $L\mathbf{x}=\mathbf{e}_v-\frac{\vec{1}}{n}$ \cite{pseudoinverse17}. However, deterministically solving such a system, even equipped with the fastest Laplacian solver \cite{LaplacianSolver}, is still very costly. To overcome this problem, we design an approximate algorithm for a column of $L^\dagger$ based on sampling spanning trees.

\stitle{Approximate the electrical flow via sampling spanning trees.} According to the "spanning tree" interpretation of $L_v^{-1}$, the current flows on graph $G$ are the average of the current flows on all spanning trees of $G$. Based on Lemma~\ref{lemma:l-dagger-electrical-system}, in order to compute the $v$-th column of $L^\dagger$, we need to send $\frac{1}{n}$ current flow in through node every node $u\in V$, and $1$ current flow out through node $v$ in the graph. Here, we can also send such flows on all spanning trees. Then, the current flows on the original graph is expected to be the average of the flows on spanning trees. Fig.~\ref{fig:flow-example} illustrates an example. There are $8$ spanning trees of $G$. In each spanning tree, we send $\frac{1}{4}$ current flow in through each node in $G$, and send $1$ current flow out through $v_2$. The current flow on each spanning tree is depicted in Fig,~\ref{fig:flow-example}(b). It is easy to verify that the current flows on $G$ is exactly the average of the current flows on all spanning trees. This gives us an interesting Monte Carlo algorithm: we can sample a small number of spanning trees, send flows on these spanning trees and take the average. Then, we get an approximate current flow of the desired electrical system. 

As we have discussed, we can utilize the Wilson algorithm \cite{wilson1996generating} to efficiently sample a spanning tree. Notice that computing the current flow on a spanning tree is relatively easy. We only need to traverse the paths from all nodes $u$ to a node root $v$ in the spanning tree, and add the current that injects node $u$ to each edge in the path. Since every node is only the endpoint of one edge in a spanning tree, we denote $\ps[u]$ the path support of edge $(u,\cdot)$, which is the total current through the edge out-edge of $u$ in the spanning tree. This can be processed in at most $n\Delta_G$ time, where $\Delta_G$ is the diameter of graph $G$. This process is often very efficient in practice, as real-world graphs typically exhibit a small-world property \cite{newman2018networks}.

\begin{algorithm}[t]
    \scriptsize
	\caption{\spanningtree (\ecc)}
	\label{algo:spanningtree-ecc}
	\LinesNumbered
	\KwIn{Graph $G$, a landmark node $v$, sample size $\omega$}
	\KwOut{$(\widetilde{L^\dagger})_{uu}$ as an unbiased estimation of $(L^\dagger)_{uu}$ for $u\in V$}
	Pre-compute a \bfs tree $\mathcal{T}_{\bfs}$ with root $v$, $\mathcal{P}_{\mathcal{T}_{\bfs}}[u]$ is the path from $u$ to $v$ in $\mathcal{T}_{\bfs}$;\\
	\For{$i=1:\omega$}{
		$\mathbf{x}_1(u)\leftarrow0$, $\mathbf{x}_2(u)\leftarrow0$ for $u\in V$;\\
	    Uniformly sample a spanning tree $\mathcal{T}$ using the Wilson algorithm, $\mathcal{T}$ stores in a vector $\nextarray$ where $\nextarray[u]$ records the endpoint of the edge $(u,\cdot)$ in $\mathcal{T}$, $\mathcal{P}_\mathcal{T}[u]$ is the path from $u$ to $v$ in $\mathcal{T}$;\\
	    \For{each $u\in V$}{
	    	\For{each edge $e=(e_1,e_2)\in \mathcal{P}_\mathcal{T}[u]$}{
	    		$\ps[e_1]\leftarrow \ps[e_1]+\frac{1}{n}$ \tcp*{Compute path support $\ps$ in $\mathcal{T}$;}
	    	}
	    }
	    Perform \dfs on $\mathcal{T}$ from node $v$, compute the visit time $\vis[u]$ and the finish time $\fin[u]$ for all nodes $u\in V$;\\
	    \For{each $u\in V$}{
	    	\For{each edge $e=(e_1,e_2)\in\mathcal{P}_{\mathcal{T}_{\bfs}}[u]$}{
	    		\If{$\nextarray[e_1]=e_2$ and $[\vis[e_1],\fin[e_1]]\subseteq[\vis[u],\fin[u]]$}{$\mathbf{x}_1(u)\leftarrow\mathbf{x}_1(u)+\frac{1}{\omega}$, $\mathbf{x}_2(u)\leftarrow \mathbf{x}_2(u)+\frac{\ps[e_1]}{\omega}$;}
	    		\If{$\nextarray[e_2]=e_1$ and $[\vis[e_2],\fin[e_2]]\subseteq[\vis[u],\fin[u]]$}{$\mathbf{x}_1(u)\leftarrow\mathbf{x}_1(u)-\frac{1}{\omega}$, $\mathbf{x}_2(u)\leftarrow \mathbf{x}_2(u)-\frac{\ps[e_2]}{\omega}$;}
	    	}
    	}
	}
	\For{each $u\in V$}{$(\widetilde{L^\dagger})_{uu}\leftarrow\mathbf{x}_1(u)-2\mathbf{x}_2(u)+\frac{1}{n}\sum_{u\in V}\mathbf{x}_2(u)$;}
	\Return $(\widetilde{L^\dagger})_{uu}$ for $u\in V$;
\end{algorithm}

\stitle{Algorithm details.} Algorithm~\ref{algo:spanningtree-ecc} illustrates the pseudo-code of our spanning tree sampling algorithm for approximating \ecc. The algorithm takes a landmark node $v$ as input, which is the "0" voltage node. We need to compute the electrical flows in the electrical system in Lemma~\ref{lemma:electrical-network} and Lemma~\ref{lemma:l-dagger-electrical-system}. According to the Kirchhoff's law, in order to compute the voltage of node $u$, we should compute the current flow along a path from $u$ to the "0" voltage node $v$. Thus, we fix a \bfs tree with root $v$ at first, which stores a path $\mathcal{P}_{\bfs}[u]$ from each node $u$ to the root $v$ (Line 1). Then we sample a number of spanning trees using the Wilson algorithm (Line 4). The results of the spanning tree will be stored in a vector "$\nextarray$" where $\nextarray[u]$ records the out-edge of $u$ in the spanning tree \cite{wilson1996generating}. For each spanning tree, we first compute the current flow on that spanning tree. Specifically, we traverse all the paths from each node $u\in V$ to the root node $v$, add the current flows injects each node $u$ on every edge in that path. The total path support $\ps[e_1]$ on each edge $(e_1,\cdot)$ is the current flow on that edge (Line 5-7). Note that the current flow for estimating $(L_v^{-1})_{uu}$ for $u\in V$ (unit current flows in through $u$, unit current flows out through $v$) is $1$ on each edge, so we do not need to compute the path support. Then, we compute the voltage of all nodes by traversing the paths stored in $\mathcal{T}_{\bfs}$. A question here is: how can we update all the voltages in linear time? To solve this problem, we perform a \dfs traversal on the sampled spanning tree $\mathcal{T}$ with root $v$ (Line 8). By computing the visit time "$\vis$" and finish time "$\fin$" in the \dfs procedure, we can answer whether a node is the ancestor of another node in a tree in constant time \cite{introduction}. Thus, we can also determine whether an edge is sampled in $\mathcal{T}$ in constant time. Next, for each node $u\in V$, we traverse the path $\mathcal{P}_{\mathcal{T}_{\bfs}}[u]$ from $u$ to $v$ in the pre-computed \bfs tree. For each edge $(e_1,e_2)$ in the path, we examine whether the spanning tree path from $u$ to $v$ by comparing the visit time and finish time. If $[\vis[e_1],\fin[e_1]]\subseteq[\vis[u],\fin[u]]$, then $e_1$ is the ancestor of $u$, thus the spanning tree path from $u$ to $v$ contains the edge $(e_1,e_2)$. We update the two voltages $\mathbf{x}_1(u)$ and $\mathbf{x}_2(u)$ accordingly (Line 9-14), where $\mathbf{x}_1(u)$ is the voltage of node $u$ in an electrical network that unit current comes in through $u$ and comes out through $v$, we have $E[\mathbf{x}_1(u)]=(L_v^{-1})_{uu}$; $\mathbf{x}_2(u)$ is the voltage on node $u$ in an electrical network where $\frac{1}{n}$ current comes in through each node $u\in V$ and comes out through node $v$. We have $E[\mathbf{x}_2(u)]=\frac{1}{n}\mathbf{e}_u^TL_v^{-1}\vec{1}$. Finally, we approximate $(L^\dagger)_{uu}$ for $u\in v$ by the formula $(\widetilde{L^\dagger})_{uu}=\mathbf{x}_1(u)-2\mathbf{x}_2(u)+\frac{1}{n}\sum_{u\in V}\mathbf{x}_2(u)$ (Line 15-16). According to Equation~(\ref{equation:l-sv}) and Equation~(\ref{equation:l-vv}), $(\widetilde{L^\dagger})_{uu}$ is an unbiased estimator of $(L^\dagger)_{uu}$.

\stitle{Algorithm analysis.} We first analyze the time complexity of Algorithm~\ref{algo:spanningtree-ecc}.

\begin{lemma}
	Let $\Delta_G$ denote the diameter of graph $G$. The time complexity of Algorithm~\ref{algo:spanningtree-ecc} is $O(Tr(I-P_v)^{-1}+n\Delta_G)$ per sample, with an $O(m+n)$ pre-processing cost.
	\comment{
	\begin{proof}
		 Precomputing a \bfs tree costs time $O(n+m)$. In each sample, sampling a spanning tree using the Wilson algorithm costs time $Tr((I-P_v)^{-1})$. Suppose $\Delta_G$ is the diameter of $G$, the complexity of computing the path support can be bounded by $O(n\Delta_G)$. Then, a \dfs on the spanning tree costs time $O(n)$. traverse each in the path $\mathcal{P}_b[u]$ for all nodes $u\in V$ also needs time $O(n\Delta_G)$. In practice, as we have discussed in Section~\ref{subsec:new-interpretation-st}, $O(Tr(I-P_v)^{-1})$ can be as small as $O(n)$. Since real-world graphs often has an $O(\log n)$ diameter, the complexity can be simplified to $O(n\log n)$. As a result, the time complexity of the algorithm is $O(n\log n)$ on real-life large graphs.
	\end{proof}}
\end{lemma}
In practice, as we have discussed in Section~\ref{subsec:new-interpretation-st}, $O(Tr(I-P_v)^{-1})$ can be as small as $O(n)$. Since real-world graphs often has an $O(\log n)$ diameter (small world \cite{newman2018networks}), the complexity can be simplified to $O(n\log n)$. As a result, the time complexity of the algorithm is $O(n\log n)$ on real-life large graphs. 

Next, we utilize the classical Hoeffding's inequality to show that Algorithm~\ref{algo:spanningtree-ecc} achieves an absolute error with a probability at least $1-p_f$, where $p_f$ is a small failure probability.
		\begin{lemma}\label{lemma:hoeffding-bound}(Hoeffding's inequality)
			Let $X_1,\cdots,X_\omega$ be independent random variables in $[a,b]$, where $-\infty <a\leq b<\infty$, for any $0<\epsilon<1$, we have
			\begin{equation*}
				Pr[|\frac{1}{\omega}\sum_{i=1}^\omega X_i-E[X]|\geq\epsilon]\leq 2exp(-\frac{2\epsilon^2\omega}{(b-a)^2}).
			\end{equation*}
		\end{lemma}
	\begin{theorem}\label{lemma:error-guarantee-spanning-tree-ecc}
		If the sample size $\omega\geq\frac{32\Delta_G^2\log (\frac{2n}{p_f})}{\epsilon^2}$, Algorithm~\ref{algo:spanningtree-ecc} outputs $(\widetilde{L^\dagger})_{uu}$ that satisfies $|(\widetilde{L^\dagger})_{uu}-(L^\dagger)_{uu}|\leq\epsilon$ for all $u\in V$ with a probability at least $1-p_f$.
	\end{theorem}
\comment{
		\begin{proof}
			Let $X_1$, $X_2$ and $X_3$ be the random variables which represent the amount we update $\mathbf{x}_1(u)$, $\mathbf{x}_2(u)$ and $\mathbf{x}_2(v)$ in Algorithm~\ref{algo:spanningtree-ecc}. We have $(L^\dagger)_{uu}=E[X_1+2X_2-X_3]$ and $(\widetilde{L^\dagger})_{uu}=\frac{X_1}{\omega}+2\frac{X_2}{\omega}-\frac{X_3}{\omega}$. Since the length of the path in the BFS tree can be bounded by $\Delta_G$, $|X_1|\leq\Delta_G$, $|X_3|\leq\Delta_G$. At the same time, the path support we computed in Line 6-8 is at most $1$, we also have $|X_2|\leq\Delta_G$. Thus, we have $|X_1+2X_2-X_3|\leq4\Delta_G$. Let $X=X_1+2X_2-X_3$, we have $E[X]=(L^\dagger)_{uu}$ and $(\widetilde{L^\dagger})_{uu}=\frac{\sum_{i=1}^\omega X_{(i)}}{\omega}$. By applying the Hoeffding's inequality, if $\omega\geq\frac{32\Delta_G^2\log(\frac{2n}{p_f})}{\epsilon^2}$, we can obtain that
				$Pr(|\frac{\sum_{i=1}^\omega X_{(i)}}{\omega}-(L^\dagger)_{uu}|\geq \epsilon)\leq 2exp(-\frac{2\epsilon^2\omega}{(8\Delta_G)^2})\leq 2exp(-\frac{2\epsilon^2\frac{32\Delta_G^2\log(\frac{2n}{p_f})}{\epsilon^2}}{64\Delta_G^2})\leq \frac{p_f}{n}$. By union bound, the probability that all nodes $u\in V$ satisfy such a condition is at least $1-p_f$.
	\end{proof}}
\subsection{A loop-erased walk sampling algorithm}\label{subsec:ecc-lewalk}
In this subsection, we propose a novel algorithm for approximating \ecc based on loop-erased walk sampling. According to Lemma~\ref{lemma:loop-erased-walk}, $(L_v^{-1})_{uu}$ for $u\in V$ and $u\neq v$ is the expected normalized number of passes to $u$ in a loop-erased walk $\lepath_v$. We also need to approximate $\mathbf{e}_u^TL_v^{-1}\vec{1}$ for $u\in V$ and $u\neq v$. We first state that $\mathbf{e}_u^TL_v^{-1}\vec{1}=\sum_{i\in V,\ i\neq v}{(L_v^{-1})_{ui}}$ is the expected normalized length of the loop-included path $\mathcal{P}_{\lepath_v}[u]$:
\begin{lemma}
	Let $|\mathcal{P}_{\lepath_v}[u]|_n$ denote the normalized length of the loop-included path $\mathcal{P}_{\lepath_v}[u]$, defined as the sum of $\frac{1}{d_u}$ for each node $u$ the path has passed (a node may exist for many times), then we have: $E[|\mathcal{P}_{\lepath_v}[u]|_n]=\mathbf{e}_u^TL_v^{-1}\vec{1}$.
	\comment{
	\begin{proof}
		According to Lemma~\ref{lemma:loop-erased-walk}, $(L_v^{-1})_{ui}=\tilde{\tau}_{\lepath_v}[u,i]=\frac{\tau_{\lepath_v}[u,i]}{d_i}$ is the expected normalized number of passes to a node $i$ in a loop-included path $\mathcal{P}_{\lepath_v}[u]$. Since a path length equals the sum of the number of passes to each node $u\in V$, $\mathbf{e}_u^TL_v^{-1}\vec{1}=\sum_{i\in V,\ i\neq v}{(L_v^{-1})_{ui}}=\sum_{i\in V,\ i\neq v}\tilde{\tau}_{\lepath_v}[u,i]=E[|\mathcal{P}_{\lepath_v}[u]|_n]$ is the expected normalized length of the loop-included path.
	\end{proof}}
\end{lemma}
\begin{algorithm}[t]
    \scriptsize
	\caption{\lewalk (\ecc)}
	\label{algo:lewalk-ecc}
	\LinesNumbered
	\KwIn{Graph $G$, a landmark node $v$, sample size $\omega$}
	\KwOut{$(\widetilde{L^\dagger})_{uu}$ as an unbiased estimation of $(L^\dagger)_{uu}$ for $u\in V$}
	Fix an arbitrary ordering $(v_1,\cdots,v_{n-1})$ of $V\setminus\{v\}$;\\
	$\mathbf{y}_1(u)\leftarrow0$, $\mathbf{y}_2(u)\leftarrow0$ for $u\in V$;\\
	\For{$i=1:\omega$}{
	    $\intree[u]\leftarrow \false$, $\nextarray[u]\leftarrow -1$ for $u\in V$;\\
	    $\intree[v]\leftarrow \true$;\\
	    \For{$j=1:n$}{
	        $u\leftarrow v_j$;\\
	        \While{$!\intree[u]$}{
	        	$\mathbf{y}_1(u)\leftarrow\mathbf{y}_1(u)+\frac{1}{\omega d_u}$;\\
	            $\nextarray[u]\leftarrow \randomneighbor(u)$;\\
	            $u\leftarrow \nextarray[u]$;\\
	        }
	        $u\leftarrow v_i$;\\
	        \While{$!\intree[u]$}{
	            $\intree[u]\leftarrow \true$, $u\leftarrow \nextarray[u]$;
	        }
	    }
	}
	\For{$i=1:\omega$}{
		Uniformly sample a node $s\in V$;\\
		$u\leftarrow s$;\\
		\While{u != v}{
			$\mathbf{y}_2(u)\leftarrow\mathbf{y}_2(u)+\frac{1}{\omega d_u}$, $u\leftarrow \randomneighbor(u)$;
		}
	}
	\For{each $u\in V$}{$(\widetilde{L^\dagger})_{uu}\leftarrow\mathbf{y}_1(u)-2\mathbf{y}_2(u)+\frac{1}{n}\Vert\mathbf{y}_2\Vert_1$;}
	\Return $(\widetilde{L^\dagger})_{uu}$ for $u\in V$;
\end{algorithm}
As a result, we can obtain an estimation of $\mathbf{e}_u^TL_v^{-1}\vec{1}$, $u\in V$ by traversing all the loop-included paths $\mathcal{P}_{\lepath_v}[u]$ and compute the normalized length $|\mathcal{P}_{\lepath_v}[u]|_n$ for an estimation of $\mathbf{e}_u^TL_v^{-1}\vec{1}$ for all $u\in V$. Recall that after we have sampled a loop-erased walk $\lepath_v$, we can traverse a loop-included path $\mathcal{P}_{\lepath_v}[u]$ in time $h(s,v)=\mathbf{e}_s^TL_v^{-1}\Vec{\mathbf{d}}_v$ if we store the stack representation of $\lepath_v$. However, traversing all the paths still requires time $\vec{1}^T(I-P_v)^{-1}\vec{1}$ which is much larger than the spanning tree sampling cost $Tr(I-P_v)^{-1}$. In practice, we can sample from the whole set of the loop-included paths.

\stitle{Sample from the paths.} First, we give a new result about $\mathbf{e}_u^TL_v^{-1}\vec{1}$ for $u\in V$:
\begin{lemma}\label{lemma:sample-path}
	For a loop-erased walk $\lepath_v$, let $\tilde{\tau}_{\lepath_v}^n[u]$ denote the expected normalized number of passes of node $u$ in a loop-included path $\mathcal{P}_{\lepath_v}[s]$ from a node $s$ sampled uniformly from $V$. We have: $\tilde{\tau}_{\lepath_v}^n[u]=\frac{1}{n}\mathbf{e}_u^TL_v^{-1}\vec{1}$.
	\comment{
	\begin{proof}
		According to Lemma~\ref{lemma:loop-erased-walk}, in a loop-erased walk $\lepath_v$, $(L_v^{-1})_{ui}=(L_v^{-1})_{iu}=\tilde{\tau}_{\lepath_v}[i,u]=\frac{\tau_{\lepath_v}[i,u]}{d_u}$ is the expected normalized number of passes to a node $u$ in a loop-included path $\mathcal{P}_{\lepath_v}[i]$ from $i$. Since $\tilde{\tau}_{\lepath_v}[s,s]=0$, the expected normalized number of passes of node $u$ in a loop-included path $\mathcal{P}_{\lepath_v}[s]$ from a node $s$ sampled uniformly from $V$ is $\sum_{s\in V}{Pr[s=u]\tilde{\tau}_{\lepath_v}[s,u]}=\sum_{s\in V,\ s\neq v}{Pr[s=u]\tilde{\tau}_{\lepath_v}[s,u]}=\sum_{s\in V,\ s\neq v}{\frac{1}{n}(L_v^{-1})_{su}}=\frac{1}{n}\mathbf{e}_u^TL_v^{-1}\vec{1}$.
	\end{proof}}
\end{lemma}

According to Lemma~\ref{lemma:sample-path}, we can sample a small number of nodes uniformly from $V$, and traverse a small fraction of loop-included paths from those nodes. By Lemma~\ref{lemma:sample-path}, it is easy to check that the average normalized passes to a node $u$ is an unbiased estimator of $\mathbf{e}_u^TL_v^{-1}\vec{1}$. On real-life graphs, since we only sample a small fraction of nodes, we do not need to store the whole stack representation of loop-erased walks. Instead, we can sample $v$-absorbed walks online, since the loop-included path $\mathcal{P}_{\lepath_v}[u]$ has the same distribution as $v$-absorbed path $\mathcal{P}_v[u]$. When choosing the node $v$ as an easy-to-hit node, the cost of sampling $v$-absorbed walks is very small compared to sampling loop-erased walks, as verified in our experiments.

\stitle{Algorithm details.} The resulting algorithm is illustrated in Algorithm~\ref{algo:lewalk-ecc}. It takes a landmark node $v$ and sample size $\omega$ as inputs. Following the Wilson's implementation of sampling loop-erased walks, Algorithm~\ref{algo:lewalk-ecc} first fixes an ordering of $V\setminus\{v\}$ (Line 1), then it samples $\omega$ loop-erased walks with root $v$ (Line 3-14). Specifically, for each loop-erased walk, it initializes two vectors $\intree$, which records the nodes that have been added into the loop-erased trajectory, and $\nextarray$, which records the next node in the loop-erased trajectory (Line 4). First, node $v$ is added into the trajectory (Line 5). Then, a random walk starts from nodes that have not been added into the trajectory until it hits nodes that are already in the trajectory (Line 6-11). After that, we retrace the random walk trajectory to erase loops, and add it into the loop-erased trajectory (Line 12-14). For each node $u$ passed, it adds $\mathbf{y}_1(u)$ by $\frac{1}{\omega d_u}$ (Line 9). We have $E[\mathbf{y}_1(u)]=(L_v^{-1})_{uu}$. After sampling the loop-erased walk, it samples $\omega$ $v$-absorbed walks from a uniformly sampled source node $s$ (Line 15-19). For each time it passes a node $u$, we add $\mathbf{y}_1(u)$ by $\frac{1}{\omega d_u}$ (Line 19). Finally, we obtain an estimation of $(L^\dagger)_{uu}$ based on the formula $(L^\dagger)_{uu}=(L_v^{-1})_{uu}-2\mathbf{e}_u^TL_v^{-1}\vec{1}+\frac{1}{n^2}\vec{1}^TL_v^{-1}\vec{1}$ (Line 21).

\stitle{Algorithm analysis.} The complexity of Algorithm~\ref{algo:spanningtree-ecc} is similar to that of Algorithm~\ref{algo:lewalk-ecc}, while it avoids the complicated operations on trees. Thus, it is more efficient in practice.
\begin{lemma}
	The time complexity of Algorithm~\ref{algo:lewalk-ecc} is $O(Tr(I-P_v)^{-1}+\sum_{u\in V}{\frac{1}{n}h(u,v)})$ for each sample.
	\comment{
	\begin{proof}
		In the first phase of sampling loop-erased walks (Line 3-14 in Algorithm~\ref{algo:lewalk-ecc}, the time complexity is $Tr(I-P_v)^{-1}$. In the second phase of sampling $v$-absorbed walks (Line 15-19 in Algorithm~\ref{algo:lewalk-ecc}), the expected length of a $v$-absorbed walk is $\sum_{u\in V}Pr[u=s]h(s,v)=\frac{1}{n}\vec{1}^T(I-P_v)^{-1}\vec{1}$. In real life datasets, there is always an easy-to-hit node $v$ (e.g., the highest-degree node). For such $v$, $\frac{1}{n}\vec{1}^T(I-P_v)^{-1}\vec{1}$ is much smaller than $Tr(I-P_v)^{-1}$. As a result, the time complexity of \lewalk is $O(Tr(I-P_v)^{-1})$ for each sample.
	\end{proof}}
\end{lemma}
In real-life graphs, there often exists an easy-to-hit node $v$ (e.g., the highest-degree node). For such $v$, $\sum_{u\in V}{\frac{1}{n}h(u,v)}$ is much smaller than $Tr(I-P_v)^{-1}$. As a result, the time complexity of \lewalk is $O(Tr(I-P_v)^{-1})$ for each sample, which is often $O(n)$ in practice. 

Since the normalized number of passes of a loop-erased walk cannot be bounded, concentration inequalities cannot be applied to analyze Algorithm~\ref{algo:lewalk-ecc}. However, we find that Algorithm~\ref{algo:lewalk-ecc} performs better than Algorithm~\ref{algo:spanningtree-ecc} in practice (see Section~\ref{subsec:exp-performance}). This is perhaps because when the landmark node $v$ is chosen as an easy-to-hit node, the probability that a node is passed twice in a loop-included path is very small, which makes the variance of the algorithm also small. As it is very hard to characterize the distribution of the number of passes to a node in a loop-erased walk, we leave the variance analysis of the loop-erased walk sampling algorithm as an open question.

\section{Algorithms for KC computation}\label{sec:kcc}
In this section, we focus on the problem of Kemeny's constant computation. Recall that Kemeny's constant is the trace of the pseudo-inverse of the normalized Laplacian $Tr(\mathcal{L}^\dagger)$. Based on Equation~(\ref{equation:ln-ss-lv}) and Equation~(\ref{equation:lv-vv-lv}), we have the following formula:
\begin{lemma}
	$Tr(\mathcal{L}^\dagger)=Tr((I-P_v)^{-1})-\frac{(\mathcal{L}^\dagger)_{vv}}{\bm{\pi}(v)}$.
	\comment{
	\begin{proof}
		According to Equation~(\ref{equation:ln-ss-lv}), $(\mathcal{L}^\dagger)_{ss}=\bm{\pi}(s)(2m\mathbf{e}_s^T L_v^{-1}\mathbf{e}_s-2\mathbf{e}_s^TL_v^{-1}\Vec{\mathbf{d}}_v+\frac{1}{2m}\Vec{\mathbf{d}}_v^TL_v^{-1}\Vec{\mathbf{d}}_v)$ for $s\neq v$; According to Equation~(\ref{equation:lv-vv-lv}), $(\mathcal{L}^\dagger)_{vv}=\bm{\pi}(v)(\frac{1}{2m}\Vec{\mathbf{d}}_v^TL_v^{-1}\Vec{\mathbf{d}}_v)$. Since $\bm{\pi}(u)=\frac{d_u}{2m}$ and $(I-P_v)^{-1}=L_v^{-1}D_v$, the trace can be formulated as:
		\begin{equation*}
			\begin{split}
				Tr(\mathcal{L}^\dagger) &= \sum_{u\in V}{(\mathcal{L}^\dagger)_{uu}} \\
				&= \sum_{u\in V,\ u\neq v}{2m\bm{\pi}(u)\mathbf{e}_u^T L_v^{-1}\mathbf{e}_u}-\sum_{u\in V,\ u\neq v}{2\bm{\pi}(u)\mathbf{e}_u^TL_v^{-1}\Vec{\mathbf{d}}_v}\\
				&\quad+\sum_{u\in V}\frac{\bm{\pi}(u)}{2m}\Vec{\mathbf{d}}_v^TL_v^{-1}\Vec{\mathbf{d}}_v\\
				&= Tr((I-P_v)^{-1}) - \frac{1}{2m}\Vec{\mathbf{d}}_v^TL_v^{-1}\Vec{\mathbf{d}}_v\\
				&= Tr((I-P_v)^{-1})-\frac{(\mathcal{L}^\dagger)_{vv}}{\bm{\pi}(v)}.
			\end{split}\end{equation*}\end{proof}}
\end{lemma}

According to this formula, we can also divide the task of computing Kemeny's constant into two stages: (i) computing the sum of $(L_v^{-1})_{uu}d_u$ (also $(I-P_v)^{-1}_{uu}$) for all $u\in V$ and $u\neq v$; (ii) computing $\frac{(\mathcal{L}^\dagger)_{vv}}{\bm{\pi}(v)}$ (also $\frac{1}{2m}\Vec{\mathbf{d}}_v^TL_v^{-1}\Vec{\mathbf{d}}_v$). Then, the Kemeny's constant can be easily obtained. In the following, we first review existing algorithms for computing the Kemeny's constant. Then, we develop two efficient algorithms based on sampling spanning trees and loop-erased walks. We also give analysis on the proposed algorithms.

\subsection{Existing solutions}
The direct method to compute Kemeny's constant is to apply eigen-decomposition on $\mathcal{L}$. This costs $O(n^3)$ time which is prohibitive. Zhang et. al. proposes a method \approxkemeny \cite{WWW20ApproxKemeny} based on solving Laplacian linear systems. Specifically, they utilize the Hutchinson's estimator that $Tr(\mathcal{L}^\dagger)=\frac{1}{\omega}\sum_{i=1}^\omega\mathbf{x}_i^T\mathcal{L}^\dagger\mathbf{x}_i$, where $\mathbf{x}_i$ is a random vector with the probability distribution $Pr[\mathbf{x}(i)\pm1]=\frac{1}{2}$, it is known that $E[\mathbf{x}_i^T\mathcal{L}^\dagger\mathbf{x}_i]=Tr(\mathcal{L}^\dagger)$ \cite{hutchinson1990stochastic}. Then, by arithmetic operation, they prove that $\frac{1}{\omega}\sum_{i=1}^\omega\mathbf{x}_i^T\mathcal{L}^\dagger\mathbf{x}_i=\frac{1}{\omega}\sum_{i=1}^\omega\Vert BL^\dagger\mathbf{y}_i\Vert_2^2$, where $\mathbf{y}_i=D^{\frac{1}{2}}(I-\frac{1}{2m}D^{\frac{1}{2}}\vec{1}\vec{1}^TD^{\frac{1}{2}})\mathbf{x}_i$. Here, instead of directly computing $L^\dagger$, they compute $L^\dagger\mathbf{y}_i$ by solving $\omega$ Laplacian linear system $L\mathbf{z}_i=\mathbf{y}_i$. For solving the Laplacian systems, they choose the state-of-the-art fastest Laplacian solver proposed in \cite{LaplacianSolver}. However, although the Laplacian solver is highly optimized, solving a number of Laplacian systems is still costly in large graphs. Recently, Li et. al. proposes a Monte Carlo algorithm \rw \cite{KDD21Kemeny} for approximating Kemeny constant. Their algorithm is based on the observation that $Tr(\mathcal{L}^\dagger)=n-1+\sum_{k=0}^\infty{(Tr(P^k)-1)}$. Then, they simulate random walks from each node $u$ with length $l$ for an estimation of the diagonal elements of $P^k$. If $k<l$, then $(P^k)_{uu}$ is the probability that a random walk from $u$ returns back to $u$ at the $k$-th step. After that, the Kemeny's constant can be approximated by adding up all the $(P^k)_{uu}$ terms. However, to achieve a small estimation error, $l$ is supposed to be very large (in \cite{KDD21Kemeny}, $l=25791$ in their experiments). The time complexity of \rw is $O(ln)$ per sample, which is costly with such a large $l$. In the following, we show that Kemeny's constant can also be approximated by sampling spanning trees and sampling loop-erased walks. The cost of sampling a spanning tree and sampling loop-erased walk with a root chosen as the landmark node $v$ is much lower than that of sampling simple random walks.

\subsection{A spanning tree sampling algorithm}\label{subsec:kc-spanningtree}
\begin{figure}[t!]
	\begin{center}
		\vspace*{-0.2cm}
		\begin{tabular}[t]{c}
			\hspace*{0cm}
			\subfigure[$G$, network flow]{
				\raisebox{0.15\height}{
					\includegraphics[width=0.205\columnwidth, height=2.6cm]{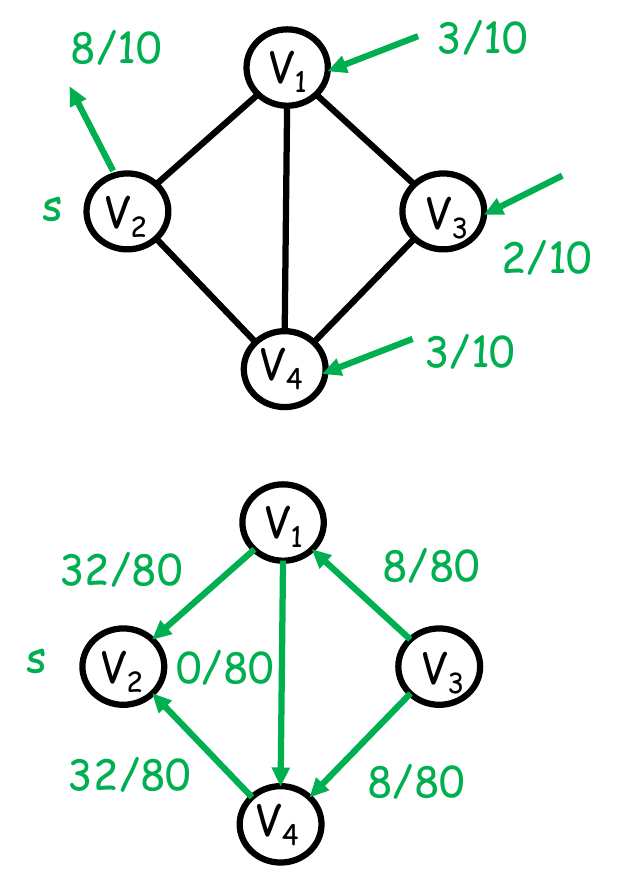}}
			}\hspace{0.08cm}
			\subfigure[flows on spanning trees]{
				\raisebox{0.1\height}{
					\includegraphics[width=0.7\columnwidth, height=2.6cm]{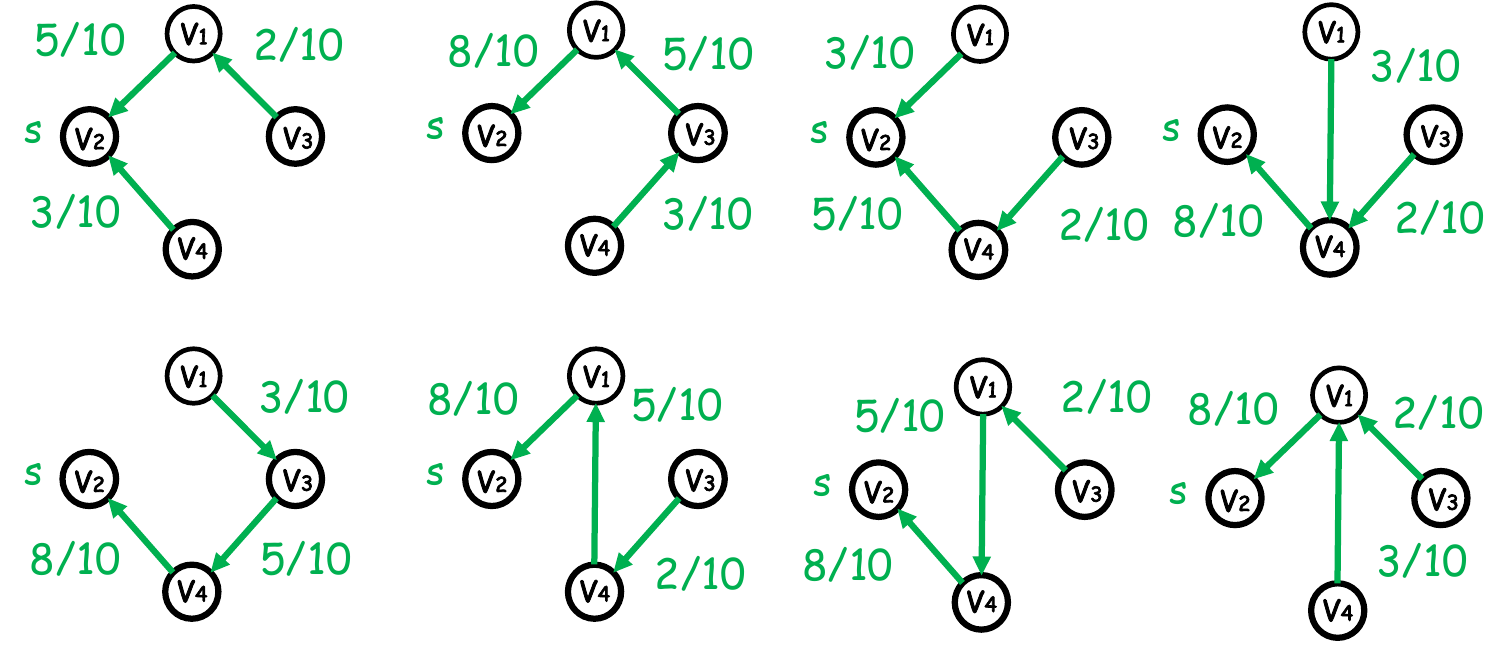}}
			}
		\end{tabular}
	\end{center}
	\vspace*{-0.5cm}
	\caption{\small Illustration of approximating $\mathcal{L}^\dagger$ by spanning trees. (a) A graph $G$ and an electrical network flow where $\frac{d_u}{2m}$ current flow comes in through every node $u\in V$ and a unit current flow comes out through $v_2$; (b) There are $8$ spanning trees. The current flows along an edge $(u, v)$ in $G$ equals the average current flows through $(u, v)$ on all spanning trees.}
	\vspace*{-0.4cm}
	\label{fig:flow-example-2}
\end{figure}

In this subsection, we propose a novel Monte Carlo algorithm to approximate Kemeny's constant based on spanning tree sampling. As mentioned in Section~\ref{subsec:ecc-spanningtree}, $(L_v^{-1})_{uu}$ for $u\in V$ and $u\neq v$ can be estimated using the spanning tree sampling algorithms proposed in \cite{22resistance, angriman2020approximation}. We can estimate $Tr((I-P_v)^{-1})=\sum_{u\in V,\ u\neq v}{(L_v^{-1})_{uu}d_u}$ by simply adding these elements. The problem remained is to approximate $\frac{1}{2m}\Vec{\mathbf{d}}_v^TL_v^{-1}\Vec{\mathbf{d}}_v$.  For $\frac{1}{2m}\Vec{\mathbf{d}}_v^TL_v^{-1}\Vec{\mathbf{d}}_v$, as it is a weighted sum of $L_v^{-1}$, we can estimate it in a similar way as estimating $\vec{1}^TL_v^{-1}\vec{1}$ based on the "spanning tree" interpretation of $L_v^{-1}$. First, we find that we can represent $\frac{1}{2m}\Vec{\mathbf{d}}_v^TL_v^{-1}\Vec{\mathbf{d}}_v$ in terms of electrical systems:
\begin{corollary}
	Consider an electrical network that $\frac{d_u}{2m}$ current flow comes in through every node $u\in V$, and unit current flow comes out through node $v$. Let $\mathbf{x}(u)$ be the voltage at node $u$. Then $\frac{(\mathcal{L}^\dagger)_{vv}}{\bm{\pi}(v)}$ is the weighted sum of voltage $\sum_{u\in V}\mathbf{x}(u)d_u$.
	\comment{
	\begin{proof}
		$\frac{(\mathcal{L}^\dagger)_{vv}}{\bm{\pi}(v)}=\frac{1}{2m}\Vec{\mathbf{d}}_v^TL_v^{-1}\Vec{\mathbf{d}}_v$. Consider the electrical system as a linear combination of $n$ electrical systems that $\frac{d_u}{2m}$ flow comes in through node $u\in V$, comes out through node $v$. According to Lemma~\ref{lemma:electrical-network}, the voltage at node a node $s\neq v$ is $\frac{d_u}{2m}(L_v^{-1})_{us}$. When combining $n$ such electrical systems, according to Kirchhoff's current law and Ohm's law, the voltage at node $s$ is $\mathbf{x}(s)=\frac{1}{2m}\sum_{u\in V}{(L_v^{-1})_{su}d_u}=\frac{1}{2m}\mathbf{e}_s^TL_v^{-1}\Vec{\mathbf{d}}_v$, this is based on the setting $\mathbf{x}(v)=0$. As a result, the weighted sum of the voltages is exactly $\sum_{s\in V}\mathbf{x}(s)d_s=\frac{1}{2m}\Vec{\mathbf{d}}_v^TL_v^{-1}\Vec{\mathbf{d}}_v$. Thus, the corollary is established.
	\end{proof}}
\end{corollary}
Based on the electrical system interpretation of $\frac{1}{2m}\Vec{\mathbf{d}}_v^TL_v^{-1}\Vec{\mathbf{d}}_v$, we can also approximate $\frac{1}{2m}\Vec{\mathbf{d}}_v^TL_v^{-1}\Vec{\mathbf{d}}_v$ by sampling spanning trees as directly solving these linear systems is hard. For example, as illustrated in Fig.~\ref{fig:flow-example-2}(a), $\frac{(\mathcal{L})_{22}}{\bm{\pi}(2)}=1.48$ is exactly the weighted sum of all voltages in the electrical network, which is $\frac{32*3}{80}+\frac{32*3}{80}+\frac{40*2}{80}$. And the electrical network flow is the average of the network flow on each spanning trees, as illustrated in Fig.~\ref{fig:flow-example-2}(b).
\begin{algorithm}[t]
	\scriptsize
	\caption{\spanningtree (\kc)}
	\label{algo:spanningtree-kc}
	\LinesNumbered
	\KwIn{Graph $G$, a landmark node $v$, sample size $\omega$}
	\KwOut{$\tilde{\kappa}(G)$ as an unbiased estimation of $\kappa(G)$}
	Pre-compute a \bfs tree $\mathcal{T}_{\bfs}$ with root $v$;\\
	\For{$i=1:\omega$}{
		$x_1\leftarrow0$, $x_2\leftarrow0$;\\
		Uniformly sample a spanning tree $\mathcal{T}$ (using the Wilson algorithm);\\
		\For{each $u\in V$}{
			\For{each edge $e=(e_1,e_2)\in \mathcal{P}_\mathcal{T}[u]$}{
				$\ps[e_1]\leftarrow \ps[e_1]+\frac{d_u}{2m}$ \tcp*{Compute path support $\ps$ in $\mathcal{T}$;}
			}
		}
		Do \dfs on $\mathcal{T}$ from node $v$, compute the visit time $\vis[u]$ and the finish time $\fin[u]$ for all nodes $u\in V$;\\
		\For{each $u\in V$}{
			\For{each edge $e=(e_1,e_2)\in\mathcal{P}_{\mathcal{T}_{BFS}}[u]$}{
				\If{$(e_1,e_2)\in \mathcal{T}$ and $[\vis[e_1],\fin[e_1]]\subseteq[\vis[u],\fin[u]]$}{$x_1\leftarrow x_1+\frac{d_u}{\omega}$, $x_2\leftarrow x_2+\frac{\ps[e_1] d_u}{\omega}$;}
				\If{$(e_2,e_1)\in \mathcal{T}$ and $[\vis[e_2],\fin[e_2]]\subseteq[\vis[u],\fin[u]]$}{$x_1\leftarrow x_1-\frac{d_u}{\omega}$, $x_2\leftarrow x_2-\frac{\ps[e_2] d_u}{\omega}$;}
			}
		}
	}
	$\tilde{\kappa}(G)\leftarrow x_1-x_2$;\\
	\Return $\tilde{\kappa}(G)$ as an estimation of $\kappa(G)$;
\end{algorithm}
It suggests that we can also sample spanning trees, send flows, and take the average (weighted sum of) voltages as an approximation of Kemeny's constant. The resulting algorithm is outlined in Algorithm~\ref{algo:spanningtree-kc}. This algorithm is very similar to Algorithm~\ref{algo:spanningtree-ecc}. The difference is that the current flows we send on spanning trees are different (Line 5-7), and in the case of approximating Kemeny's constant, we only need the (weighted) sum of the voltages instead of each one (Line 9-14). Finally, the two terms $x_1$ and $x_2$ are simply combined (Line 15). Since $E[x_1]=Tr(I-P_v)^{-1}$, $E[x_2]=\frac{1}{2m}\Vec{\mathbf{d}}_v^TL_v^{-1}\Vec{\mathbf{d}}_v$, we have $E[x_1-x_2]=\kappa(G)$. Similar to the analysis of Algorithm~\ref{algo:spanningtree-ecc}, the time complexity of Algorithm~\ref{algo:spanningtree-kc} is also $O(Tr(I-P_v)^{-1}+n\Delta_G)$ per sample. We can also give an absolute error guarantee by properly setting the sample size $\omega$.
\begin{theorem}\label{lemma:error-guarantee-spanning-tree-kc}
	If the sample size $\omega\geq\frac{8m^2\Delta_G^2\log(\frac{2}{p_f})}{\epsilon^2}$, Algorithm~\ref{algo:spanningtree-kc} outputs $\tilde{\kappa}(G)$ that satisfies $|\tilde{\kappa}(G)-\kappa(G)|\leq\epsilon$ with a probability at least $1-p_f$.
\end{theorem}
\comment{
\begin{proof}
	Let $X_1$ and $X_2$ be the random variables which represent the amount we update $x_1$ and $x_2$ in Algorithm~\ref{algo:spanningtree-kc} (Line 10-14). We have $\kappa(G)=E[X_1-X_2]$ and $\tilde{\kappa}(G)=\frac{X_1}{\omega}-\frac{X_2}{\omega}$. Notice that the length of the path $\mathcal{P}_{\mathcal{T}_{\bfs}}$ in the \bfs tree can be bounded by $\Delta_G$. The \bfs tree stores paths from each node $u\in V$ to the root. In the path from $\mathcal{P}_{\mathcal{T}_{\bfs}}$, if an edge is sampled in a random spanning tree, it updates $x_1$ by $d_u$, we have $|X_1|\leq \sum_{u\in V}d_u\Delta_G=m\Delta_G$. At the same time, the path support $\ps[u]$ we compute (Line 5-7) is at most $1$, we also have $|X_2|\leq m\Delta_G$. Thus, we have $|X_1-X_2|\leq2m\Delta_G$. Let $X=X_1-X_2$, we have $E[X]=\kappa(G)$ and $\tilde{\kappa}(G)=\frac{\sum_{i=1}^\omega X_{(i)}}{\omega}$. By applying the Hoeffding's inequality, if $\omega\geq\frac{8m^2\Delta_G^2\log(\frac{2}{p_f})}{\epsilon^2}$, we can obtain that
	$Pr(|\frac{\sum_{i=1}^\omega X_{(i)}}{\omega}-\kappa(G)|\geq \epsilon)\leq 2exp(-\frac{2\epsilon^2\omega}{(4m\Delta_G)^2})\leq 2exp(-\frac{2\epsilon^2\frac{8m^2\Delta_G^2\log(\frac{2}{p_f})}{\epsilon^2}}{16m^2\Delta_G^2})\leq p_f$. Thus, the theorem is established.
\end{proof}}

\subsection{A loop-erased walk sampling algorithm}
In this subsection, we propose the second novel Monte Carlo algorithm for approximating Kemeny's constant based on loop-erased walk sampling. By the "loop-erased walk" interpretation of $L_v^{-1}$, $Tr((I-P_v)^{-1})$ is exactly the expected number of walk lengths of a loop-erased walk $\lepath_v$ with root $v$. Thus,  $Tr((I-P_v)^{-1})$ can be approximated by sampling a number of loop-erased walks and take the average of the random walk lengths. On the other hand, $\frac{(\mathcal{L}^\dagger)_{vv}}{\bm{\pi}(v)}=\frac{1}{2m}\Vec{\mathbf{d}}_v^TL_v^{-1}\Vec{\mathbf{d}}_v$ can also be interpreted by loop-erased walks. Since $\frac{1}{2m}\Vec{\mathbf{d}}_v^TL_v^{-1}\Vec{\mathbf{d}}_v=\sum_{u\in V,\ u\neq v}{\bm{\pi}(u)\mathbf{e}_u^TL_v^{-1}\Vec{\mathbf{d}}_v}=\sum_{u\in V}\bm{\pi}(u)h(u,v)$, it can be represented as the expected length of a loop-included path $\mathcal{P}_{\lepath_v}[u]$ with node $u$ sampled from the stationary distribution $\bm{\pi}$. To approximate $\frac{1}{2m}\Vec{\mathbf{d}}_v^TL_v^{-1}\Vec{\mathbf{d}}_v$, we can traverse the stack representation of the loop-erased walks sampled for approximating $Tr((I-P_v)^{-1})$. Again, such an algorithm requires to store the stack representation of the loop-erased walks, which consumes $Tr((I-P_v)^{-1})$ space per sample. In order to save memory consumption, we can approximate $\frac{1}{2m}\Vec{\mathbf{d}}_v^TL_v^{-1}\Vec{\mathbf{d}}_v$ by sampling $v$-absorbed walks online (with little additional cost) since the loop-included path $\mathcal{P}_{\lepath_v}[u]$ has the same distribution as the $v$-absorbed walk path $\mathcal{P}_{v}[u]$.

The resulting algorithm is outlined in Algorithm~\ref{algo:lewalk-kc}. For each sample, it first simulates a loop-erased walk with root $v$ and record the walk length $y_1$ (Line 2). We have $E[y_1]=Tr((I-P_v)^{-1})$. Then, it simulates a $v$-absorbed walk from a node sampled from the stationary distribution, record the walk length $y_2$  (Line 3). We have $E[y_2]=\frac{1}{2m}\Vec{\mathbf{d}}_v^TL_v^{-1}\Vec{\mathbf{d}}_v$. After that, we can update the approximated Kemeny's constant $\tilde{\kappa}(G)$ accordingly (Line 4). Since $Tr(\mathcal{L}^\dagger)=Tr((I-P_v)^{-1}) - \frac{1}{2m}\Vec{\mathbf{d}}_v^TL_v^{-1}\Vec{\mathbf{d}}_v$, we have $E[y_1-y_2]=Tr(\mathcal{L}^\dagger)$.

The running time of Algorithm~\ref{algo:lewalk-kc} consists two parts: (i) the time of sampling loop-erased walks ($Tr((I-P_v)^{-1})$) and (ii)  the time of sampling $v$-absorbed walks ($\sum_{u\in V}\bm{\pi}(u)h(u,v)$). For real-world graphs, there often exists a high-degree node which is easy-to-hit. The value of $\sum_{u\in V}\bm{\pi}(u)h(u,v)$ is much smaller than $Tr((I-P_v)^{-1})$. Thus, the expected running time of Algorithm~\ref{algo:lewalk-kc} is $O(Tr((I-P_v)^{-1}))$ per sample. Similar to Algorithm~\ref{algo:lewalk-ecc}, concentration inequality such as Hoeffding's inequality cannot be applied to analyze Algorithm~\ref{algo:lewalk-kc}, because the random walk length of a loop-erased walk cannot be bounded. Nevertheless, the empirical performance of Algorithm~\ref{algo:lewalk-kc} in real-life graphs is very good (even much better than Algorithm~\ref{algo:spanningtree-kc}), as evidenced in our experiments.
\begin{algorithm}[t]
	\scriptsize
	\caption{\lewalk (\kc)}
	\label{algo:lewalk-kc}
	\LinesNumbered
	\KwIn{Graph $G$, a landmark node $v$, sample size $\omega$}
	\KwOut{$\tilde{\kappa}(G)$ as an unbiased estimation of $\kappa(G)$}
	\For{$i=1:\omega$}{
		Run a loop-erased random walk with root $v$, record the walk length $y_1$;\\
		Sample a node $s$ from the stationary distribution $\bm{\pi}$, run a $v$-absorbed walk from $s$, record the walk length $y_2$;\\
		$\tilde{\kappa}(G)\leftarrow\tilde{\kappa}(G)+\frac{y_1}{\omega}-\frac{y_2}{\omega}$;
	}
	\Return $\tilde{\kappa}(G)$;
\end{algorithm}

\stitle{Discussions.} Although our main focus in this study is on approximating the electrical closeness centrality and Kemeny's constant, it is worth noting our approaches can also be applied to approximate other graph metrics, such as the Kirchhoff index \cite{kirchhoff08} and random walk betweenness centrality \cite{RandomWalkBetweeness}. Specifically, the Kirchhoff index \cite{kirchhoff08} is defined as $Tr(L^\dagger)$. In Section~\ref{sec:ecc}, we have already estimated the diagonal elements $(L^\dagger)_{uu}$ for $u\in V$. To obtain an estimation of the Kirchhoff index, we simply need to sum up all these diagonal elements. The random walk betweenness centrality $\mathbf{c}_b$ \cite{RandomWalkBetweeness} is defined as $\mathbf{c}_b(u)=\frac{1}{n}+\frac{Tr(L^\dagger)}{(n-1)(Tr(L^\dagger)+n(L^\dagger)_{uu})}$ for $u\in V$ \cite{angriman2020approximation}. Similarly, to approximate this quantity, we can sum up the diagonal elements $(L^\dagger)_{uu}$ for $u\in V$ using the proposed techniques.

\section{Experiments}\label{sec:exp}
\comment{
In this section, we conduct experiments to evaluate the performance of our algorithms for approximating \ecc and \kc. We also conduct case studies to verify the effectiveness of \ecc (\kc) as a graph centrality metric (graph invariant metric).}
\subsection{Experimental settings}\label{subsec:exp-setting}
\stitle{Datasets.} We use $5$ real-life datasets: $2$ small graphs and $3$ large graphs. All datasets can be obtained from \cite{snapnets} and the detailed statistics is shown in Table~\ref{tab:datasets}. Note that we also estimate $\frac{Tr((I-P_v)^{-1})}{n}$ for each dataset by simulating $10^4$ loop-erased walks with root $v$ as the highest-degree node and take the average steps. We can observe that real-life datasets often have a small value of $\frac{Tr((I-P_v)^{-1})}{n}$. This confirms our analysis that sampling a loop-erased walk has an $O(n)$ time complexity on real-life datasets. For \ecc approximation, the result is a vector $\mathbf{c}$ with $\mathbf{c}(u)$ represents the \ecc of node $u$ for all $u\in V$. For \kc approximation, the result is a constant $\kappa(G)$. On two small graphs, we apply eigen-decomposition to compute the exact value of \ecc and \kc. On three large graphs, it is hard to derive the exact values. Thus, we use our best algorithm \lewalk with $\omega=10^6$ for both \ecc approximation and \kc approximation to obtain "ground-truths". As verified in the experiments (see Section~\ref{subsec:exp-performance}), the variance of our algorithm is very small with that setting of $\omega$, therefore we can expect that the estimated "ground-truth" values are very close to their true values.
\begin{table}[t]\vspace*{-0.2cm}
	\scriptsize
	\centering
	\caption{\small Datasets ($\Bar{d}$: average degree; $\Delta_G$: diameter of the graph)} \label{tab:datasets}
	\vspace{-0.3cm}
	\begin{tabular}{cccccccc}
		\toprule
		\bf Type & \bf Dataset & $n$ & $m$ & \bf $\Bar{d}$ & $\Delta_G$ & $\frac{Tr((I-P_v)^{-1})}{n}$\cr\midrule
		Small & \astroph & 17,903 & 196,972 & 22 & 14 & 1.33 \cr
		graphs & \emailenron & 33,696 & 180,811 & 10.73 & 13 & 1.43 \cr \hline
		& \youtube & 1,134,890 & 2,987,624 & 5.27 & 24 & 1.55 \cr
		Large & \pokec & 1,632,803 & 22,301,964 & 27.32  & 14 & 1.07 \cr
		graphs & \orkut & 3,072,441 & 117,184,899 & 76.28 & 10 & 1.02 \cr
		\bottomrule
	\end{tabular}
	\vspace*{-0.5cm}
\end{table}
\begin{figure*}
	\vspace*{-0.8cm}
	\begin{center}
		\begin{tabular}[t]{c}
			\subfigure[{\scriptsize \astroph}]{
				\includegraphics[width=0.38\columnwidth, height=2.2cm]{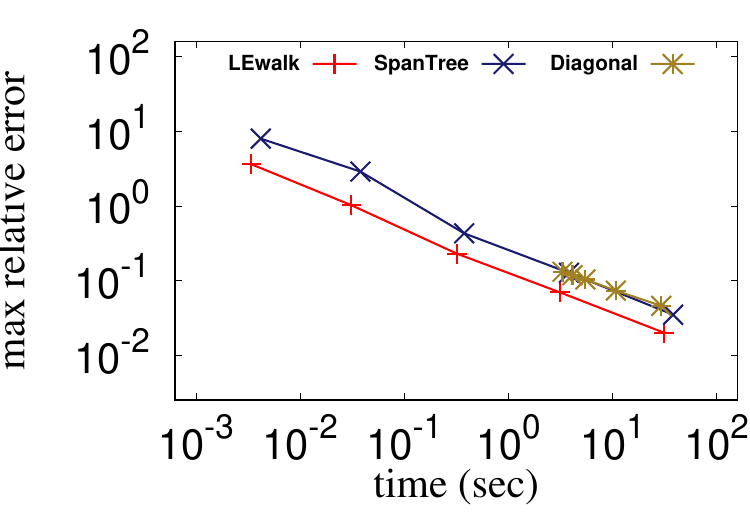}
			}
			\subfigure[{\scriptsize \emailenron}]{
				\includegraphics[width=0.38\columnwidth, height=2.2cm]{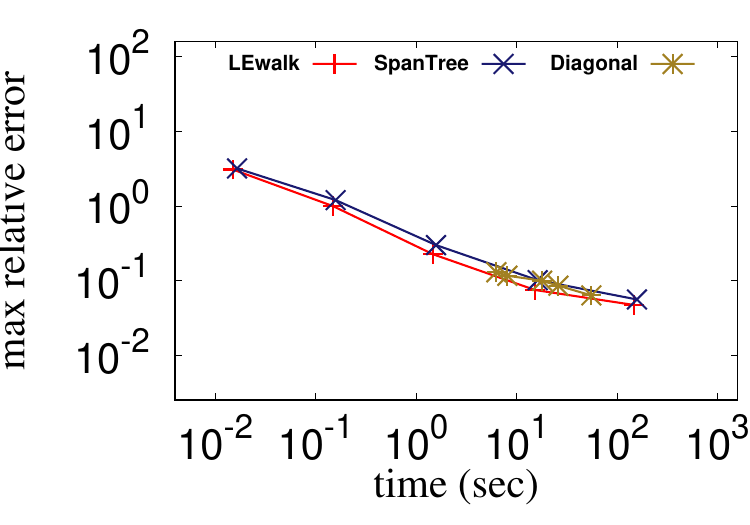}
			}
			\subfigure[{\scriptsize \youtube}]{
				\includegraphics[width=0.38\columnwidth, height=2.2cm]{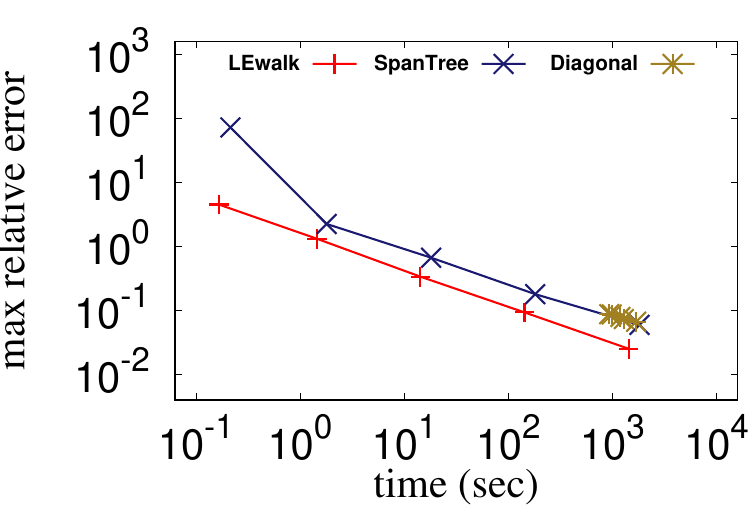}
			}
			\subfigure[{\scriptsize \pokec}]{
				\includegraphics[width=0.38\columnwidth, height=2.2cm]{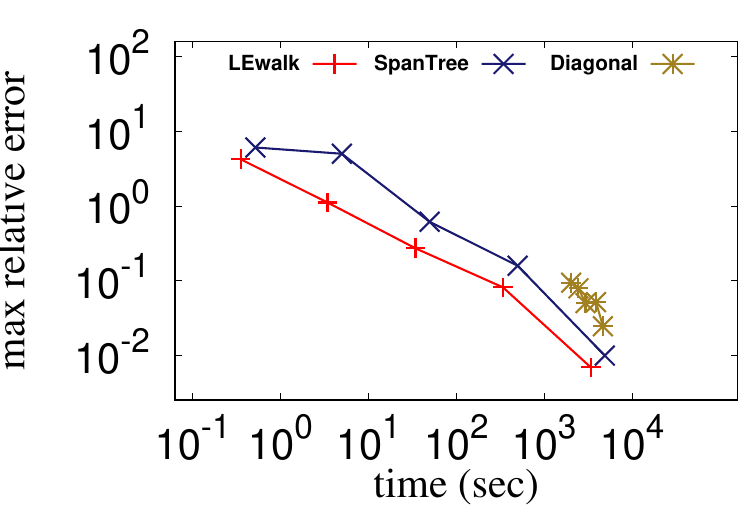}
			}
			\subfigure[{\scriptsize \orkut}]{
				\includegraphics[width=0.38\columnwidth, height=2.2cm]{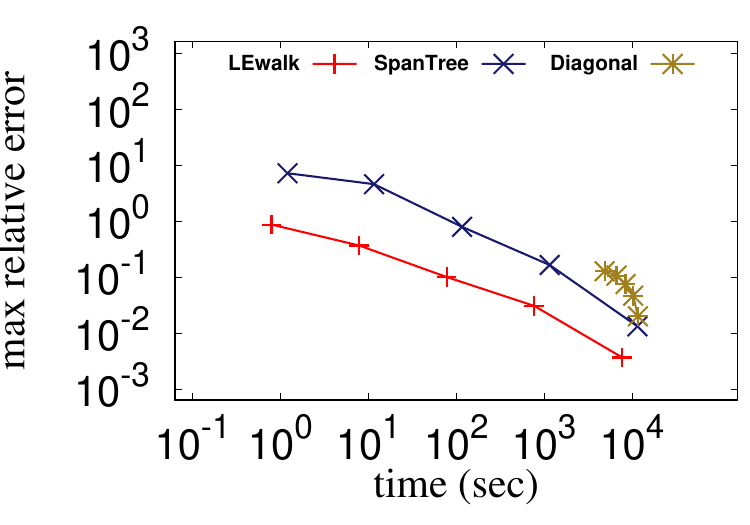}
			}
		\end{tabular}
	\end{center}
	\vspace*{-0.5cm}
	\caption{\small Maximum relative error v.s. runtime of different algorithms for approximating electrical closeness centrality}
	\vspace*{-0.2cm}
	\label{fig:relative-err-time-ecc}
\end{figure*}
\begin{figure}
	\vspace*{-0.5cm}
	\begin{center}
		\begin{tabular}[t]{c}
			\subfigure[{\scriptsize \youtube}]{
				\includegraphics[width=0.38\columnwidth, height=2.2cm]{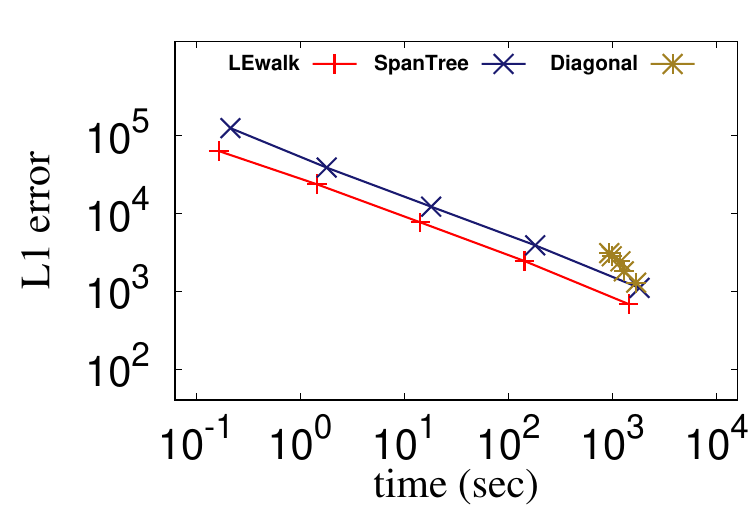}
			}
			\subfigure[{\scriptsize \orkut}]{
				\includegraphics[width=0.38\columnwidth, height=2.2cm]{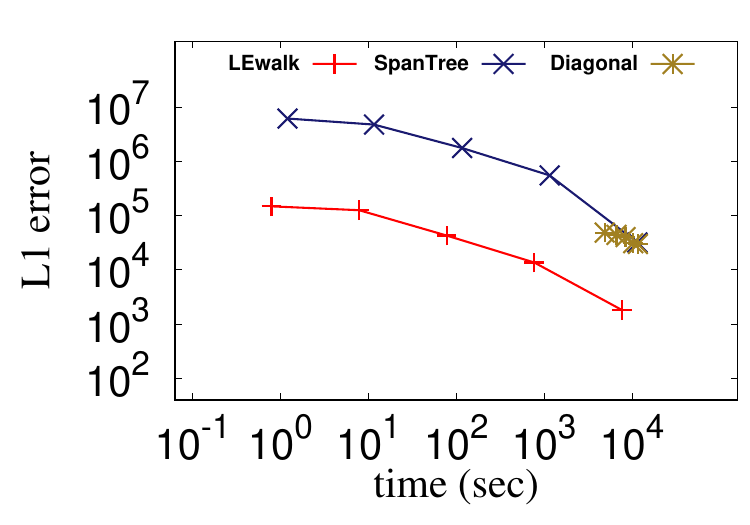}
			}
		\end{tabular}
	\end{center}
	\vspace*{-0.5cm}
	\caption{\small $L1$-error v.s. runtime of different algorithms for approximating electrical closeness centrality}
	\vspace*{-0.4cm}
	\label{fig:l1-err-time-ecc}
\end{figure}
\begin{figure*}
	\vspace*{-0.6cm}
	\begin{center}
		\begin{tabular}[t]{c}
			\subfigure[{\scriptsize \astroph}]{
				\includegraphics[width=0.38\columnwidth, height=2.2cm]{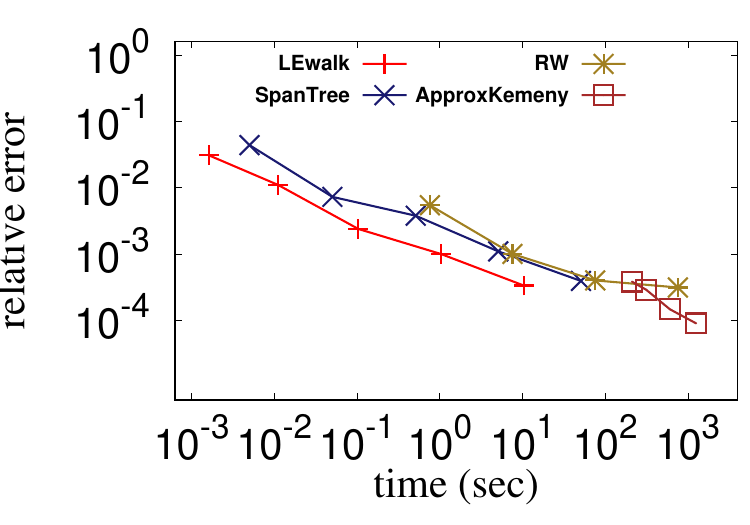}
			}
			\subfigure[{\scriptsize \emailenron}]{
				\includegraphics[width=0.38\columnwidth, height=2.2cm]{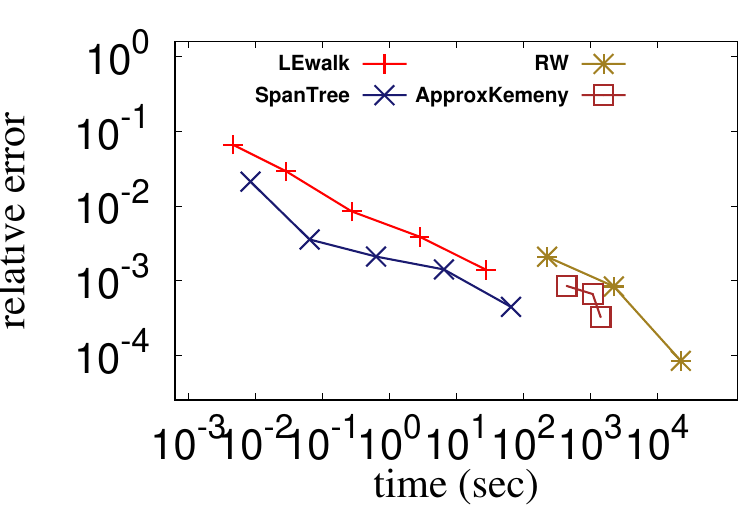}
			}
			\subfigure[{\scriptsize \youtube}]{
				\includegraphics[width=0.38\columnwidth, height=2.2cm]{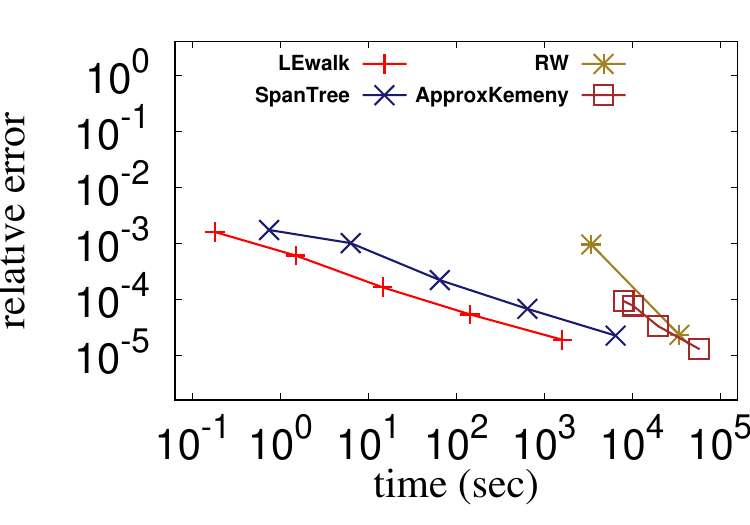}
			}
			\subfigure[{\scriptsize \pokec}]{
				\includegraphics[width=0.38\columnwidth, height=2.2cm]{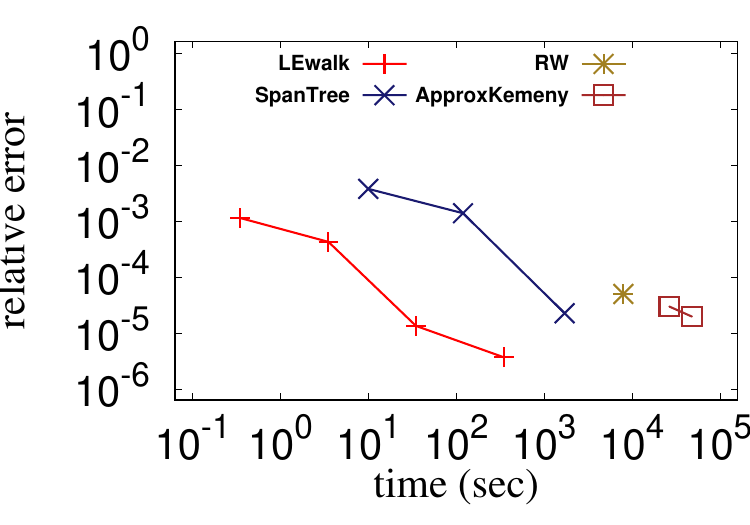}
			}
			\subfigure[{\scriptsize \orkut}]{
				\includegraphics[width=0.38\columnwidth, height=2.2cm]{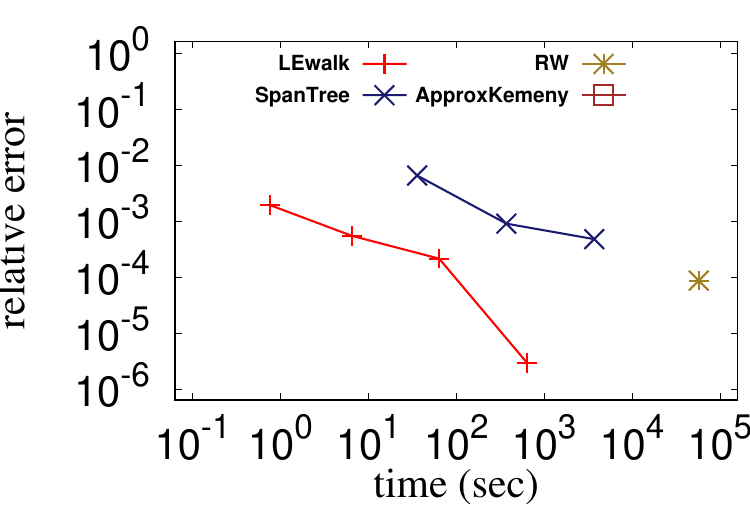}
			}
		\end{tabular}
	\end{center}
	\vspace*{-0.5cm}
	\caption{\small Relative error v.s. runtime of different algorithms for approximating Kemeny's constant}
	\vspace*{-0.2cm}
	\label{fig:relative-err-time-kc}
\end{figure*}
\begin{figure}
	\vspace*{-0.5cm}
	\begin{center}
		\begin{tabular}[t]{c}
			\subfigure[{\scriptsize \youtube}]{
				\includegraphics[width=0.38\columnwidth, height=2.2cm]{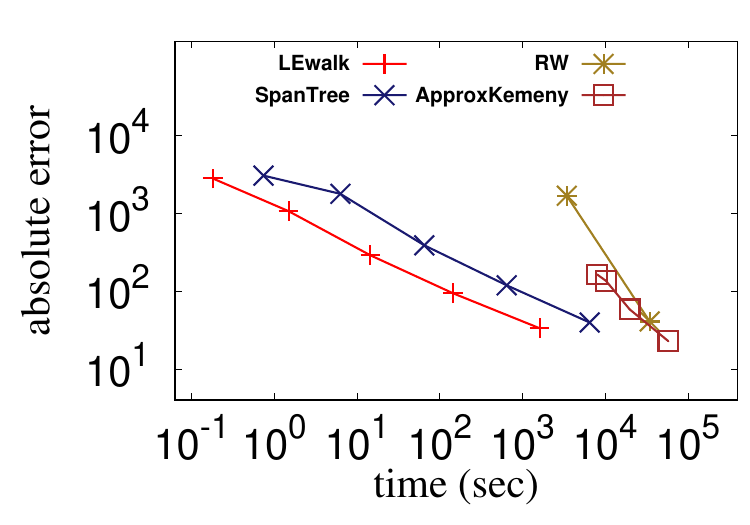}
			}
			\subfigure[{\scriptsize \orkut}]{
				\includegraphics[width=0.38\columnwidth, height=2.2cm]{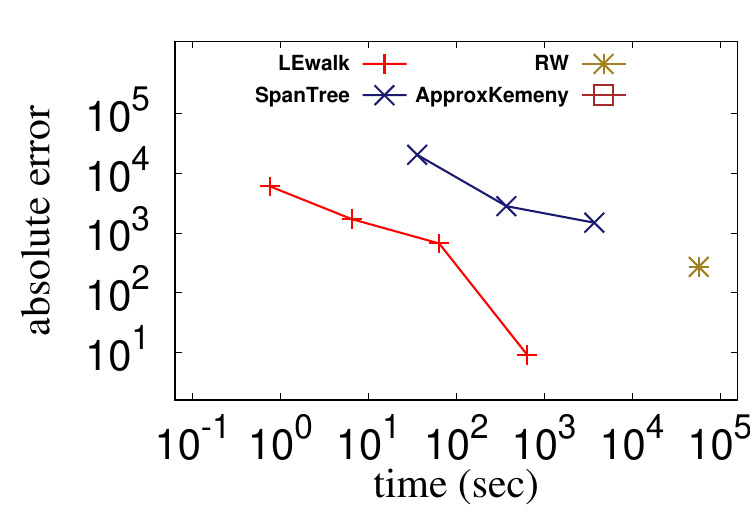}
			}
		\end{tabular}
	\end{center}
	\vspace*{-0.5cm}
	\caption{\small Absolute error v.s. runtime of different algorithms for approximating Kemeny's constant}
	\vspace*{-0.2cm}
	\label{fig:absolute-err-time-kc}
\end{figure}
\begin{figure}
	\vspace*{-0.5cm}
	\begin{center}
		\begin{tabular}[t]{c}
			\subfigure[{\scriptsize \ecc, relative error}]{
				\includegraphics[width=0.38\columnwidth, height=2.2cm]{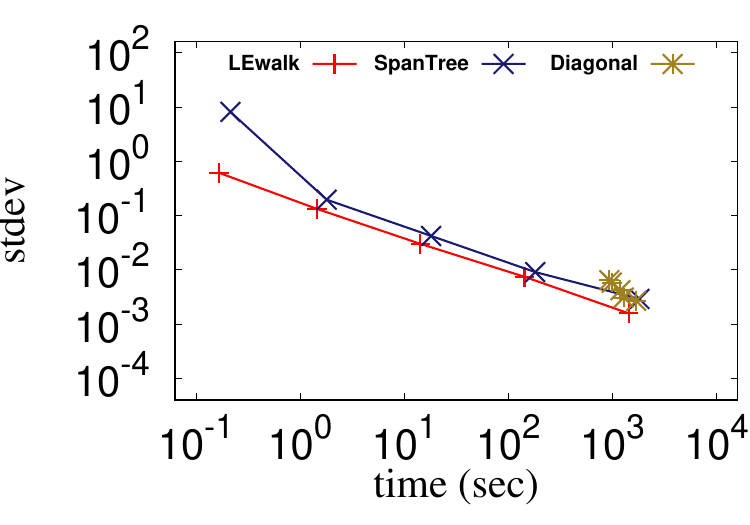}
			}
			\subfigure[{\scriptsize \ecc, $L1$-error}]{
				\includegraphics[width=0.38\columnwidth, height=2.2cm]{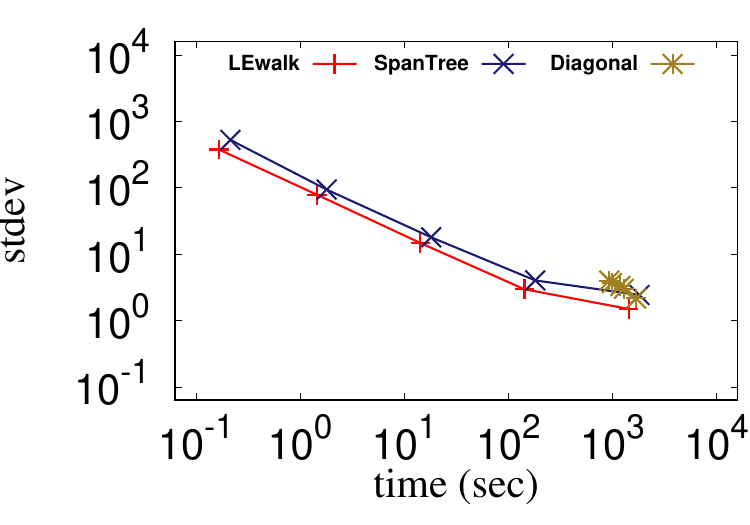}
			}\vspace{-0.2cm}\\
			\subfigure[{\scriptsize \kc, relative error}]{
				\includegraphics[width=0.38\columnwidth, height=2.2cm]{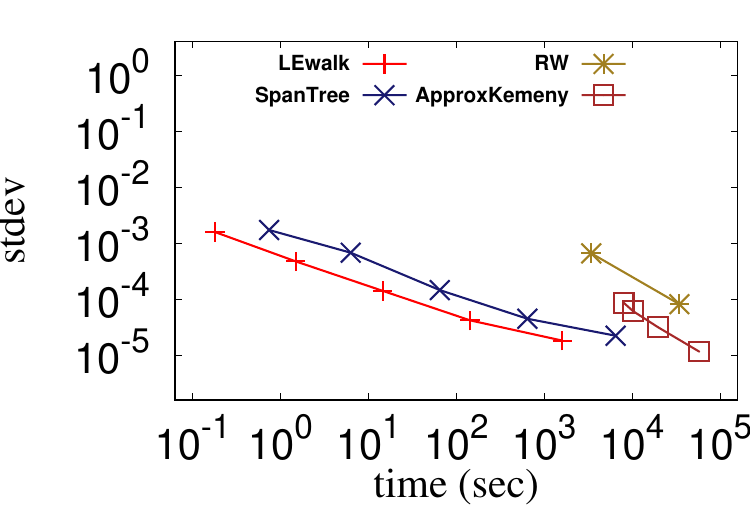}
			}
			\subfigure[{\scriptsize \kc, absolute error}]{
				\includegraphics[width=0.38\columnwidth, height=2.2cm]{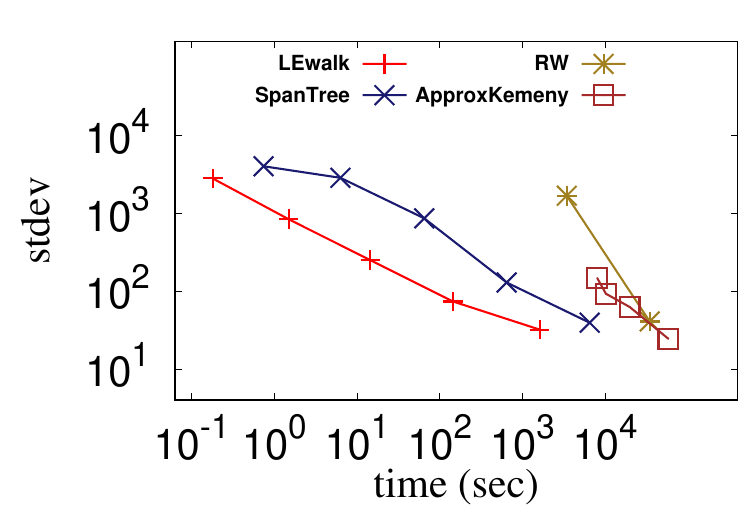}
			}
		\end{tabular}
	\end{center}
	\vspace*{-0.5cm}
	\caption{\small Standard deviation of error v.s. average runtime of different algorithms on \youtube}
	\vspace*{-0.2cm}
	\label{fig:stdev}
\end{figure}

\stitle{Parameters.} For \ecc approximation, we implement two algorithms, \spanningtree and \lewalk. There is one parameter $\omega$ which controls the sample size. We compare the proposed algorithms with the state-of-the-art (SOTA) method \diagonal \cite{angriman2020approximation}. There are two parameters $\epsilon$, $\kappa$ in \diagonal which controls the accuracy of the algorithm. We set $\kappa=0.3$ as default and vary $\epsilon$ following \cite{angriman2020approximation}. We do not include other algorithms because they are all outperformed by \diagonal \cite{angriman2020approximation}. For \ecc approximation, we also implement two our algorithms \spanningtree and \lewalk. There is also one parameter $\omega$ that controls the sample size. We compare the two proposed algorithms with two SOTA methods, \rw \cite{KDD21Kemeny} and \approxkemeny \cite{WWW20ApproxKemeny}. \rw has one parameter $\omega$ to control the sample size. \approxkemeny has one parameter $\epsilon$ to control the accuracy. For all the proposed algorithms, there is also a landmark node $v$ that we take as input. We set $v$ as the highest-degree node by default, and we will evaluate the other choices of $v$ in Section~\ref{subsec:exp-performance}.

\stitle{Experimental environment.} All our algorithms are implemented in C++. All experiments are run on a Linux 20.04 server with Intel 2.0 GHz CPU and 128GB memory. All algorithms used in our experiments are complied using GCC9.3.0 with -O3 optimization.

\subsection{Performance studies}\label{subsec:exp-performance}
\stitle{Results of \ecc computation.}
In this experiment, we evaluate the performance of algorithms for approximating \ecc. We plot the error v.s. time results for the two proposed algorithms \spanningtree and \lewalk, as well as the SOTA method \diagonal. For \spanningtree and \lewalk, we vary the sample size $\omega$ from $10^0$ to $10^5$. For \diagonal, we vary $\epsilon$ from $0.9$ to $0.3$ following the settings in \cite{angriman2020approximation}. We evaluate the error of \ecc vector by the maximum relative error defined as $\max_{u\in V}\frac{|\mathbf{c}(u)-\tilde{\mathbf{c}}(u)|}{\mathbf{c}(u)}$. We also evaluate the error by the $L1$-error which is defined as $\sum_{u\in V}{|\tilde{\mathbf{c}}(u)-\mathbf{c}(u)|}$. We run each algorithm for $100$ times and report the average results. The results are shown in Fig.~\ref{fig:relative-err-time-ecc} and Fig.~\ref{fig:l1-err-time-ecc}. For the $L1$-error metric, we only show the results on \youtube and \orkut, and the results on the other datasets are consistent (details can be found in \cite{fullversion}). As can be seen, \spanningtree can achieve comparable and slightly better relative error and $L1$-error compared to \diagonal within similar time. For example, on \pokec, when taking $4000$ seconds to compute the \ecc, \diagonal can achieve a max relative error $0.01$ while \spanningtree can achieve a max relative error as small as $0.006$. Note that \spanningtree avoids the calling of highly-optimized Laplacian solvers as invoked in \diagonal. It is a "pure" spanning tree sampling algorithm which is much easier to implement. \lewalk can achieve a certain accuracy in terms of both relative error and $L1$-error in much less time compared to \spanningtree and \diagonal. For example, on \orkut, \lewalk takes $774$ seconds while \spanningtree takes $12,540$ seconds to achieve a max relative error $0.01$. These results demonstrate the high efficiency of our loop-erased walk sampling approach.

\stitle{Results of \kc computation.} In this experiment, we study the performance of algorithms for approximating \kc. We also plot the error v.s. time results for two proposed algorithms \spanningtree and \lewalk, as well as two SOTA methods \rw \cite{KDD21Kemeny} and $\approxkemeny$ \cite{WWW20ApproxKemeny}. For \spanningtree, \lewalk, and \rw, we vary the sample size $\omega$ from $10^0$ to $10^5$, following a similar experimental setting in \cite{KDD21Kemeny}. For \approxkemeny, we vary the accuracy parameter $\epsilon$ from $0.9$ to $0.3$ following \cite{WWW20ApproxKemeny}. We measure the quality of results by the relative error defined as $\frac{|\tilde{\kappa}(G)-\kappa(G)|}{\kappa(G)}$, as well as the absolute error defined as $|\tilde{\kappa}(G)-\kappa(G)|$. We run each algorithm for $100$ times and report the average results. We omit the results when the running time of an experiment exceeds $24$ hours. The results are shown in Fig.~\ref{fig:relative-err-time-kc} and Fig.~\ref{fig:absolute-err-time-kc}. For the absolute error metric, we show the results of \youtube and \orkut, and the results on other datasets are consistent (details can be found in  \cite{fullversion}). As can be seen, the running time of our algorithms are orders of magnitude faster than the SOTA algorithms. For example, on \youtube, to achieve a relative error $10^{-4}$, \rw takes $34,637$ seconds, $\approxkemeny$ takes $45,609$ seconds, while the proposed algorithm \lewalk takes only $1,615$ seconds, which is $22\times$ faster than the SOTA algorithms. For the proposed algorithms, \lewalk is slightly faster than \spanningtree. On large datasets \youtube and \orkut, \lewalk can achieve an absolute error as small as $10^1$ in hundreds of seconds, and its relative error is also extremely small ($10^{-4}$). These results demonstrate that our loop-erased walk sampling based algorithms are very efficient in computing Kemeny's constant.

\stitle{Variance analysis of our algorithms.} The standard deviation of Monte Carlo algorithms is an important metric to evaluate the stability, thus the quality of the results. In former experiments, we only report the average results. Here we evaluate the standard deviation of the proposed algorithms for both \ecc approximation and \kc approximation. We conduct the experiments following the former experimental settings. Since the running time of the algorithms is quite consistent (the standard deviation is very small), we plot the standard deviation of error (maximum relative error and $L1$-error for \ecc approximation; relative error and absolute error for \kc approximation) v.s. average runtime of all the experiments. Fig.~\ref{fig:stdev} shows the results on \youtube. Similar results can also be observed on the other datasets. We can see that as the time grows, the standard deviation becomes smaller together with the average error. In similar running time, our algorithms have smaller standard deviation compared to the SOTA algorithms.

\begin{figure}
	\vspace*{-0.5cm}
	\begin{center}
		\begin{tabular}[t]{c}
			\subfigure[{\scriptsize \emailenron}]{
				\includegraphics[width=0.38\columnwidth, height=2.0cm]{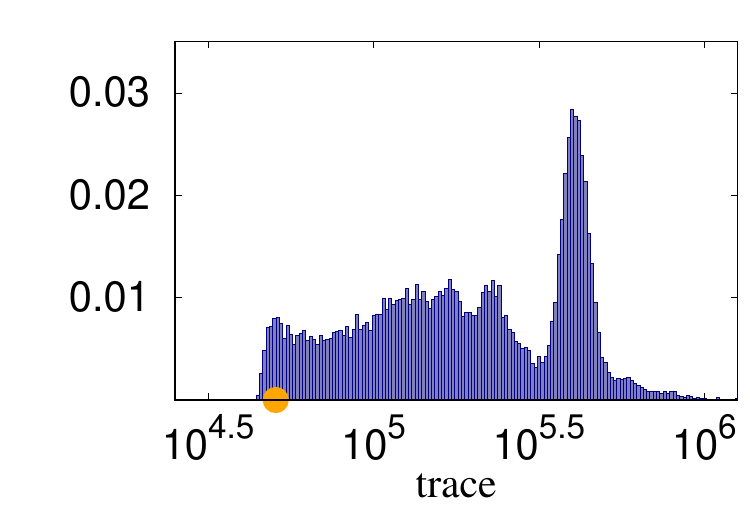}
			}
			\subfigure[{\scriptsize \youtube}]{
				\includegraphics[width=0.38\columnwidth, height=2.0cm]{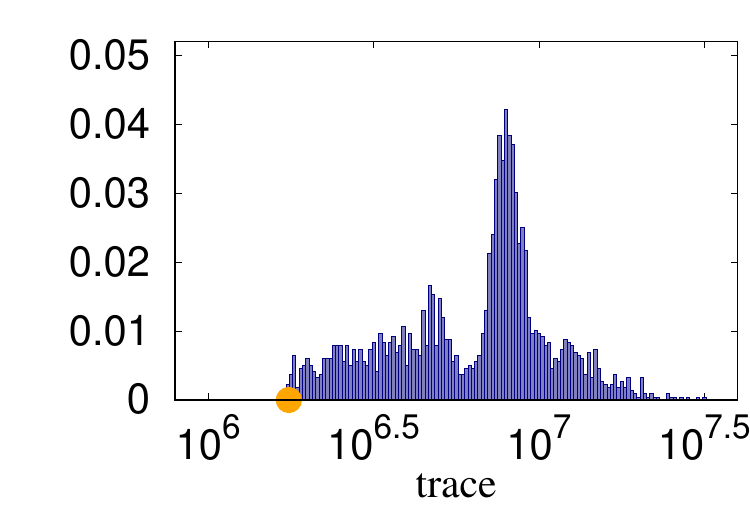}
			}\vspace{-0cm}
		\end{tabular}
	\end{center}
	\vspace*{-0.5cm}
	\caption{\small Distribution of $Tr((I-P_v)^{-1})$ on different datasets (the orange circle denotes the value of $Tr((I-P_v)^{-1}$ with $v$ chosen as the highest-degree node)}
	\vspace*{-0.4cm}
	\label{fig:landmark-trace}
\end{figure}

\begin{figure}
	\vspace*{-0.5cm}
	\begin{center}
		\begin{tabular}[t]{c}
			\subfigure[{\scriptsize \emailenron, time}]{
				\includegraphics[width=0.38\columnwidth, height=2.0cm]{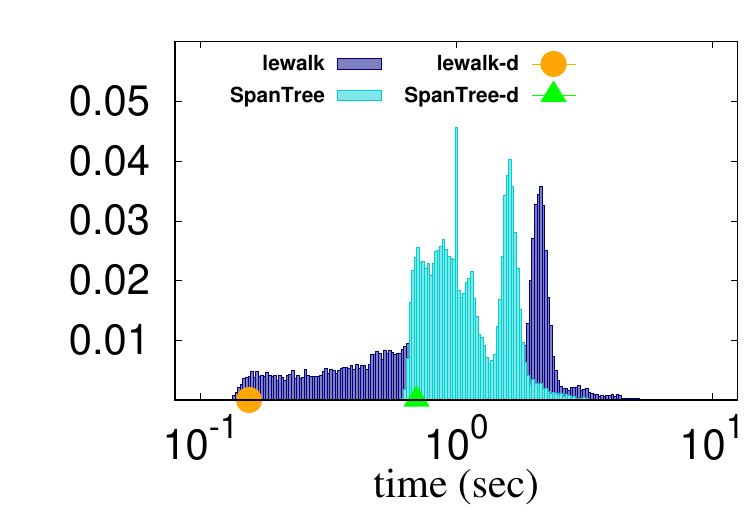}
			}
			\subfigure[{\scriptsize \emailenron, error}]{
				\includegraphics[width=0.38\columnwidth, height=2.0cm]{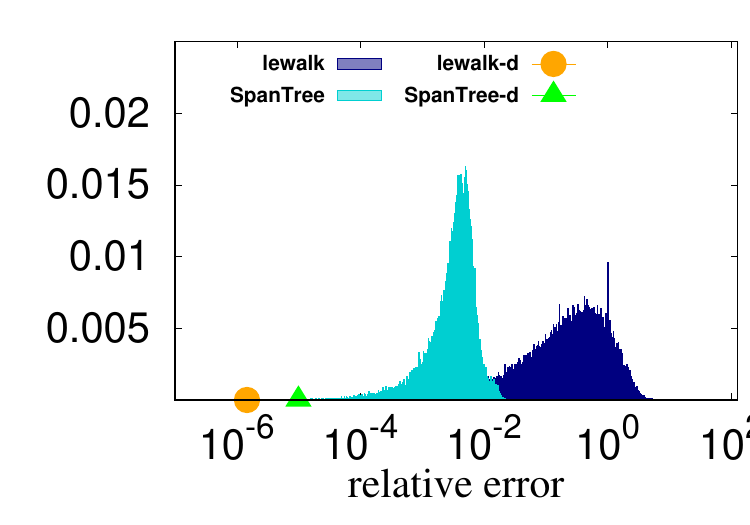}
			}\vspace{-0.2cm}\\
			\subfigure[{\scriptsize \youtube, time}]{
				\includegraphics[width=0.38\columnwidth, height=2.0cm]{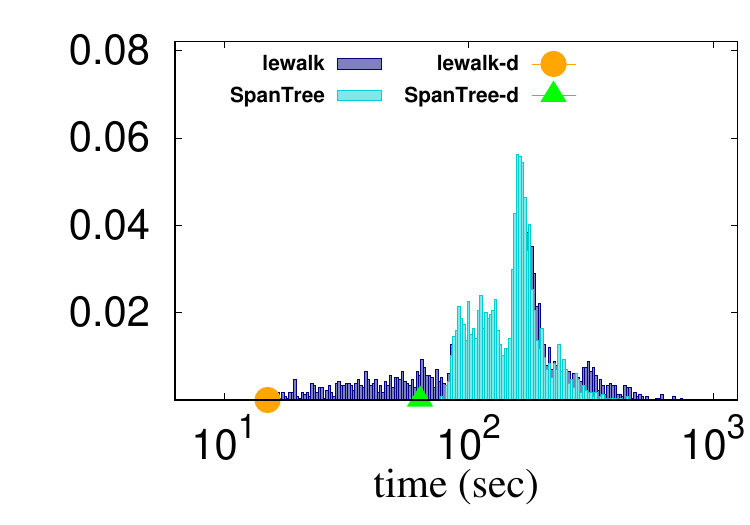}
			}
			\subfigure[{\scriptsize \youtube, error}]{
				\includegraphics[width=0.38\columnwidth, height=2.0cm]{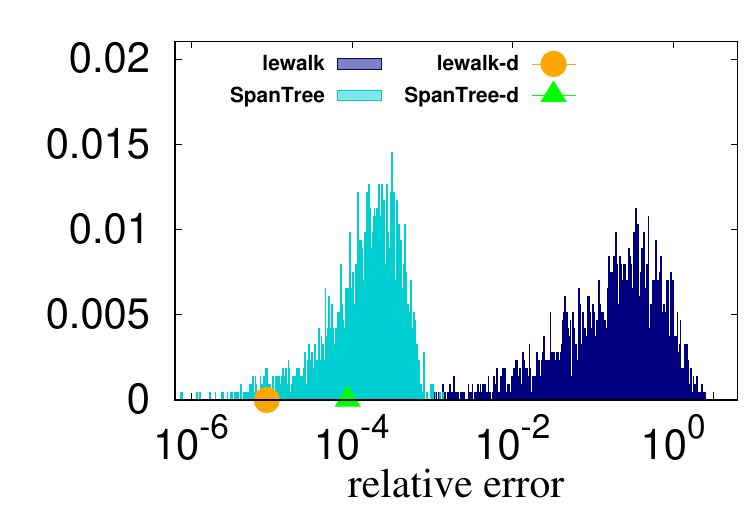}
			}
		\end{tabular}
	\end{center}
	\vspace*{-0.6cm}
	\caption{\small Distribution of performance of algorithms for different choices of landmark node (Kemeny's constant)}
	\vspace*{-0.4cm}
	\label{fig:lamark-kc}
\end{figure}

\stitle{The effect of different landmark selection methods.} In former experiments, we select the landmark node $v$ as the highest-degree node by default. In this experiment, we investigate the effect of various landmark selection strategies. We conduct experiments on both small graphs and large graphs. For small graphs, we run \lewalk and \spanningtree for \ecc approximation and \kc approximation with all nodes in graph chosen as the landmark node $v$. For large graphs, we sample $10^4$ nodes uniformly, and run \lewalk and \spanningtree with those nodes chosen as landmarks. We run all algorithms for $100$ times and report the average results. We first plot the distribution of $Tr((I-P_v)^{-1})$, since it is the expected running time of sampling a loop-erased walk $\lepath_v$, which is used in all proposed algorithms. The results are shown in Fig.~\ref{fig:landmark-trace}. We show the results of a small graph \emailenron and a large graph \youtube, and the results on the other datasets are consistent. It can be seen that $Tr((I-P_v)^{-1})$ varies when $v$ changes, the highest-degree node has (nearly) smallest $Tr((I-P_v)^{-1})$ value. Then, we plot the runtime and error (relative error) distribution with different $v$. The results of \kc computation are shown Fig.~\ref{fig:lamark-kc}, and the results of \ecc computation are similar (details can be found in \cite{fullversion}). Again, we can clearly see that the highest-degree node is almost the best choice for achieving short time and low error. The reason is that the highest-degree node is often easy-to-hit, thus the length of the loop-included path in $\lepath_v$ may be short, leading to both low traversal cost and small variance. These results confirm our theoretical analysis in Section~\ref{sec:new-results}.

\comment{Specifically, we find that if $v$ is not chosen as the highest-degree node, \lewalk may preforms poorly by both the runtime and estimation error. This is because when $v$ is not easy-to-hit, the length of the loop-included path in $\lepath_v$ may be very long, which leads to both high traversal cost and large variance. This result confirms our theoretical analysis in Section~\ref{sec:new-results}.
}

\comment{
	\begin{figure}
		\vspace*{-0.5cm}
		\begin{center}
			\begin{tabular}[t]{c}
				\subfigure[{\scriptsize \emailenron, time}]{
					\includegraphics[width=0.38\columnwidth, height=2.2cm]{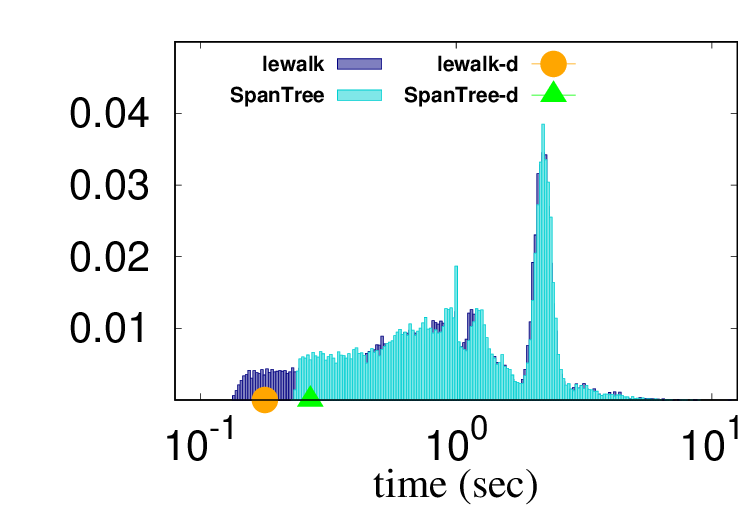}
				}
				\subfigure[{\scriptsize \emailenron, error}]{
					\includegraphics[width=0.38\columnwidth, height=2.2cm]{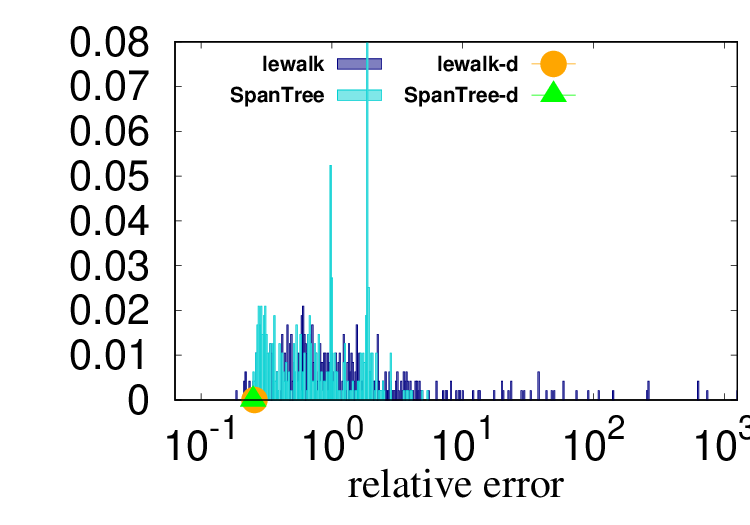}
				}\vspace{-0.2cm}\\
				\subfigure[{\scriptsize \youtube, time}]{
					\includegraphics[width=0.38\columnwidth, height=2.2cm]{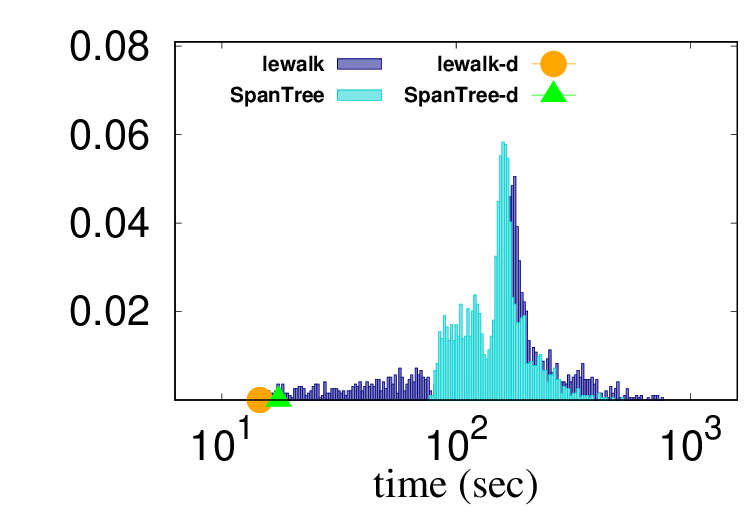}
				}
				\subfigure[{\scriptsize \youtube, error}]{
					\includegraphics[width=0.38\columnwidth, height=2.2cm]{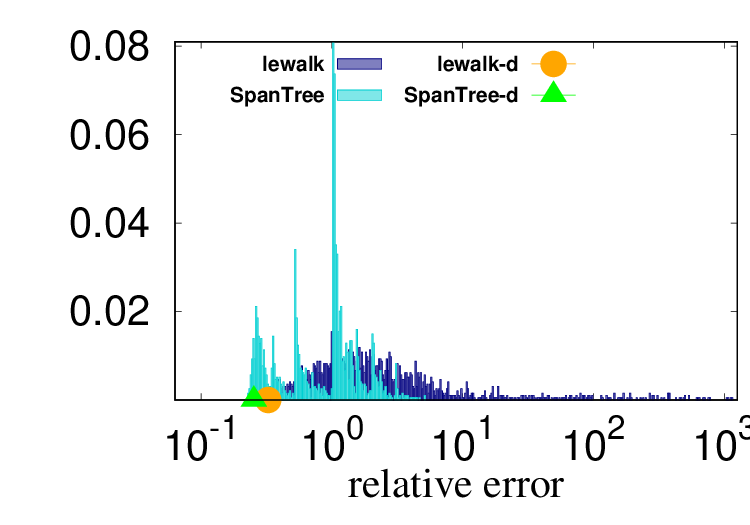}
				}
			\end{tabular}
		\end{center}
		\vspace*{-0.5cm}
		\caption{Distribution of performance of algorithms for different choices of landmark node (electrical closeness centrality)}
		\vspace*{-0.2cm}
		\label{fig:landmark-ecc}
\end{figure}}
\subsection{Case studies}\label{subsec:exp-casestudy}
\comment{\begin{table*}[t!]\vspace*{0cm}
	\vspace*{-0.1cm}
	\scriptsize
	\centering
	\caption{Top-$10$ centrality results on \dblp and \wordnet} \label{tab:casestudy}
	\vspace{-0.3cm}
	\begin{tabular}{|c|ccccc|ccccc|}
		\hline
		& \multicolumn{5}{c}{\dblp} & \multicolumn{5}{|c|}{\wordnet} \cr\hline
		\bf Rank & \degree & \pagerank & \bc & \cc & \ecc &  \degree & \pagerank & \bc & \cc & \ecc \cr\hline
		1  & Jiawei Han           & Jiawei Han           & Philip S. Yu        & Jiawei Han          & Jiawei Han          & FOOD   & FOOD   & MONEY  & MONEY  & FOOD   \\
		2  & Philip S. Yu         & Philip S. Yu         & Jiawei Han          & Philip S. Yu        & Philip S. Yu        & MONEY  & MONEY  & FOOD   & GOOD   & MONEY  \\
		3  & Christos Faloutsos   & Christos Faloutsos   & Christos Faloutsos  & Beng Chin Ooi       & Christos Faloutsos  & WATER  & WATER  & WATER  & CAR    & WATER  \\
		4  & Michael J. Franklin  & Jian Pei             & Gerhard Weikum      & Christos Faloutsos  & Michael J. Franklin & CAR    & CAR    & GOOD   & FOOD   & CAR    \\
		5  & Jian Pei             & Gerhard Weikum       & Beng Chin Ooi       & Gerhard Weikum      & Raghu Ramakrishnan  & GOOD   & GOOD   & CAR    & WORK   & GOOD   \\
		6  & Gerhard Weikum       & Michael Stonebraker  & Jian Pei            & Michael J. Franklin & Beng Chin Ooi       & BAD    & BAD    & BAD    & BAD    & BAD    \\
		7  & Michael Stonebraker  & Hector Garcia-Molina & Michael J. Franklin & H. V. Jagadish      & Jian Pei            & WORK   & WORK   & WORK   & SCHOOL & WORK   \\
		8  & Raghu Ramakrishnan   & Michael J. Franklin  & Raghu Ramakrishnan  & Divesh Srivastava   & Gerhard Weikum      & SCHOOL & HOUSE  & HOUSE  & HOUSE  & SCHOOL \\
		9  & Beng Chin Ooi        & Beng Chin Ooi        & Haixun Wang         & Kenneth A. Ross     & Jeffrey F. Naughton & HOUSE  & SCHOOL & MAN    & WATER  & HOUSE  \\
		10 & Hector Garcia-Molina & Haixun Wang          & H. V. Jagadish      & Raghu Ramakrishnan  & Michael Stonebraker & LOVE   & LOVE   & SCHOOL & MAN    & LOVE \\
		\hline
	\end{tabular}
	\vspace*{-0.2cm}
\end{table*}
\begin{table*}[t!]\vspace*{0cm}
	\vspace*{-0.1cm}
	\scriptsize
	\centering
	\caption{Top-$10$ centrality results on \dblp and \wordnet after removing $30\%$ edges} \label{tab:casestudy-remove}
	\vspace{-0.3cm}
	\begin{tabular}{|c|ccccc|ccccc|}
		\hline
		& \multicolumn{5}{c}{\dblp} & \multicolumn{5}{|c|}{\wordnet} \cr\hline
		\bf Rank & \degree & \pagerank & \bc & \cc & \ecc &  \degree & \pagerank & \bc & \cc & \ecc \cr\hline
		1  & Christos Faloutsos   & Christos Faloutsos    & Christos Faloutsos  & Divesh Srivastava   & Christos Faloutsos  & FOOD  & FOOD  & FOOD  & BIG   & FOOD  \\
		2  & Jiawei Han           & Philip S. Yu          & Philip S. Yu        & Beng Chin Ooi       & Jiawei Han          & MONEY & MONEY & MONEY & UP    & MONEY \\
		3  & Philip S. Yu         & Jiawei Han            & Jiawei Han          & Michael J. Franklin & Philip S. Yu        & WATER & WATER & CAR   & CAR   & WATER \\
		4  & Michael J. Franklin  & Jian Pei              & Jian Pei            & Christos Faloutsos  & Michael J. Franklin & CAR   & CAR   & WATER & BLACK & CAR   \\
		5  & Gerhard Weikum       & Gerhard Weikum        & Gerhard Weikum      & Gerhard Weikum      & Raghu Ramakrishnan  & GOOD  & GOOD  & GOOD  & TIME  & GOOD  \\
		6  & Jian Pei             & Michael Stonebraker   & Michael J. Franklin & Raghu Ramakrishnan  & Jian Pei            & BAD   & HOT   & TIME  & DOG   & BAD   \\
		7  & Michael Stonebraker  & W. Bruce Croft        & ChengXiang Zhai     & H. V. Jagadish      & Gerhard Weikum      & HOT   & GREEN & BIG   & LIFE  & HOT   \\
		8  & Raghu Ramakrishnan   & Michael J. Franklin   & Beng Chin Ooi       & Kenneth A. Ross     & Beng Chin Ooi       & GREEN & BAD   & DOG   & MONEY & MEAN  \\
		9  & Beng Chin Ooi        & Elke A. Rundensteiner & Raghu Ramakrishnan  & AnHai Doan          & Surajit Chaudhuri   & GIRL  & GIRL  & BLACK & WATER & GREEN \\
		10 & Hector Garcia-Molina & Hector Garcia-Molina  & Charu C. Aggarwal   & Jian Pei            & Michael Stonebraker & MEAN  & TIME  & BAD   & MAN   & GIRL \\
		\hline
	\end{tabular}
	\vspace*{-0.2cm}
\end{table*}}
\begin{table*}[t!]\vspace*{0cm}
	\vspace*{0cm}
	\scriptsize
	\centering
	\caption{\small Top-$10$ centrality results on \dblp (the original graph and the graph obtained by removing 30\% edges)} \label{tab:casestudy-dblp}
	\vspace{-0.3cm}
	\scalebox{0.77}{
	\begin{tabular}{|c|ccccc|ccccc|}
		\hline
		& \multicolumn{5}{c}{the original graph} & \multicolumn{5}{|c|}{the graph obtained by removing 30\% edges} \cr\hline
		\bf Rank & \degree & \pagerank & \bc & \cc & \ecc &  \degree & \pagerank & \bc & \cc & \ecc \cr\hline
		1  & Jiawei Han           & Jiawei Han           & Philip S. Yu        & Jiawei Han          & Jiawei Han          & Christos Faloutsos   & Christos Faloutsos    & Christos Faloutsos  & Divesh Srivastava   & Christos Faloutsos  \\
		2  & Philip S. Yu         & Philip S. Yu         & Jiawei Han          & Philip S. Yu        & Philip S. Yu        & Jiawei Han           & Philip S. Yu          & Philip S. Yu        & Beng Chin Ooi       & Jiawei Han          \\
		3  & Christos Faloutsos   & Christos Faloutsos   & Christos Faloutsos  & Beng Chin Ooi       & Christos Faloutsos  & Philip S. Yu         & Jiawei Han            & Jiawei Han          & Michael J. Franklin & Philip S. Yu        \\
		4  & Michael J. Franklin  & Jian Pei             & Gerhard Weikum      & Christos Faloutsos  & Michael J. Franklin & Michael J. Franklin  & Jian Pei              & Jian Pei            & Christos Faloutsos  & Michael J. Franklin \\
		5  & Jian Pei             & Gerhard Weikum       & Beng Chin Ooi       & Gerhard Weikum      & Raghu Ramakrishnan  & Gerhard Weikum       & Gerhard Weikum        & Gerhard Weikum      & Gerhard Weikum      & Raghu Ramakrishnan  \\
		6  & Gerhard Weikum       & Michael Stonebraker  & Jian Pei            & Michael J. Franklin & Beng Chin Ooi       & Jian Pei             & Michael Stonebraker   & Michael J. Franklin & Raghu Ramakrishnan  & Jian Pei            \\
		7  & Michael Stonebraker  & Hector Garcia-Molina & Michael J. Franklin & H. V. Jagadish      & Jian Pei            & Michael Stonebraker  & W. Bruce Croft        & ChengXiang Zhai     & H. V. Jagadish      & Gerhard Weikum      \\
		8  & Raghu Ramakrishnan   & Michael J. Franklin  & Raghu Ramakrishnan  & Divesh Srivastava   & Gerhard Weikum      & Raghu Ramakrishnan   & Michael J. Franklin   & Beng Chin Ooi       & Kenneth A. Ross     & Beng Chin Ooi       \\
		9  & Beng Chin Ooi        & Beng Chin Ooi        & Haixun Wang         & Kenneth A. Ross     & Jeffrey F. Naughton & Beng Chin Ooi        & Elke A. Rundensteiner & Raghu Ramakrishnan  & AnHai Doan          & Surajit Chaudhuri   \\
		10 & Hector Garcia-Molina & Haixun Wang          & H. V. Jagadish      & Raghu Ramakrishnan  & Michael Stonebraker & Hector Garcia-Molina & Hector Garcia-Molina  & Charu C. Aggarwal   & Jian Pei      & Michael Stonebraker \\
		\hline
	\end{tabular}}
	\vspace*{-0.2cm}
\end{table*}
In this subsection, we conduct case studies to show the applications of \ecc as a graph centrality metric, and \kc as a graph invariant.

\stitle{Comparison of different centrality measures on \dblp.} We first compare different centrality measures on one real-life dataset \dblp. \dblp is a collaboration network for researchers in database and data mining with $37,177$ nodes and $131,715$ edges, where each node represents an author, each edge represents a collaboration relationship. We compare four commonly used centrality measures, degree centrality, PageRank centrality, betweenness centrality (\bc) \cite{core20} and closeness centrality (\cc) \cite{closeness08} with the electrical closeness centrality (\ecc). For PageRank centrality, we set $\alpha=0.15$, which is the commonly-used choice \cite{15pagerankbeyond}. The top-$10$ centrality nodes in \dblp measured by these centralities are shown in Table~\ref{tab:casestudy-dblp}. As can be seen, all the top-$10$ results are well-known authors. The results of \ecc are close to that of other widely-used centrality measures. This result verifies that \ecc is indeed a useful centrality metric. Furthermore, we also rank the centrality values after randomly removing $30\%$ edges from the original graphs. The results are also shown in Table~\ref{tab:casestudy-dblp}. As discussed before, \ecc considers all paths between nodes, while both \cc and \bc only consider the shortest paths. Thus, \ecc is often more robust with respect to edge perturbations. We can observe in Table~\ref{tab:casestudy-dblp} that after removing a fraction of edges, there are more drastic changes in the rankings of \bc and \cc compared to \ecc. These results suggest that \ecc is an effective and robust metric for detecting the important nodes in real-life networks. We also conduct a case study on another real-life dataset \wordnet, and the results are consistent. The details can be found in the full version of this paper \cite{fullversion}.
\comment{
\begin{figure}
	\vspace*{-0.3cm}
	\begin{center}
			\includegraphics[width=0.85\columnwidth, height=2.6cm]{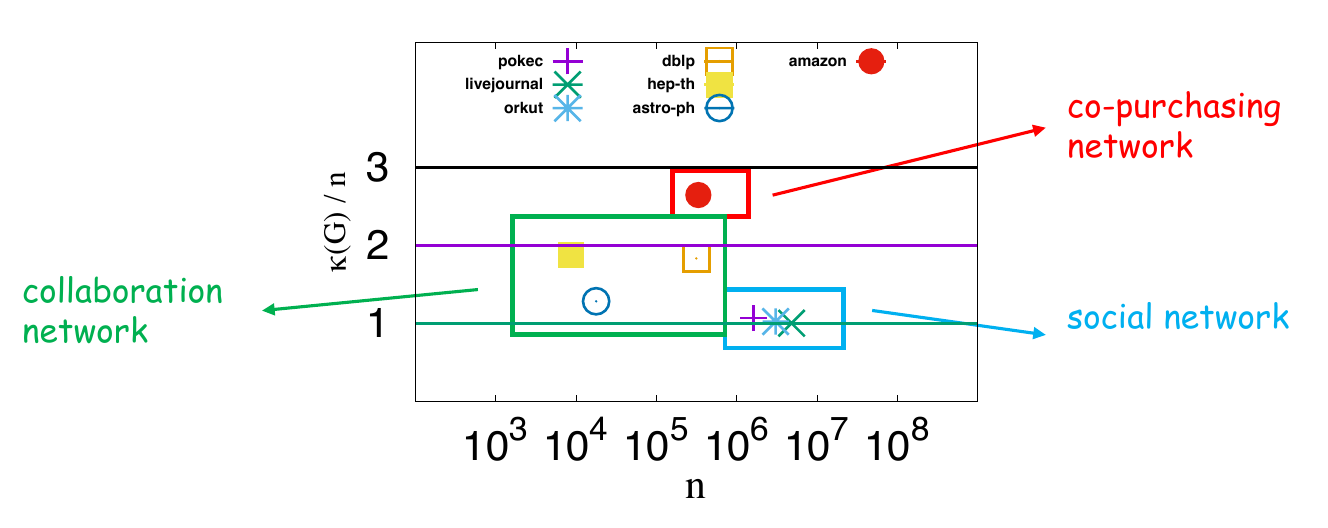}
	\end{center}
	\vspace*{-0.5cm}
	\caption{Kemeny's constant for various networks}
	\vspace*{-0.4cm}
	\label{fig:case-study}
\end{figure}}
\begin{figure}
	\vspace*{-0.2cm}
	\begin{center}
		\begin{tabular}[t]{c}\hspace{-0.12\columnwidth}
			\subfigure[{\scriptsize real-world datasets}]{
				\includegraphics[width=0.66\columnwidth, height=2.0cm]{exp/exp5/casestudy.pdf}
			}\hspace{-0.02\columnwidth}
			\subfigure[{\scriptsize Barabasi-Albert graphs}]{
				\includegraphics[width=0.36\columnwidth, height=2.0cm]{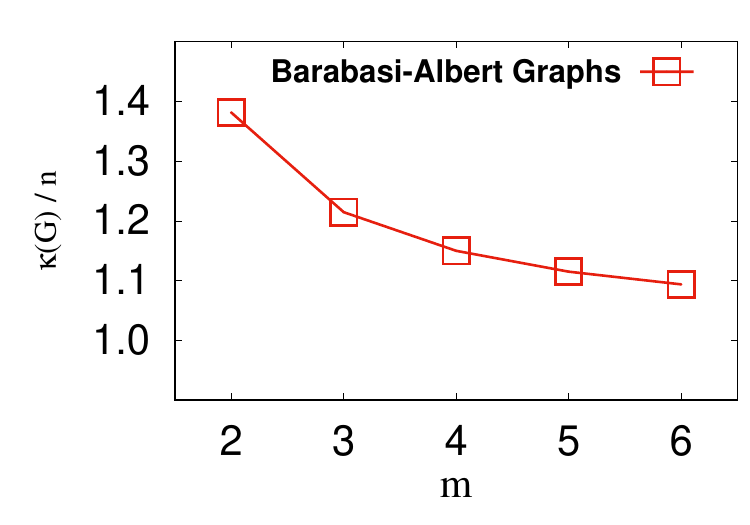}
			}\vspace{-0.2cm}\\
			\subfigure[{\scriptsize Erdos-Renyi graphs}]{
				\includegraphics[width=0.36\columnwidth, height=2.0cm]{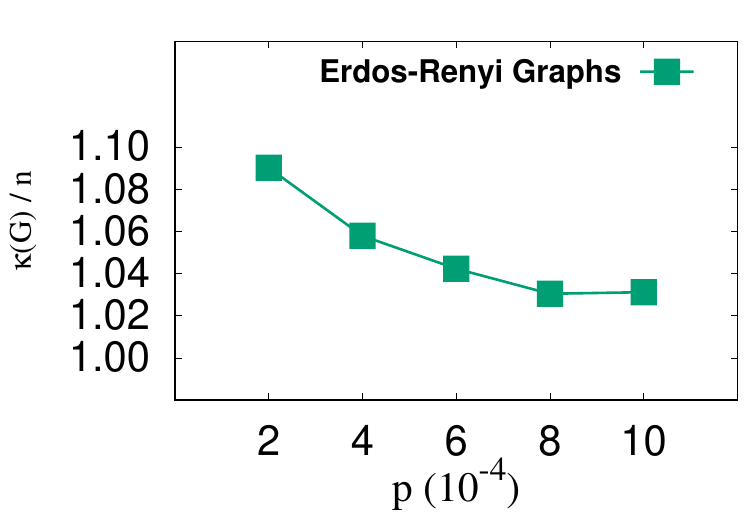}
			}\hspace{0.16\columnwidth}
			\subfigure[{\scriptsize Hyperbolic graphs}]{
				\includegraphics[width=0.36\columnwidth, height=2.0cm]{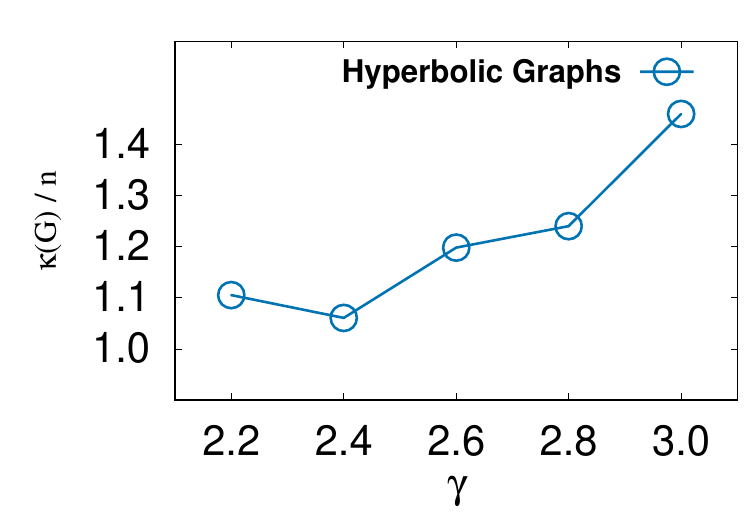}
			}
		\end{tabular}
	\end{center}
	\vspace*{-0.5cm}
	\caption{\small Kemeny's constant for various networks}
	\vspace*{-0.6cm}
	\label{fig:case-study}
\end{figure}

\stitle{Kemeny's constant for real-life networks.} We compute \kc for $7$ real-life networks. All the datasets are available at \cite{snapnets}. \comment{The detailed statistics of the datasets can be found in the full version of the paper.} In summary, they can be divided into $3$ categories: social networks, collaboration networks and co-purchasing networks. We plot the results with node number $n$ on the $x$ axis, and $\frac{\kappa(G)}{n}$ on the $y$ axis. The results are shown in Fig.~\ref{fig:case-study}(a). We can clearly see that the social network \pokec, \livejournal, \orkut has $\kappa(G)$ very close to the node number $n$, while the $3$ citation networks \dblp, \hepth and \astroph have relatively larger $\kappa(G)$. The co-purchasing network \amazon has the largest $\frac{\kappa(G)}{n}$ value. It suggests that social network connects more closely compared to collaboration networks and co-purchasing networks. However, all these real-life networks have $O(n)$ Kemeny constants. In \cite{KDD21Kemeny}, the authors show that real-world power-law graphs often has a \kc value close to the optimal value $1+\frac{(n-1)^2}{n}$, which is also the \kc value for complete graphs. This result implies that Kemeny's constant can reveal the connectivity property of a network, and real-life networks often have a small and near-optimal Kemeny's constant.

\stitle{Kemeny's constant for synthetic networks.} We also compute \kc for synthetic networks. There are many models to generate networks that are similar to real-world graphs. We select three popular models: the Barabasi-Albert graph (\ba), the Erdos-Renyi graph (\er) and the hyperbolic random graph (\hy). Specifically, \ba model generates networks based on a small network with $m_0\geq m$ edges. For each new coming node, it add $m$ new edges based on the degree of the current nodes. \ba model generates graphs with a power-law distribution. \er model generates the famous $G(n,p)$ Erdos-Renyi graphs with $n$ nodes and $p$ is the probability that there exists an edge between any pair of nodes. The \hy model generates points in hyperbolic space and add edges between them according to the distance between them. The generated graphs have power-law degree distribution, small diameter and high clustering coefficient which is similar to real-world graphs \cite{hyperbolic12}. There is one parameter to control the power-law degree $\gamma$, and the other parameter average degree $\Bar{d}$, which we set as $40$ by default. We generate $100$ networks with $10^6$ nodes, and vary $m$ in \ba as $m=\{2,3,4,5,6\}$; $p$ in \er as $p=\{2,4,6,8,10\}\times10^{-4}$ and $\gamma$ as $\gamma=\{2.2,2.4,2.6,2.8.3.0\}$. The results are shown in Fig.~\ref{fig:case-study}(b)-(d). We plot the parameter on the x-axis, and $\frac{\kappa(G)}{n}$ on the y-axis. As can be seen, as $m$ grows larger (as $p$ becomes smaller), $\frac{\kappa(G)}{n}$ decreases. Also, as $\gamma$ becomes larger, $\frac{\kappa(G)}{n}$ becomes larger. However, the \kc value is always very close to the node number $n$. These findings can help us design and generate networks, which can be applied in robot surveillance \cite{20robot} and web search \cite{02surfer} where networks with low Kemeny's constants (with low communication / navigation cost) are desired. This result further confirms that real-world graphs often have a near optimal \kc value.

\section{Further Related Work}\label{sec:related-work}
\stitle{Spanning tree sampling.} Spanning tree sampling is a powerful tool for efficient approximation of graph-related quantities. It is first used by Hayashi et al. \cite{SpanningEdgeCentrality} for approximating spanning edge centrality, which is $(e_s-e_t)^TL^\dagger(e_s-e_t)$ (also $(L_s^{-1})_{tt}$) if there is an edge between $s$ and $t$. Angriman et al. \cite{angriman2020approximation,22resistance} generalized the technique to approximate resistance distance $(e_s-e_t)^TL^\dagger(e_s-e_t)$ (also $(L_s^{-1})_{tt}$) for arbitrary $s$ and $t$. However, how to use spanning tree sampling to approximate $L^\dagger$ or the arbitrary elements of $L_v^{-1}$ still remains unresolved. Yusuf et al. \cite{GraphSignal21, RSF20,inversetrace19} used spanning forests sampling for graph signal processing tasks. Their objective is to approximate $(L+qI)^{-1}qI$ and $Tr((L+qI)^{-1}qI)$. Recently, Liao et al. \cite{SpanningForestPageRank} applied spanning forests sampling to approximate personalized PageRank, which is $(L+\beta D)^{-1}\beta D$. These studies can be regarded as a special case of our techniques for $L_v^{-1}$, because the matrix $L+qI$ and $L+\beta D$ are submatrix of the Laplacian of an extended graph with an additional dummy node. \comment{Efficient sampling a spanning tree is also a well-studied problem in theoretical computer science. Aldous-Broder algorithm \cite{aldous1990random} simulates random walks from a node until all nodes are covered. Russo et al. propose an algorithm \cite{russo2018linking} based on Makrov chain built over all spanning trees, but the algorithm requires complex data structure such as link-cut trees. For theoretical results, the algorithm \cite{schild2018almost} by Schild is the best with time complexity $O(m^{1+o(1)})$ . For real-life large datasets, the Wilson algorithm that utilizing loop-erased walk sampling \cite{wilson1996generating} is the most efficient algorithm, and it is chosen for spanning tree sampling algorithms in previous work \cite{SpanningEdgeCentrality, angriman2020approximation, 22resistance}. The time complexity of the Wilson algorithm is $Tr((I-P_v)^{-1})$ \cite{HamiltonCycle}.}

\stitle{Loop-erased walk sampling.} Compared to spanning tree sampling, loop-erased walk sampling itself is less studied. Wilson et al. \cite{wilson1996generating} proved that the expected running time of the Wilson algorithm with root $v$ is $\sum_{u\in V}(\bm{\pi}(u)h(u,v)+h(v,u)\bm{\pi}(u))$. Marchal et al. \cite{HamiltonCycle} further proved that the time complexity can be formulated as $Tr((I-P_v)^{-1})$. Recently, Liao et al. \cite{22resistance} utilized loop-erased walk to approximating single source resistance distance, which is $(L_v^{-1})_{uu}$ for $u\in V$ and $u\neq v$. However, it is still unknown how to approximate all elements of $L_v^{-1}$ and $L^\dagger$ using loop-erased walk sampling.

\stitle{Other random walk related quantities computation.} In addition to Laplacian pseudo-inverse, there also exist some other random walk based quantities, such as personalized PageRank (PPR) and resistance distance. Push-based methods \cite{FOCS06,pagerankWAW07,22resistance} and random walk sampling based methods \cite{MonteCarloOneIteration,KDDlocal21,22resistance,ResistanceYang} were developed to compute PPR and resistance distance. The advantage of such methods is that they are local, meaning that they often only need to visit a small portion of the graph instead of the entire graph. In comparison, Laplacian pseudo-inverse is a global quantity which cannot be approximated locally. Random walk sampling and push were combined to further improve the efficiency of PPR computation \cite{FORA17,SPEEDPPR21,Bipprs,PageRankVariance23} and resistance distance computation \cite{22resistance,ResistanceYang}. However, when it comes to deriving the Laplacian pseudo-inverse, it is necessary to compute the entire diagonal of $L_v^{-1}$. This requires to invoke the random walk sampling and push algorithms for $n$ times, which is clearly very expensive for large graphs. 

\section{Conclusion}\label{sec:conclusion}
In this paper, we study the problem of efficient computation of two kinds of Laplacian pseudo-inverse, with applications to electrical closeness centrality approximation and Kemeny's constant approximation. We first give new formulas to express the pseudo-inverses of two types of Laplacians, including the combinatorial Laplacian and the normalized Laplacian, by $L_v^{-1}$. Then, we propose two new combinatorial interpretations of $L_v^{-1}$ in terms of two interesting objects: spanning trees and loop-erased walks. Based on these theoretical results, we develop two novel Monte Carlo algorithms for both electrical closeness centrality and Kemeny's constant computations. The results of extensive experiments on 5 real-life datasets demonstrate that our proposed algorithms substantially outperform the state-of-the-art algorithms in terms of both runtime and accuracy. We also conduct two case studies to demonstrate the effectiveness of electrical closeness centrality and Kemeny's constant.


\balance
\bibliographystyle{ACM-Reference-Format}
\bibliography{main}
\end{document}